\documentclass[%
 reprint,
superscriptaddress,
 amsmath,amssymb,
 aps,
]{revtex4-2}

\usepackage{graphicx}% Include figure files
\usepackage{svg}

\usepackage{dcolumn}% Align table columns on decimal point
\usepackage{bm}% bold math
\usepackage{chngcntr}
\usepackage{amsmath,amsthm,amssymb,mathrsfs}
\usepackage{mathtools}
\usepackage{physics}
\usepackage{textgreek}
\usepackage{array}
\usepackage{graphicx}
\usepackage{float}
\usepackage{rotating}  %sidewaystable, sidewaysfigure
\usepackage{dcolumn}
\usepackage{tablefootnote}
\usepackage{cancel}

\usepackage{tikz}
\usepackage{pgfplots}
\pgfplotsset{compat=1.18}
\usepackage[american]{circuitikz}
\usetikzlibrary{circuits}
\usetikzlibrary{arrows.meta}
\usetikzlibrary{shapes,arrows}
\tikzset{every picture/.style={line width=1pt}}
\ctikzset{bipoles/thickness=1}

\DeclareMathAlphabet{\mathscrbf}{OMS}{mdugm}{b}{n} % simulate bold script font 

\DeclareFontFamily{U}{BOONDOX-calo}{\skewchar\font=45 }
\DeclareFontShape{U}{BOONDOX-calo}{m}{n}{
  <-> s*[1.05] BOONDOX-r-calo}{}
\DeclareFontShape{U}{BOONDOX-calo}{b}{n}{
  <-> s*[1.05] BOONDOX-b-calo}{}
\DeclareMathAlphabet{\mathcalboondox}{U}{BOONDOX-calo}{m}{n}
\SetMathAlphabet{\mathcalboondox}{bold}{U}{BOONDOX-calo}{b}{n}
\DeclareMathAlphabet{\mathbcalboondox}{U}{BOONDOX-calo}{b}{n}

\newcommand{\ci}{\mathrm{i}}

\usepackage{scalerel,stackengine}
\newcommand\Bigs[1]{\scalerel*[5.5pt]{\Big#1}{%
  \ensurestackMath{\addstackgap[1.5pt]{\big#1}}}}
\newcommand\Bigsl[1]{\mathopen{\Bigs{#1}}}
\newcommand\Bigsr[1]{\mathclose{\Bigs{#1}}}

\begin{document}

\preprint{APS/123-QED}
\title{Stochastic modeling of superconducting qudits in the dispersive regime}
% \title{Analysis of qutrit measurement:\\ Measurement-induced dephasing and quantum trajectories}% Force line breaks with \\

\author{Kangdi Yu}
    \email{kangdiyu98@ucla.edu}
\author{Murat C. Sarihan}
\author{Jin Ho Kang}
\author{Madeline Taylor}
\author{Cody S. Fan}
\author{Ananyo Banerjee}
\affiliation{Department of Electrical and Computer Engineering, University of California, Los Angeles, California 90095, USA}%

\author{Jonathan L. DuBois}
\author{Yaniv J. Rosen}
\affiliation{Lawrence Livermore National Laboratory, Livermore, California 94550, USA}%

\author{Chee Wei Wong}
    \email{cheewei.wong@ucla.edu}
\affiliation{Department of Electrical and Computer Engineering, University of California, Los Angeles, California 90095, USA}

% \affiliation{%
%  Authors' institution and/or address\\
%  This line break forced with \textbackslash\textbackslash
% }%

% \collaboration{MUSO Collaboration}%\noaffiliation

%  \homepage{http://www.Second.institution.edu/~Charlie.Author}
% \affiliation{
%  Second institution and/or address\\
%  This line break forced% with \\
% }%
% \affiliation{
%  Third institution, the second for Charlie Author
% \affiliation{%
%  Authors' institution and/or address\\
%  This line break forced with \textbackslash\textbackslash
% }%

% \collaboration{CLEO Collaboration}%\noaffiliation

\date{\today}% It is always \today, today,

\begin{abstract}
The field of superconducting quantum computing, based on Josephson junctions, has recently seen remarkable strides in scaling the number of logical qubits. In particular, the fidelities of one- and two-qubit gates have reached the breakeven point with the novel error mitigation and correction methods. Parallel to these advances is the effort to expand the Hilbert space within a single junction or device by employing high-dimensional qubits, otherwise known as qudits. Research has demonstrated the possibility of driving higher-order transitions in a transmon or designing innovative multimode superconducting circuits, termed multimons. These advances can significantly expand the computational basis while simplifying the interconnects in a large-scale quantum processor. In this work we extend the measurement theory of a conventional superconducting qubit to that of a qudit, focusing on modeling the dispersive quadrature measurement in an open quantum system. Under the Markov assumption, the qudit Lindblad and stochastic master equations are formulated and analyzed; in addition, both the ensemble-averaged and the quantum-jump approach of decoherence analysis are detailed with analytical and numerical comparisons. We verify our stochastic model with a series of experimental results on a transmon-type qutrit, demonstrating the validity
of our high-dimensional formalism.

% \begin{description}
% Secondary publications and information retrieval purposes.
% You may use the \texttt{description} environment to structure your abstract;
% use the optional argument of the \verb+\item+ command to give the category of each item. 
% \end{description}
\end{abstract}

\maketitle

%\tableofcontents
\section{\label{sec:1}Introduction}
Superconducting quantum computation based on single- and two-qubit gates \cite{nature08121, PhysRevA.93.060302, PhysRevLett.127.130501} has recently demonstrated milestone successes in comparison to classical computation \cite{doi:10.1146/annurev-conmatphys-031119-050605, s41586_023_06096_3}. While the error correction scheme can overcome the noise in the qubit-qubit interaction and correct the undesired decoherence due to coupling to the environment, the required hardware resources, chip footprint, and peripheral connections will soon face scalability issues. To create a more hardware-efficient quantum process with fewer quantum gates and interconnects, an emerging effort has been placed into developing high-dimensional superconducting quantum computing. In particular, the hope is that both the size of the quantum algorithm and the error rate induced in a long gate sequence can be reduced by using the naturally available higher energy levels of the localized artificial atoms, also termed qudits.

Experiments involving the usage of higher energy levels already in the transmon qubits have been examined \cite{PhysRevLett.125.170502, PhysRevLett.114.010501}. Furthermore, a more complex Josephson-junction-based network with multiple nonlinear modes coupled longitudinally \cite{PhysRevApplied.7.054025, PhysRevA.98.052318} can also be developed. Despite the recent experimental efforts on higher dimensional quantum computation, the theory of dispersive measurement used in inferring a qudit state under sources of decoherence still mostly relies on a heuristic extension from the qubit formalism. In this work, we adopt both Lindblad master equations and the method of quantum trajectories to analyze the state evolution of a qudit when measured dispersively.

In circuit or cavity quantum electrodynamics (QED), continuous measurement and other unwanted coupling to the environment can be modeled by master equations of the Lindblad form under the Born-Markov assumption. Analysis of the qubit-resonator coupling in the dispersive regime has relied on this formalism, aided by the good agreement between measurement and the predicted measurement-induced dephasing and number splitting \cite{PhysRevA.74.042318, PhysRevA.69.062320}. Our work first generalizes the measurement of the observable $\hat{\sigma}_z$ in the qubit case to a generic measurement in the longitudinal direction of the qudit, thus extending the notion of measurement-induced dephasing. Due to the longitudinal coupling appearing in the dispersive regime, we show analytically that each energy eigenstate of the qudit is entangled with a coherent state of the resonator; consequently, measurement of the quadrature fields of the resonator can be used to infer the qudit state.

However, since the qudit Lindblad master equation only describes the ensemble-averaged time evolution of the state, it does not make use of the information leaking out of the system, part of which is, of course, what one measures in real experiments. To capture the update of an observer's knowledge during a continuous measurement, in this work we subsequently apply the quantum trajectory theory \cite{PhysRevA.77.012112, gardiner2004quantum, WISEMAN200191, PhysRevA.47.642} in which our knowledge of the quantum state can be modified based on the measurement record. Given that the measurement outcomes are stochastic as required by quantum mechanics, the entire measurement record forms a random process that can be modeled as a diffusive process when the measurement is continuous and weak. In particular, under the diffusive limit, we derive an effective heterodyne stochastic master equation (SME) of the qudit subspace (in the It\^{o} sense) which is then compared with the unconditioned master equation after taking the ensemble average. Recently, the stochastic nature of qudit dispersive measurement has been theoretically studied in the absence of other decoherence processes such as qudit $T_1$-decay and dephasing, and specific examples such as the $N$-level ``clock'' system were examined \cite{PhysRevA.105.052229}. Our approach relaxes the assumptions to include other experimentally observed decoherence processes in a transmon qudit and makes an intuitive link between the stochastic trajectories solved from SME and the averaged dynamics given by the Lindblad master equation.  

In this work, we first review the dispersive coupling of a high-dimensional system to a linear resonator in Section \ref{sec:2}. In Section \ref{sec:3}, the (unconditioned) Lindblad master equation of the combined system (i.e., a qudit plus a resonator) is solved analytically using two methods -- the positive $P$-representation and the qudit-state-dependent displacement operator. Consequently, an effective master equation for the qudit with the resonator degrees of freedom traced out can be fortunately formulated as a Markovian system when the measurement-induced frequency shifts are negligible. Then, in Section \ref{sec:4}, we turn to the conditioned qudit state and derive an effective qudit SME in the diffusive limit, with the measurement outcomes described by two stochastic differential equations. Finally, in Section \ref{sec:5}, the simulated quantum trajectories are compared with our experiments on a transmon qutrit (i.e., a three-level system) coupled to a 3D cavity, demonstrating a remarkable match between our formalism and experiments.

\section{\label{sec:2}\MakeLowercase{c}QED for Dispersive Measurement of a Qudit}
For a general analysis, we consider a weakly anharmonic oscillator coupled to a linear resonator for readout and control. For future reference, we denote the anharmonic oscillator, also referred to as the system or qudit, by $\mathcal{S}$ and the resonator by $\mathcal{R}$. Hence, the combined system lives in the Hilbert space $\mathscr{H}_{\mathcal{SR}} = \mathscr{H}_{\mathcal{S}} \otimes \mathscr{H}_{\mathcal{R}}$. The anharmonic oscillator model can be used to describe a transmon-type superconducting qudit while the resonator, modeled by a harmonic oscillator, can be realized as a 3D cavity or a planar transmission-line resonator \cite{PRXQuantum.2.040202, nature02851, doi:10.1063/1.3010859}. On the one hand, the size of a typical 3D cavity, and thus the associated mode volume, is much larger than that of a superconducting qudit, allowing one to apply the dipole approximation to describe the qudit-resonator interaction. On the other hand, even though a planar resonator has a more confined mode profile and is coupled to the qubit locally (e.g., capacitive coupling), one can still derive a similar dipole interaction at the circuit level. Consequently, for both the 3D and 2D resonators, the Hamiltonian of the combined system under the rotating-frame approximation (RWA) is given by \cite{PhysRevA.76.042319}
\begin{multline} \label{eq:transmon_res_hamiltonian}
    \hat{H}/\hbar 
    = \omega_{\text{q}} \hat{a}_{\text{q}}^{\dagger} \hat{a}_{\text{q}}
        + \frac{\alpha_{\text{q}}}{2} 
            \hat{a}_{\text{q}}^{\dagger} \hat{a}_{\text{q}}^{\dagger} 
            \hat{a}_{\text{q}} 
            \hat{a}_{\text{q}}
\\
        + \omega_{\text{r}} \! 
            \left( 
                \hat{a}_{\text{r}}^{\dagger} \hat{a}_{\text{r}} 
                + \frac{1}{2} 
            \right)
        - \Big( 
                g \hat{a}_{\text{r}} 
                    \hat{a}_{\text{q}}^{\dagger} 
                + g^* \hat{a}_{\text{r}}^{\dagger} 
                    \hat{a}_{\text{q}}     
            \Big),
\end{multline}
where $\omega_{\text{q}}$ is the qubit frequency (i.e., the transition frequency between the ground and first excited states), $\alpha_{\text{q}}$ the anharmonicity of the qudit, $\omega_{\text{r}}$ the resonator frequency, and $g$ the qudit-resonator coupling coefficient. We also define $\Delta_{\text{qr}} = \omega_{q} - \omega_{r}$ as the detuning between the qudit and the resonator.

In Eq.(\ref{eq:transmon_res_hamiltonian}), we have assumed the fourth-order expansion of the transmon Hamiltonian, which ignores the fact that a transmon can only support a finite number of bound energy eigenstates. More generally, we can replace the fourth-order expansion with the Hamiltonian of a general qudit such that 
\begin{multline} \label{eq:qudit_res_hamiltonian}
    \hat{H}/\hbar 
    = \sum_{j = 0}^{D-1} \omega_{j} \ket{j} \! \bra{j}
        + \omega_{\text{r}} \! 
            \left( 
                \hat{a}_{\text{r}}^{\dagger} \hat{a}_{\text{r}} 
                + \frac{1}{2} 
            \right)
\\
        - \sum_{j,k=0}^{D-1}\Big( 
                g_{jk} 
                    \ket{j}\!\bra{k}
                    \hat{a}_{\text{q}}^{\dagger} 
                + g_{jk} \ket{k}\!\bra{j}
                    \hat{a}_{\text{q}}     
            \Big)
\end{multline}
with $\omega_{j}$ and $\ket{j}$ representing the energy and state vector of the $j$th energy level of the qudit. The weakly anharmonic model corresponds to the case where
\begin{equation} \label{eq:weakly_anharmonic_coupling_coefficient_multimon}
    g_{jk} 
    \approx 
    \begin{cases}
        \sqrt{j+1} \, g & \text{if } \ \  j - k = 1,  \\
        0 & \text{otherwise}, 
    \end{cases}
\end{equation}
for all $j, k=0,..., D-1$ ($g$ is the same coupling coefficient defined in Eq.(\ref{eq:transmon_res_hamiltonian})).

To perform a quantum non-demolition (QND) measurement (justified in the following text), we set the resonator frequency to be detuned from (and usually higher than) the qubit frequency. In the dispersive regime where $|g| \ll |\Delta_{\text{qr}}|$, Eq.(\ref{eq:transmon_res_hamiltonian}) can be approximated as
\begin{multline} \label{eq:disperive_Hamiltonian_main}
    \hat{H}^{\text{disp}} /\hbar 
    = \Tilde{\omega}_{\text{q}} 
            \hat{a}_{\text{q}}^{\dagger} 
            \hat{a}_{\text{q}}
        + \frac{\alpha_{\text{q}}}{2} 
            \hat{a}_{\text{q}}^{\dagger} \hat{a}_{\text{q}}^{\dagger} 
            \hat{a}_{\text{q}} 
            \hat{a}_{\text{q}}
\\
        + \omega_{\text{r}} \! 
            \left( 
                \hat{a}_{\text{r}}^{\dagger} \hat{a}_{\text{r}} 
                + \frac{1}{2} 
            \right)
        + \chi_{\text{qr}} 
            \hat{a}_{\text{q}}^{\dagger} \hat{a}_{\text{q}}
            \hat{a}_{\text{r}}^{\dagger} \hat{a}_{\text{r}},
\end{multline}
where $\Tilde{\omega}_{\text{q}} = \omega_{\text{q}} + |g|^2/\Delta_{\text{qr}}$ is the Lamb-shifted qubit frequency and
\begin{equation}
    \chi_{\text{qr}} 
    = \frac{2 \alpha_{\text{q}} |g|^2}{\Delta_{\text{qr}} (\Delta_{\text{qr}} + \alpha_{\text{q}})}
\end{equation}
is the dispersive shift, also known as the (fourth-order) cross-Kerr coefficient \cite{PhysRevA.76.042319, s41534_021_00461_8}. By lumping the last term in Eq.(\ref{eq:disperive_Hamiltonian_main}) into the resonator Hamiltonian, one observes a qudit-state-dependent shift in the resonator frequency. In particular, if the qudit is in the energy eigenstate $\ket{j}$, the resonator will experience a dispersive shift $j \chi_{\text{qr}}$. By determining this frequency shift via a resonator transmission or reflection measurement, one should be able to infer the qudit state.

Moreover, if Eq.(\ref{eq:qudit_res_hamiltonian}) is adopted instead, one finds
\begin{multline}
\label{eq:dispersive_Hamiltonian_general_qudit}
    \hat{H}^{\text{disp}} /\hbar 
    = 
        \sum_{j=0}^{D-1} 
            (\omega_{j} + \Lambda_j) 
            \ket{j}\!\bra{j}
\\
        + \omega_{\text{r}} \! 
            \left( 
                \hat{a}_{\text{r}}^{\dagger} \hat{a}_{\text{r}} 
                + \frac{1}{2} 
            \right)
        + \sum_{j=0}^{D-1} 
            \chi_{j}
            \hat{a}^{\dagger}
            \hat{a} 
            \ket{j} \! \bra{j}.
\end{multline}
to the second order in $|g_{jk}/(\omega_{j}-\omega_{k}-\omega_{\text{r}})|$ in the dispersive regime. The Lamb shift $\Lambda_j$ and the dispersive shift $\chi_j$ of the $j$th energy level of the qudit are given, respectively, by \cite{RevModPhys.93.025005}
\begin{equation} \label{eq:general_lambda_shifts}
    \Lambda_j
    = \sum_{k=0}^{D-1} \chi_{jk}
    = \sum_{k=0}^{D-1}
        \frac{|g_{jk}|^2 }{\omega_j - \omega_k - \omega_{\text{r}}},
\end{equation}
\begin{equation}
    \chi_{j}
    = \sum_{k=0}^{D-1} 
        (\chi_{jk} - \chi_{kj}),
\end{equation}
where
\begin{equation} \label{eq:general_chi_shifts}
    \chi_{jk} 
    = \frac{|g_{jk}|^2 }{\omega_j - \omega_k - \omega_{\text{r}}}.
\end{equation}

The main goal of this work is to quantify the qudit dispersive measurement in terms of the rate at which the information leaks out from the resonator. In addition, we need to answer to what extent a weak and continuous dispersive measurement is QND. The following analysis does not rely on the particular form of the dispersive shift nor on the relationships among the dispersive shifts of different energy levels; for simplicity, we will mainly use Eq.(\ref{eq:disperive_Hamiltonian_main}). Nevertheless, it should be mentioned that Eq.(\ref{eq:disperive_Hamiltonian_main}) and (\ref{eq:dispersive_Hamiltonian_general_qudit}) are valid only when the resonator photon number is low \cite{RevModPhys.93.025005}. In particular, the Schrieffer–Wolff transformation used in deriving Eq.(\ref{eq:disperive_Hamiltonian_main}) and (\ref{eq:dispersive_Hamiltonian_general_qudit}) assumes that the strength of the qudit-resonator interaction is much smaller than $\Delta_{\text{qr}}$. Due to the creation and annihilation operators that appear in the qudit-resonator coupling term, there is a $\sqrt{n_{\text{r}}}$-scaling of the interaction strength, where $n_{\text{r}}$ is the resonator photon number. Hence, we constrain to a low readout power in both the derivation and experiment.

\section{\label{sec:3}Unconditioned Master Equation}
We first consider the unconditioned master equation where information leaking out from the combined system is averaged unconditionally. Later, we derive a stochastic differential equation in which the density operator at time $t$ is conditioned on the heterodyne measurement at time $s<t$. Since only the ensemble-averaged state is examined in the unconditioned master equation, the solution to the differential equation is deterministic given an initial condition. 

The master equation of a qubit coupled to a resonator dispersively has been studied extensively \cite{PhysRevA.74.042318, PhysRevA.79.013819}. It has been shown that each energy eigenstate (i.e., the $z$-basis vectors) of the qubit is entangled with a coherent state of the resonator; in addition, a qubit in a superposition state subject to a continuous readout pulse will dephase without changing the population in the $z$-basis. Here we generalize the solution to an arbitrary qudit in the dispersive-coupling regime. To avoid writing down too many equations when $D$ is large, we will explicitly show the derivation for a qutrit measured dispersively while experiencing the $T_1$-decay and pure dephasing; nevertheless, the approaches used for solving the master equation can be easily extended to higher dimensional systems by adding terms of similar forms.

To set up the problem, we let the combined system be a qutrit (labeled as $\mathcal{S}$) coupled to a resonator ($\mathcal{R}$) dispersively. Note that the environment is not a part of the composite system, i.e., we have already traced out the environment to write down a master equation. The (reduced) state of the combined system, denoted by $\hat{\rho}_{\mathcal{SR}}$, lives in the Hilbert space $\mathscr{H}_{\mathcal{SR}}$. For clarity, we denote the energy levels of the qutrit by $\ket{g}$, $\ket{e}$, and $\ket{f}$ (ordered with increasing energy) to differentiate from the Fock states of the resonator. We study the time evolution of the composite state under the usual Born and Markov approximations \cite{cohen1992atom, schlosshauer2007decoherence} in which the state transition maps form a quantum dynamical semigroup and are described by a Lindblad master equation.

\subsection{Master Equation for the Composite System in the Laboratory Frame}
Suppose the resonator is coupled to the environment with a total decay rate of $\kappa$. For a 3D microwave cavity with two ports, $\kappa$ is the sum of the input decay rate $\kappa_{\text{in}}$, output decay rate $\kappa_{\text{out}}$, and the internal decay rates $\kappa_{\text{int}}$ due to material losses. If the resonator is configured in the reflection mode (i.e., one port), then $\kappa_{\text{in}} = \kappa_{\text{out}} \doteq \kappa_{\text{ext}}$ and the total decay rate is $\kappa = \kappa_{\text{int}} + \kappa_{\text{ext}}$. In reality, the resonator supports many modes; here we focus on only one mode (usually the fundamental mode) of the resonator with frequency $\omega_{\text{r}}$ which captures the dispersive measurement succinctly. The resonator-environment interaction is modeled as a harmonic oscillator coupled to a continuum of bath oscillators. At superconducting temperature, we assume that the bath is in the vacuum state (i.e., the mean photon number at the resonator frequency is $\bar{N}(\omega_{\text{r}}) = 0$) so that the usual terms in the quantum optical master equation \cite{breuer2002theory}, i.e., 
\begin{align}
    &\kappa 
            \big[ 
                \bar{N}(\omega_{\text{r}}) + 1 
            \big]
            \mathcalboondox{D}
            \big[ 
                \hat{a}
            \big]
            \hat{\rho}_{\mathcal{R}}(t)
        + \kappa 
            \bar{N}(\omega_{\text{r}})
            \mathcalboondox{D}
            \Bigsl[ 
                \hat{a}^{\dagger}
            \Bigsr]
            \hat{\rho}_{\mathcal{R}}(t),
\end{align}
reduces to $\mathcalboondox{D} \big[ \hat{a} \big] \hat{\rho}_{\mathcal{R}}$, where $\mathcalboondox{D}\Bigsl[ \hat{L} \Bigsr]$ is the dissipation superoperator associated with the Lindblad operator $\hat{L}$ defined via the action
\begin{equation}
    \mathcalboondox{D}
        \Bigsl[ 
            \hat{L}
        \Bigsr]
        \hat{\rho}
    = \hat{L}
            \hat{\rho}
            \hat{L}^{\dagger}
        - \frac{1}{2}
            \hat{L}^{\dagger} 
            \hat{L}
            \hat{\rho}
        - \frac{1}{2}
            \hat{\rho}
            \hat{L}^{\dagger} 
            \hat{L}
\end{equation}
on any density operator $\hat{\rho}$ \cite{PhysRevA.75.032329, walls2008quantum_optics, gardiner2004quantum, breuer2002theory}.

For the qutrit, we study both spontaneous decay and pure dephasing. Without imposing any selection rule, we assume the qutrit can decay from $\ket{f}$ to $\ket{e}$, from $\ket{f}$ to $\ket{g}$, and from $\ket{e}$ to $\ket{g}$ with decay rates $\gamma_{1,ef}$, $\gamma_{1,gf}$, and $\gamma_{1,ge}$, respectively. We also include the pairwise pure dephasing with rates $\gamma_{\phi,ge}, \gamma_{\phi,gf},$ and $\gamma_{\phi,ef}$ to study the coherence time of superposition states. One can show that the three pairwise dephasing terms are equivalent to a single dephasing term with three energy levels included; that is, given any set $\{\gamma_{\phi,g},\gamma_{\phi,e},\gamma_{\phi,f} \}$, there exists $\{\gamma_{\phi,ge}, \gamma_{\phi,gf}, \gamma_{\phi,ef} \}$ such that
\begin{multline}\label{eq:dephasing_equivalence}
    \mathcalboondox{D}
        \left[ 
            \sum_{a\in \{g,e,f\}}
                \sqrt{\gamma_{\phi,a}}
                \ket{a} \! \bra{a}
        \right]
\\ 
    = \sum_{a\in \{g,e,f\}} \sum_{b > a} 
        \mathcalboondox{D}
        \left[ 
                \sqrt{\frac{\gamma_{\phi,ab}}{2}} 
                \left(
                    \ket{a} \! \bra{a}
                    - \ket{b} \! \bra{b}
                \right)
        \right].
\end{multline}
Therefore, we stick with the three pairwise dephasing rates without loss of generality. For a general qudit with $D$ levels, we will need $D(D-1)/2$ pairwise dephasing terms; Eq.(\ref{eq:dephasing_equivalence}) can also be extended to a $D$-level system easily.

By including the decoherence channels mentioned above, we can write down the Lindblad master equation of the composite system 
\begin{widetext}
\begin{align}
    \dot{\hat{\rho}}_{\mathcal{SR}}(t)
    &= - \frac{\ci}{\hbar} \!
            \left[ 
                \hat{H}_{\text{eff}}(t), 
                \hat{\rho}_{\mathcal{SR}}(t)
            \right]
        + \kappa 
            \mathcalboondox{D}\big[\hat{a}\big] 
            \hat{\rho}_{\mathcal{SR}}(t)
        + \gamma_{1,ge}
            \mathcalboondox{D}\big[\hat{\sigma}_{ge}\big]
            \hat{\rho}_{\mathcal{SR}}(t)
        + \gamma_{1,gf}
            \mathcalboondox{D}\big[\hat{\sigma}_{gf}\big]
            \hat{\rho}_{\mathcal{SR}}(t)
        + \gamma_{1,ef}
            \mathcalboondox{D}\big[\hat{\sigma}_{ef}\big]
            \hat{\rho}_{\mathcal{SR}}(t)
\nonumber \\
        &\ \ \    \ \ \ \ 
        + \frac{\gamma_{\phi,ge}}{2} 
            \mathcalboondox{D}\big[\hat{\sigma}_{z,ge}\big]
            \hat{\rho}_{\mathcal{SR}}(t)
        + \frac{\gamma_{\phi,gf}}{2} 
            \mathcalboondox{D}\big[\hat{\sigma}_{z,gf}\big]
            \hat{\rho}_{\mathcal{SR}}(t)
        + \frac{\gamma_{\phi,ef}}{2} 
            \mathcalboondox{D}\big[\hat{\sigma}_{z,ef}\big]
            \hat{\rho}_{\mathcal{SR}}(t)
\nonumber \\ \label{eq:general_qutrit_cavity_master_equation_main}
    &\approx - \frac{\ci}{\hbar} \!
            \left[ 
                \hat{H}_{\text{eff}}(t), 
                \hat{\rho}_{\mathcal{SR}}(t)
            \right]
        + \left( 
                \kappa 
                    \mathcalboondox{D}\big[\hat{a}\big] 
                + \frac{\tilde{\gamma}_{2,ge}}{2} 
                    \mathcalboondox{D}\big[\hat{\sigma}_{z,ge}\big]
                + \frac{\tilde{\gamma}_{2,gf}}{2} 
                    \mathcalboondox{D}\big[\hat{\sigma}_{z,gf}\big]
                + \frac{\tilde{\gamma}_{2,ef}}{2} 
                    \mathcalboondox{D}\big[\hat{\sigma}_{z,ef}\big]
            \right)
            \hat{\rho}_{\mathcal{SR}}(t),
\end{align}
\end{widetext}
where we adopt the notations
\begin{align}
    \hat{\sigma}_{z,ab}
    = \ket{a}\!\bra{a} - \ket{b}\!\bra{b} \ \ 
    \text{ and } \ \ \hat{\sigma}_{ab}
    = \ket{a}\!\bra{b}
\end{align}
for $a \neq b$. In addition, by assuming that $T_{1,ab} = 1/\gamma_{1,ab}$ is much longer than the other decoherence timescales, we have removed the qutrit decay terms and have lumped the extra dephasing rates $\gamma_{1,ab}/2$ with the pure dephasing rates $\gamma_{\phi,ab}$ to define 
\begin{align}
    \tilde{\gamma}_{2,ge} 
    &= \gamma_{\phi,ge} + \gamma_{1,ge} /2,
\\
    \tilde{\gamma}_{2,gf} 
    &= \gamma_{\phi,gf} + (\gamma_{1,gf} + \gamma_{1,ef}) /2,
\\
    \tilde{\gamma}_{2,ef} 
    &= \gamma_{\phi,ef} + (\gamma_{1,ge} + \gamma_{1,gf} + \gamma_{1,ef}) /2.
\end{align}
Note, for example, that $\tilde{\gamma}_{2,ge}$ is not the total decay rate of the $g$-$e$ coherence because $\tilde{\gamma}_{2,gf}$ and $\tilde{\gamma}_{2,ef}$ will also influences the $g$-$e$ coherence through the master equation. In the Appendices, two approaches to solving the master equation are discussed: The first method, which requires less algebra, is the method of positive $P$-representation, typically used when a harmonic oscillator is driven into a coherent state \cite{PhysRevA.74.042318}. The long-$T_1$ assumption is used for the derivation to have a closed-form solution. In contrast, the second method, introduced in \cite{PhysRevA.77.012112}, relies on the resonator displacement operator and allows one to solve Eq.(\ref{eq:general_qutrit_cavity_master_equation_main}) with $\gamma_{1,ab}$ included.

The effective Hamiltonian of a qutrit coupled with a resonator in the dispersive regime subject to a readout probe (under the RWA) is given by
\begin{align}
    &\hat{H}^{\text{disp}}_{\mathcal{SR}}(t) / \hbar
\nonumber \\
    &= \tilde{\omega}_{\text{q}} \ket{e}\!\bra{e}
        + (2 \tilde{\omega}_{\text{q}} + \alpha_{\text{q}}) \ket{f}\!\bra{f}
        + \omega_{\text{r}} \hat{a}^{\dagger} \hat{a} 
\nonumber \\
    & \ \ \ \ 
        + \chi_{\text{qr}} (\ket{e}\!\bra{e} + 2 \ket{f}\!\bra{f})\hat{a}^{\dagger} \hat{a} 
\nonumber \\ \label{eq:effective_hamiltonian_in_dispersive_with_drive}
    & \ \ \ \ 
        - \left(
            \sqrt{\kappa_{\text{in}}} \bar{a}_{\text{in}} 
                e^{-\ci\omega_{\text{d}} t} 
                \,\hat{a}^{\dagger} 
            + \sqrt{\kappa_{\text{in}}} \bar{a}_{\text{in}}^{*} 
                e^{\ci\omega_{\text{d}} t} 
                \, \hat{a} 
        \right),
\end{align}
where we have set the zero-energy reference to be the ground-state energy of the dressed system and used $\tilde{\omega}_{\text{q}} = \Tilde{\omega}_{e} - \Tilde{\omega}_{g}$ to denote the qubit frequency with the Lamb shift included. To address the second-excited state $\ket{f}$, we also introduce the anharmonicity $\alpha_{\text{q}} = (\Tilde{\omega}_{f} - \Tilde{\omega}_{e}) - 2 \tilde{\omega}_{\text{q}}$, which is negative for a transmon. The last term in Eq.(\ref{eq:effective_hamiltonian_in_dispersive_with_drive}) represents the readout signal sent to the resonator with frequency $\omega_{\text{d}}$ and amplitude (in terms of the square root of the average photon flux) $\bar{a}_{\text{in}}$. It is assumed that the readout signal can be modeled by a classical signal \cite{PRXQuantum.2.040202, gardiner2004quantum} (i.e., the stiff-pump limit) and the imaginary frequency added to the resonator due to the decay is negligible since $\kappa \ll \omega_{\text{r}}$. Moreover, for a transmon-type qudit, we use the fact that the dispersive shift is a linear function of the number of excitations in the qudit, i.e., the cavity frequency shifts by $\chi_{\text{qr}}$ when exited from $\ket{g}$ to $\ket{e}$ and shifts by $2\chi_{\text{qr}}$ when exited from $\ket{g}$ to $\ket{f}$. We emphasize that the two methods used to solve the master equation are still valid even if the dispersive shift scales nonlinearly.

The subsequent calculation can be simplified if we move the cavity part of the Hamiltonian to a frame that rotates at the drive frequency $\omega_{\text{d}}$. (Note that the qutrit Hamiltonian stays the same.) Then, the time-varying drive $\varepsilon_{\text{d}}(t)$ reduces to a complex scalar $\epsilon = \sqrt{\kappa_{\text{in}}} \bar{a}_{\text{in}}$ and the Hamiltonian in this rotating frame
\begin{align}
    &\hat{H}_{\mathcal{SR},\text{rot}}^{\text{disp}} / \hbar
\nonumber \\
    &= \tilde{\omega}_{\text{q}} 
            \ket{e}\!\bra{e}
        + (
                2\tilde{\omega}_{\text{q}} 
                + \alpha_{\text{q}}
            ) 
            \ket{f}\!\bra{f}
        + \Delta_{\text{rd}} 
            \hat{a}^{\dagger} \hat{a} 
\nonumber \\
    & \ \ \ \ 
        + \chi_{\text{qr}} 
            (
                \ket{e}\!\bra{e} + 2 \ket{f}\!\bra{f}
            )
            \hat{a}^{\dagger} \hat{a} 
        - \left( 
            \epsilon \hat{a}^{\dagger}
            + \epsilon^* \hat{a}
        \right),
\end{align}
with $\Delta_{\text{rd}} = \omega_{\text{r}} - \omega_{\text{d}}$,
is now time-independent. Then, the master equation of the composite system in the rotating frame is obtained by making the substitution $\hat{H}_{\text{eff}} = \hat{H}_{\mathcal{SR},\text{rot}}^{\text{disp}}$ in Eq.(\ref{eq:general_qutrit_cavity_master_equation_main}). 

\begin{figure*}
    \includegraphics[scale=0.43]{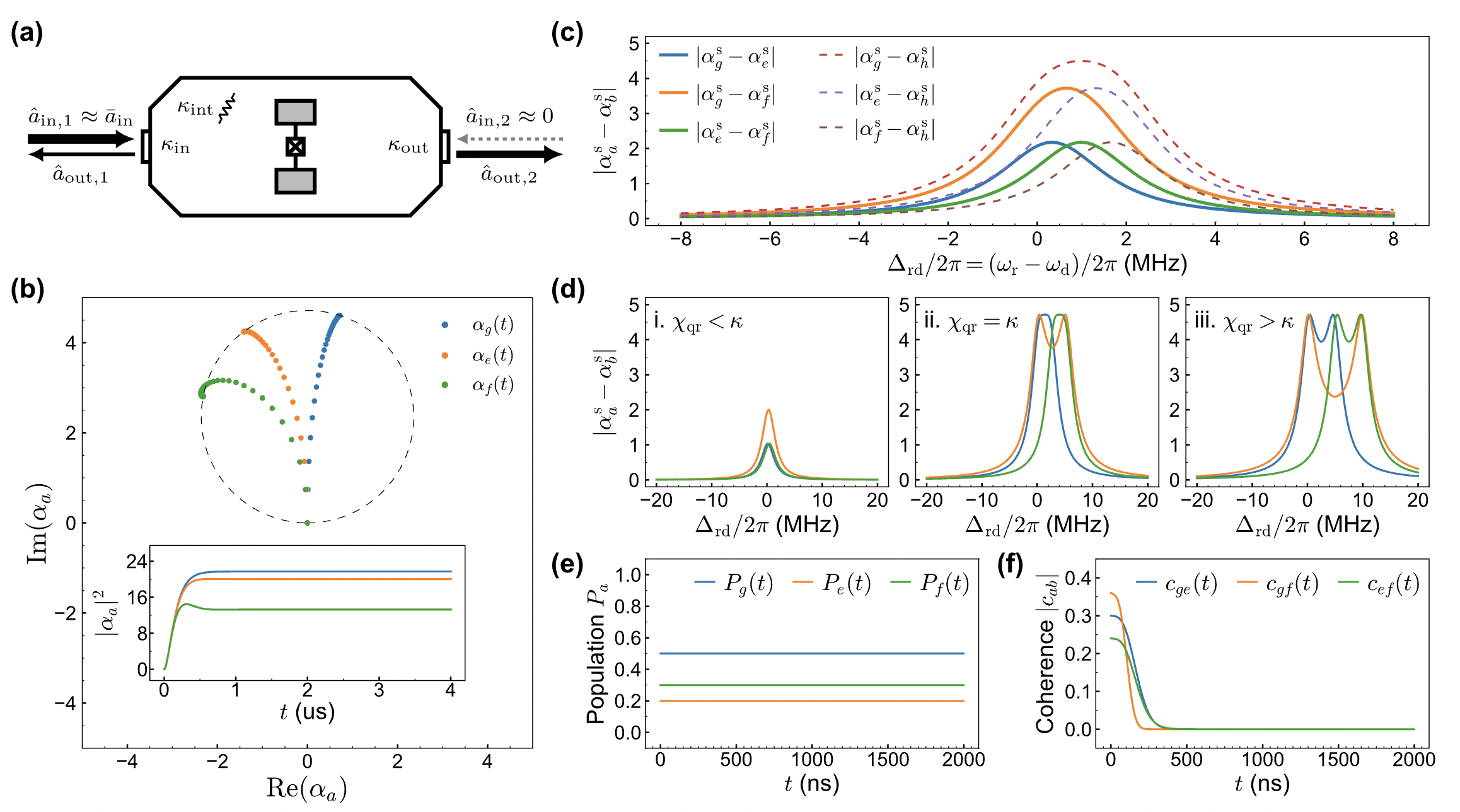}
    \caption{\textbf{Schematic of qutrit readout and solution of the composite-system master equation.} \textbf{a}. The input-output perspective of the transmission-mode dispersive measurement. The readout signal $\hat{a}_{\text{in}}$ entering the cavity from the left port (i.e., port 1) is approximated by a classical drive with complex amplitude $\Bar{a}_{\text{in}}$ while the transmitted signal at the right port (i.e., port 2) described by the traveling-wave annihilation operator $\hat{a}_{\text{out},2}$ in order to capture the quadrature uncertainty. \textbf{b}. The transient complex amplitude of the three coherent states $\ket{\alpha_a}$ of the resonator associated with the $\ket{a}$ for $a=g,e,f$. The steady state of each coherent state amplitude lies on a circle going through the origin of the phase plane. Inset: The build-up of mean photon number of $\ket{\alpha}_a$ as a function of time. \textbf{c}. Distance between two coherent state amplitudes as a function of the readout frequency. To illustrate a more general trend, we also include the fourth energy level $\ket{h}$ of the transmon. \textbf{d}. Same as \textbf{c}, but plotted with $\chi_{\text{qr}}$ smaller, equal, or larger than a fixed $\kappa$. \textbf{e}/\textbf{f}. Time evolution of the composite state as solved from the composite-system master equation in the long-$T_1$ limit.}
    \label{fig:cavity_amplitude_plot_simulations}
\end{figure*}

\subsection{Analytical Solution to the Master Equation of the Combined System} \label{subsection:analytical_soln_combined_me}
To solve Eq.(\ref{eq:general_qutrit_cavity_master_equation_main}), we first express $\hat{\rho}_{\mathcal{SR}}$ as
\begin{equation}
    \hat{\rho}_{\mathcal{SR}}(t)
    = \sum_{a,b \in \{g,e,f\}} \hat{\rho}_{ab}(t)
        \ket{a} \! \bra{b} ,
\end{equation}
where 
\begin{equation}
    \hat{\rho}_{ab}(t) 
    = \bra{a}
            \hat{\rho}_{\mathcal{SR}}(t)
        \ket{b}
\end{equation}
for $a, b \in \{ g, e, f\}$ are operators acting on the Fock space of the resonator. Now, Eq.(\ref{eq:general_qutrit_cavity_master_equation_main}) can be rewritten as a set of nine coupled operator equations in terms of $\hat{\rho}_{ab}$. As shown in Appendix \ref{A.1}, by expressing $\hat{\rho}_{ab}$ in terms of the positive $P$-representation \cite{P_D_Drummond_1980}
\begin{equation}
    \hat{\rho}_{ab}(t) 
    = \int \mathrm{d}^2 \alpha
        \int \mathrm{d}^2 \beta
            \frac{\ket{\alpha}\!\bra{\beta^*}}{\braket{\beta^*}{\alpha}} P_{ab}(\alpha, \beta, t),
\end{equation}
one can reduce the nine operator differential equations into nine scalar equations of $P_{ab}$. The technique of the positive $P$-representation has already been applied to the qubit case to reveal the measurement-induced dephasing on the qubit; Appendix \ref{Appendix:A} demonstrates the effectiveness of this method in solving the qutrit-resonator master equation. In fact, one can apply it to a general qudit system (in the long-$T_1$ limit).

For the qutrit case, the time evolution of the combined state is shown to be
\begin{align}
    &\hat{\rho}_{\mathcal{SR}}(t) 
    = \sum_{a \in \{g,e,f\}} p_a(0) \ket{a} \! \bra{a} \otimes  \ket{\alpha_a(t)} \! \bra{\alpha_a(t)}
\nonumber \\ \label{eq:general_qutrit_resonator_solution_main}
    &+ \sum_{a,b \in \{g,e,f\}} \frac{c_{ab}(t)}{\bra{\alpha_{b}(t)}\ket{\alpha_{a}(t)}} \ket{a} \! \bra{b} \otimes  \ket{\alpha_a(t)} \! \bra{\alpha_b(t)},
\end{align}
where $p_{g,e,f}(0)$ are the initial populations in $\ket{g}$, $\ket{e}$, and $\ket{f}$, respectively and $\ket{\alpha_a(t)}$ represent the coherent states of the resonator. In addition,
\begin{align} \label{eq:differential_eqn_alpha_g_main}
    \dot{\alpha}_g
    &= - \ci (\Delta_{\text{rd}} - \ci \kappa/2) \alpha_g
        + \ci \epsilon,
\\ \label{eq:differential_eqn_alpha_e_main}
    \dot{\alpha}_e
    &=  - \ci (\Delta_{\text{rd}} + \chi_{\text{qr}} - \ci \kappa/2) \alpha_e
        + \ci \epsilon,
\\ \label{eq:differential_eqn_alpha_f_main}
    \dot{\alpha}_f
    &= - \ci (\Delta_{\text{rd}} + 2 \chi_{\text{qr}} - \ci \kappa/2) \alpha_f
        + \ci \epsilon.
\end{align}
determine the time evolution of the coherent states entangled with the qutrit states whereas 
\begin{align} \label{eq:c_ge_diff_eqn_main}
    \dot{c}_{ge}
    &= \ci(\tilde{\omega}_{\text{q}} + \ci \gamma_{2,ge}) c_{ge} 
        + \ci \chi_{\text{qr}} \alpha_g \alpha_e^* c_{ge},
\\ \label{eq:c_gf_diff_eqn_main}
    \dot{c}_{gf}
    &= \ci(2\tilde{\omega}_{\text{q}} + \alpha_{\text{q}} + \ci \gamma_{2,gf}) c_{gf} 
        + \ci 2\chi_{\text{qr}} \alpha_g \alpha_f^* c_{gf},
\\ \label{eq:c_ef_diff_eqn_main}
    \dot{c}_{ef}
    &= \ci(\tilde{\omega}_{\text{q}} + \alpha_{\text{q}} + \ci \gamma_{2,ef}) c_{ef} 
        + \ci \chi_{\text{qr}} \alpha_e \alpha_f^* c_{ef},
\end{align}
with $c_{ab} = c_{ba}^{*}$ for $a \neq b$, govern the oscillation and decay of the off-diagonal terms of the density operator. The total decay rates of the coherence are given by
\begin{align}
    \gamma_{2,ge} 
    = \tilde{\gamma}_{2,ge} 
        + (\tilde{\gamma}_{2,gf} 
        + \tilde{\gamma}_{2,ef}) / 4,
\\
    \gamma_{2,gf} 
    = \tilde{\gamma}_{2,gf} 
        + (\tilde{\gamma}_{2,ge} 
        + \tilde{\gamma}_{2,ef}) / 4,
\\
    \gamma_{2,ef} 
    = \tilde{\gamma}_{2,ef} 
        + (\tilde{\gamma}_{2,ge}
        + \tilde{\gamma}_{2,gf}) / 4.
\end{align}
Since we have ignored $\gamma_{1,ab}$, we observe that the populations of the qutrit do not change over time, a critical feature of the QND measurement. However, the coherence terms will decay to zero with additional rates proportional to $\chi_{\text{qr}}$. It should be noted that $\alpha_a \alpha_b^*$ is complex in general; hence, the last term in Eq.(\ref{eq:c_ge_diff_eqn_main}), (\ref{eq:c_gf_diff_eqn_main}) or (\ref{eq:c_ef_diff_eqn_main}) contains both a decay and a frequency shift. An example of the time evolution of $\hat{\rho}_{\mathcal{SR}}$ is plotted in Figure \ref{fig:cavity_amplitude_plot_simulations}(e) and (f).

Given the general solution, of particular interest are the steady-state amplitudes of the cavity coherent states
\begin{align} \label{eq:steady_state_cavity_amplitude_alpha_g_main}
    \alpha_g (+\infty)
    &= \frac{\sqrt{\kappa_{\text{in}}} \bar{a}_{\text{in}}}{\Delta_{\text{rd}} - \ci \kappa/2},
\\ \label{eq:steady_state_cavity_amplitude_alpha_e_main}
    \alpha_e (+\infty)
    &= \frac{\sqrt{\kappa_{\text{in}}} \bar{a}_{\text{in}}}{\Delta_{\text{rd}} + \chi_{\text{qr}} - \ci \kappa/2},
\\ \label{eq:steady_state_cavity_amplitude_alpha_f_main}
    \alpha_f (+\infty)
    &= \frac{\sqrt{\kappa_{\text{in}}} \bar{a}_{\text{in}}}{\Delta_{\text{rd}} + 2\chi_{\text{qr}} - \ci \kappa/2},
\end{align}
which can be anticipated from the solutions of the quantum Langevin equation (QLE) when the resonator is driven by a classical source \cite{PhysRevA.92.012325}
, i.e.,
\begin{multline}
    \dot{\hat{a}}(t)
    = - \ci \Big[
            \Delta_{\text{rd}} 
            + \big(
                    \chi_{\text{qr}} 
                    \ket{e} \! \bra{e}
                    + 2 \chi_{\text{qr}} 
                    \ket{f} \! \bra{f}
                \big )
        \Big] \hat{a}(t)
\\ 
    - \frac{\kappa}{2} \hat{a}(t) + \ci \sqrt{\kappa_{\text{in}}} \bar{a}_{\text{in}}.
\end{multline}
What is not obvious by looking at the QLE is the dephasing rate captured in Eq.(\ref{eq:c_ge_diff_eqn_main})-(\ref{eq:c_ef_diff_eqn_main}).
Note that the solution is obtained in the rotating frame; to go back to the rest frame, we just need to restore the phase $e^{- \ci \omega_{\text{d}} t}$ in each coherent state. 

We can go one step further by tracing out the resonator part; in other words, the reduced density operator for the qutrit is given by
\begin{align} \label{eq:composite_state_long_T1_limit_main}
    \hat{\rho}_{\mathcal{S}}(t) 
    &= \Tr_{\mathcal{R}}
        \Bigsl[
            \hat{\rho}_{\mathcal{SR}}(t) 
        \Bigsr]
\nonumber \\
    &= \sum_{a} p_a(0) \ket{a} \! \bra{a} 
        + \sum_{a,b} c_{ab}(t) \ket{a} \! \bra{b};
\end{align}
hence, $c_{ab}$ are simply the coherence of the qutrit (i.e., the off-diagonal terms of the reduced density operator of the qutrit) and can be solved from Eq.(\ref{eq:differential_eqn_alpha_g_main})-(\ref{eq:c_ef_diff_eqn_main}).

Before discussing the effective qutrit master, we briefly mention the case when the thermal bath is equilibrated at a nonzero temperature. The operator differential equations of $\hat{\rho}_{ab}$ are almost the same as before except that we replace $\kappa \mathcalboondox{D}\big[\hat{a}\big] \hat{\rho}_{ab}$ with $\kappa \big(\bar{N} + 1 \big) \mathcalboondox{D}\big[\hat{a}\big] \hat{\rho}_{ab} + \kappa \bar{N} \mathcalboondox{D}\Bigsl[\hat{a}^{\dagger}\Bigsr] \hat{\rho}_{ab}$ for $\Bar{N} > 0$. Consequently, each scalar differential equations of $P_{ab}$ acquires a second partial derivative, i.e.,
\begin{multline}\label{eq:fokker_planck_P_ab_main}
    \dot{P}_{ab}
    = \big( \text{terms from the case $\bar{N} = 0$, see Appendix \ref{A.1}}\big) 
\\ 
        + \kappa \bar{N} 
            \frac{\partial^2}{\partial \alpha \partial \beta} P_{ab}.
\end{multline} 

Since Eq.(\ref{eq:fokker_planck_P_ab_main}) with $a=b$ has the same form as the classical Fokker-Planck equation \cite{breuer2002theory, risken2012fokker}, its solution is described by a 2D Gaussian distribution of finite width. In other words, instead of building up a coherent state (which is a delta function in the positive $P$-representation) in the resonator, the external drive will excite a Gaussian state with a quadrature uncertainty broadened by the thermal bath. This also means that the resonator state is now a linear combination of a continuum of coherent states with amplitudes near $\alpha_g$, $\alpha_e$, or $\alpha_f$. In contrast, if the bath is in the vacuum state, a coherent state excited in the resonator will remain coherent indefinitely. See Appendix \ref{A.2} for more details.

\subsection{Effective Qutrit/Qudit Master Equation} \label{subsection:effective_qutrit_me}
Going back to the zero-temperature assumption, the fact that Eq.(\ref{eq:composite_state_long_T1_limit_main}) describes the ensemble-averaged time evolution of a qutrit suggests that we can construct an effective qutrit master equation for dispersive measurement. To relax the long-$T_1$ assumption, we adopt the method of displacement operator used in \cite{PhysRevA.77.012112} to solve for the qutrit case. 

We move to the displaced frame by introducing a qutrit-state-dependent displacement operator
\begin{equation} \label{eq:dispaced_frame_transformation_main}
    \hat{\mathsf{P}}(t)
    = \hat{\Pi}_g \hat{D}(\alpha_g(t))
        + \hat{\Pi}_e \hat{D}(\alpha_e(t))
        + \hat{\Pi}_f \hat{D}(\alpha_f(t)),
\end{equation}
where $\hat{\Pi}_a = \ket{a} \! \bra{a}$ is the projection operator onto the qutrit state $\ket{a}$ for $a \in \{g,e,f\}$ and
\begin{equation} \label{eq:displacement_ops_def_main}
    \hat{D}(\alpha_a(t))
    = \exp
        \Bigsl [ 
            \alpha_a(t) 
                \hat{a}_{\text{r}}^{\dagger} 
            - \alpha_a^*(t) 
                \hat{a}_{\text{r}}
        \Bigsr]
\end{equation}
is the (time-dependent) resonator displacement operator which displaces a coherent state $\ket{\alpha_{a}(t)}$ to the vacuum state \cite{PhysRevA.75.032329}. In other words, by performing the transformation $\hat{\rho}_{\mathcal{SR}}^{\mathsf{P}} = \mathsf{P}^{\dagger} \hat{\rho}_{\mathcal{SR}} \mathsf{P}$ on the combined state and $\hat{O}^{\mathsf{P}} = \mathsf{P}^{\dagger} \hat{O}^{\mathsf{P}} \mathsf{P}$ for any operator $\hat{O}$ in the laboratory frame, we have effectively removed all the resonator photons entangled with the qutrit.

Let $\rho_{\mathcal{S},ab} = \bra{a} \hat{\rho}_{\mathcal{S}} \ket{b}$ with $a,b \in \{g,e,f\}$ be the matrix element of the qutrit density operator in the laboratory frame. As shown in Appendix \ref{Appendix:B} and \ref{Appendix:C}, by first solving the matrix elements of the density operator in the displaced frame and then moving back to the laboratory frame with the resonator state traced out, one arrives at 
\begin{align} \label{eq:qutrit_population_gg_main}
    \dot{\rho}_{\mathcal{S},gg}
    &= \gamma_{1, ge} \rho_{\mathcal{S},ee} 
        + \gamma_{1, gf} \rho_{\mathcal{S},ff} ,
\\ \label{eq:qutrit_population_ee_main}
    \dot{\rho}_{\mathcal{S},ee}
    &= - \gamma_{1, ge} \rho_{\mathcal{S},ee} 
        + \gamma_{1, ef} \rho_{\mathcal{S},ff} ,
\\ \label{eq:qutrit_population_ff_main}
    \dot{\rho}_{\mathcal{S},ff}
    &= - (\gamma_{1, gf} + \gamma_{1, ef}) \rho_{\mathcal{S},ff}.
\end{align}
\begin{align} \label{eq:lambda_0000_ge_main}
    \dot{\rho}_{\mathcal{S}, ge}
    &= \Big[
            \ci \Bar{\omega}_{eg}
            - \gamma_{1,ge} / 2 
            - \gamma_{\phi,ge} 
            + \Gamma_{\text{d},ge}
        \Big]
        \rho_{\mathcal{S}, ge},
\\ \label{eq:lambda_0000_gf_main}
    \dot{\rho}_{\mathcal{S}, gf}
    &= \Big[
            \ci \Bar{\omega}_{gf}
            - \gamma_{1,ge} / 2 
            - \gamma_{\phi,gf} 
            + \Gamma_{\text{d},gf}
        \Big]
        \rho_{\mathcal{S}, gf},
\\ \label{eq:lambda_0000_ef_main}
    \dot{\rho}_{\mathcal{S}, ef}
    &= \Big[
            \ci \Bar{\omega}_{ef}
            - \gamma_{1,ef} / 2 
            - \gamma_{\phi,ef} 
            + \Gamma_{\text{d},ef}
        \Big]
        \rho_{\mathcal{S}, ef},
\end{align}
where the extra dephasing rates appearing in Eq.(\ref{eq:lambda_0000_ge_main})-(\ref{eq:lambda_0000_ef_main}) are given by
\begin{align} \label{eq:measurement_induced_dephasing_ge_main}
    \Gamma_{\text{d},ge}(t) 
    = \Gamma_{\text{d},eg}(t)
    &= \chi_{\text{qr}} \Im(\alpha_g \alpha_e^*),
\\ \label{eq:measurement_induced_dephasing_gf_main}
    \Gamma_{\text{d},gf}(t)
    = \Gamma_{\text{d},fg}(t)
    &= 2 \chi_{\text{qr}} \Im(\alpha_g \alpha_f^*),
\\ \label{eq:measurement_induced_dephasing_ef_main}
    \Gamma_{\text{d},ef}(t)
    = \Gamma_{\text{d},fe}(t)
    &= \chi_{\text{qr}} \Im(\alpha_e \alpha_f^*),
\end{align}
and the time evolution of $\alpha_a(t)$ are still govened by Eq.(\ref{eq:differential_eqn_alpha_g_main})-(\ref{eq:differential_eqn_alpha_f_main}). In addition, $\bar{\omega}_{ba} = \tilde{\omega}_b - \Tilde{\omega}_a - \Delta_{\text{d},ba}$ are the new transition frequencies with the extra shifts
\begin{align}
    \Delta_{\text{d},eg} (t)
    = - \Delta_{\text{d},ge} (t)
    &= \chi_{\text{qr}} \Re(\alpha_e \alpha_g^*),
\\
    \Delta_{\text{d},fg} (t)
    = - \Delta_{\text{d},gf} (t)
    &= 2\chi_{\text{qr}} \Re(\alpha_f \alpha_g^*),
\\
    \Delta_{\text{d},fe} (t)
    = - \Delta_{\text{d},ef} (t)
    &= \chi_{\text{qr}} \Re(\alpha_f \alpha_e^*).
\end{align}
The dephasing rate $\Gamma_{\text{d},ab}$ and the frequency shift $\Delta_{\text{d},ba}$ are nothing more than the real and imaginary parts of the last terms in Eq.(\ref{eq:c_ge_diff_eqn_main})-(\ref{eq:c_ef_diff_eqn_main}).

On the one hand, since we have included the effect of $T_1$, Eq.(\ref{eq:qutrit_population_gg_main})-(\ref{eq:qutrit_population_ff_main}) simply restate the semiclassical rate equations. On the other hand, the time evolution of the coherence shown in Eq.(\ref{eq:lambda_0000_ge_main})-(\ref{eq:lambda_0000_ef_main}) is exactly the same as that in Eq.(\ref{eq:c_ge_diff_eqn_main})-(\ref{eq:c_ef_diff_eqn_main}). $\Gamma_{\text{d},ab}$ is known as the measurement-induced dephasing \cite{PhysRevA.74.042318} and is a function of the dispersive shift $\chi_{\text{qr}}$, resonator decay rate ${\kappa}$, readout detuning $\Delta_{\text{rd}}$, and readout amplitude $\epsilon$. Furthermore, the same conclusion holds for a general qudit measured dispersively: When subject to a coherent readout probe, the resonator is excited to a superposition of $D$ coherent states each entangled with a qudit eigenstate. The complex amplitude of the $j$th (with $j=0,...,D-1$) coherent state evolves according to
\begin{equation}
    \dot{\alpha}_j = - \ci (\Delta_{\text{rd}} + \chi_j - \ci \kappa / 2) \alpha_{j} + \ci \epsilon;
\end{equation}
that is, the readout probe sees a resonator with a frequency shift $\chi_j$ relative to the bare frequency $\omega_{\text{r}}$.
In addition, the time evolution of the qutrit populations, again, follows the rate equation, i.e.,
\begin{equation}
    \dot{\rho}_{\mathcal{S},jj} 
    = \sum_{k>j} \gamma_{1, jk} 
            \rho_{\mathcal{S},kk} 
        - \sum_{k<j} \gamma_{1, jk} 
            \rho_{\mathcal{S},jj} ,
\end{equation}
where $\rho_{\mathcal{S},jj}$ for $j=0,...,D-1$ is the population of the $j$th qudit eigenstate and $\gamma_{1,jk}$ with $j < k$ is the decay rate from state $\ket{k}$ to $\ket{j}$. The time evolution of coherence term $\rho_{\mathcal{S},jk}$ ($j < k$) is given by
\begin{equation}
    \dot{\rho}_{\mathcal{S}, jk}
    = \Big[
            \ci \Bar{\omega}_{kj}
            - \gamma_{1,jk} / 2 
            - \gamma_{\phi,jk} 
            \Gamma_{\text{d},jk}
        \Big]
        \rho_{\mathcal{S}, jk},
\end{equation}
where $\gamma_{\phi,jk}$ is the pure dephasing, 
\begin{equation}
    \Gamma_{\text{d},jk}(t) 
    = \Gamma_{\text{d},kj}(t)
    = (\chi_{k} - \chi_{j}) \Im(\alpha_j \alpha_k^*)
\end{equation}
the measurement-induced dephasing, and
\begin{equation}
    \Bar{\omega}_{kj} (t)
    = - \Bar{\omega}_{jk} (t)
    = (\tilde{\omega}_{k} - \tilde{\omega}_{j})  
        + (\chi_{k} - \chi_{j}) \Re(\alpha_k \alpha_j^*)
\end{equation}
the shifted transition frequencies.

At this point, one might attempt to write down an effective master equation for the qutrit based on Eq.(\ref{eq:qutrit_population_gg_main})-(\ref{eq:qutrit_population_ff_main}) and (\ref{eq:lambda_0000_ge_main})-(\ref{eq:lambda_0000_ef_main}); however, the qutrit frequencies appeared in Eq.(\ref{eq:lambda_0000_ge_main})-(\ref{eq:lambda_0000_ef_main}), in general, do not satisfy the relation
\begin{equation}
    \Bar{\omega}_{fg} - \Bar{\omega}_{fe} 
    = \Bar{\omega}_{ge}
\end{equation}
for a three-level system, so we cannot write down an exact master equation of the Lindblad form. Such a problem does not appear in the qubit case \cite{PhysRevA.77.012112} since the single transition frequency $\Bar{\omega}_{ge}$ is not subject to any constraint. Nevertheless, if the frequency shifts are much smaller than other rate parameters, we can still \textit{approximate} the qutrit alone as a simple Markovian system, thus writing down an effective master equation
\begin{widetext}
\begin{align}
    \dot{\hat{\rho}}_{\mathcal{S}}
    &= - \frac{\ci}{\hbar}
            \Big[ 
                \hat{H}_{\text{q,eff}}, 
                \hat{\rho}_{\mathcal{S}}
            \Big]
        + \gamma_{1,ge} 
            \mathcalboondox{D}
                \Bigsl[
                    \hat{\sigma}_{ge}
                \Bigsr] 
            \hat{\rho}_{\mathcal{S}}
        + \gamma_{1,gf} 
            \mathcalboondox{D}
                \Bigsl[
                    \hat{\sigma}_{gf}
                \Bigsr] 
            \hat{\rho}_{\mathcal{S}}
        + \gamma_{1,ef} 
            \mathcalboondox{D}
                \Bigsl[
                    \hat{\sigma}_{ef}
                \Bigsr] 
            \hat{\rho}_{\mathcal{S}}
\nonumber \\ \label{eq:effective_qutrit_me_main}
    & \ \ \ \
        + \frac{\gamma_{\phi,ge} + \Gamma_{\text{d},ge}}{2} 
            \mathcalboondox{D}
                \Bigsl[
                    \hat{\sigma}_{z,ge}
                \Bigsr]
            \hat{\rho}_{\mathcal{S}}
        + \frac{\gamma_{\phi,gf}+ \Gamma_{\text{d},gf}}{2} 
            \mathcalboondox{D}
                \Bigsl[
                    \hat{\sigma}_{z,gf}
                \Bigsr]
            \hat{\rho}_{\mathcal{S}}
        + \frac{\gamma_{\phi,ef}+ \Gamma_{\text{d},ef}}{2} 
            \mathcalboondox{D}
                \Bigsl[
                    \hat{\sigma}_{z,ef}
                \Bigsr]
            \hat{\rho}_{\mathcal{S}},
\end{align}
\end{widetext}
where the effective qutrit Hamiltonian is assumed to describe a self-consistent set of energy levels (e.g., by ignoring the measurement-induced frequency shifts). Moreover, for the same reason, it should be clear that we cannot write down an exact effective master equation for any qudit with $D \geq 3$. However, by ignoring the frequency shifts, the effective master equation for a qudit will take the form
\begin{align}
    \dot{\hat{\rho}}_{\mathcal{S}}
    &= - \frac{\ci}{\hbar}
            \Big[ 
                \hat{H}_{\text{q,eff}}, 
                \hat{\rho}_{\mathcal{S}}
            \Big]
        + \sum_{j=0}^{D-1} \sum_{k>j} 
            \gamma_{1,jk} 
            \mathcalboondox{D}
                \Bigsl[
                    \hat{\sigma}_{jk}
                \Bigsr] 
            \hat{\rho}_{\mathcal{S}}
\nonumber \\ \label{eq:effective_qudit_me_main}
    & \ \ \ \  \ \ \ 
        + \sum_{j=0}^{D-1} \sum_{k>j} 
            \frac{\gamma_{\phi,jk}+ \Gamma_{\text{d},jk}}{2} 
            \mathcalboondox{D}
                \Bigsl[
                    \hat{\sigma}_{z,jk}
                \Bigsr]
            \hat{\rho}_{\mathcal{S}}.
\end{align}

\subsection{Measurement-Induced Dephasing}
We have already pointed out the connection between the diagonal terms of the effective master equation and the semi-classical rate equation. The more interesting phenomenon lies in the time evolution of the coherence terms. In particular, we see that the product of the dispersive shift and the imaginary part of $\alpha_a \alpha_b^*$ induces a dephasing for each energy level of the qutrit. There are three factors that affect the measurement-induced dephasing rates: 
\begin{enumerate}
    \item [(i)] A large readout drive leads to large coherent state amplitudes and thus a stronger dephasing during the measurement. From the point of view of information theory, since the field leaks out from the resonator will have a larger amplitude as well, we are more likely to gain useful information from the measurement since the signal-to-noise ratio increases as the power of the readout signal goes up \footnote{However, remember that the dispersive coupling is valid only in the low-photon case, so the readout power is usually not a good parameter to adjust.}. Nevertheless, our measurement of the coherent states leaked out of the resonator is subject to the quadrature uncertainty, thus, the measurement result will be distributed as Gaussians centered at $\alpha_{g}$, $\alpha_{e}$, or $\alpha_{f}$. The uncertainty in the measurement will then lead to a random backaction on the qutrit conditioned on the measurement results. It is this random backaction that leads to the dephasing of the qutrit.

    \item[(ii)] Related to the coherent state amplitudes is the readout frequency. As shown in Eq.(\ref{eq:steady_state_cavity_amplitude_alpha_g_main})-(\ref{eq:steady_state_cavity_amplitude_alpha_f_main}), the field amplitudes are Lorentzian functions of the detuning. Hence, for the same drive strength $\bar{a}_{\text{in}}$, the amplitude built up inside the resonator will be the highest when the detuning is zero. However, we cannot drive all three dressed frequencies with zero detuning simultaneously, which means that we have to play with the readout frequency so that the separations among the three coherent states are maximized for state classification. Figure \ref{fig:cavity_amplitude_plot_simulations}(b) shows the steady-state amplitudes of the resonator for some arbitrary readout frequency near the bare resonator frequency; in general, the complex steady-state amplitudes lie on a circle that goes through the origin of the phase plane. In addition, Figure \ref{fig:cavity_amplitude_plot_simulations}(c) plots the distance between two coherent state amplitudes as a function of the readout frequency. As shown in Section \ref{sec:4}, the distance between two coherent states determines the measurement rate.

    \item [(iii)] A larger dispersive shift $\chi_{\text{qr}}$ will also lead to a faster decoherence time. This, again, can be argued from an information-theory point of view. The dispersive shift determines how well we can separate the three qutrit-dependent resonator frequencies; hence the larger the dispersive shifts, the easier the state classification. However, as we have seen in the analysis of the qudit-resonator coupling, $\chi_{\text{qr}}$ is proportional to the square of the coupling coefficient and is approximately inversely proportional to the square of the detuning. For the dispersive coupling to be valid, one cannot make the qutrit-resonator coupling arbitrarily large, thus limiting the amount of $\chi_{\text{qr}}$ realizable in practice. Furthermore, there is another ratio we can design to improve the state classification -- the ratio between the dispersive shift $\chi_{\text{qr}}$ and cavity decay rate $\kappa$. The effect of $\kappa$ is hidden in the expression of the steady-state amplitudes. As shown in Figure \ref{fig:cavity_amplitude_plot_simulations}(d), the distance between the coherent states can be improved by making $\chi_{\text{qr}} > \kappa$.
\end{enumerate}

\section{\label{sec:4}Effective Qutrit Stochastic Master Equation}
An unconditioned master equation can be interpreted as the stochastic trajectories ensemble-averaged over all the possible measurement outcomes \cite{PhysRevLett.70.2273}. The combined system of the qudit and the resonator can be measured either actively by us or implicitly by the environment. In Section \ref{sec:3}, we ignored the information coming out of the resonator, which is equivalent to assuming that all measurements are implicitly made by the environment. To describe an active dispersive measurement by us on a qudit, we thus need to retrieve the information that has been so far averaged out. Since measurements are probabilistic in quantum mechanics, we introduce an effective SME to model the random measurement outcomes.

\subsection{Heterodyne Detection in the Transmission Mode}
Now, we consider a qutrit coupled to the cavity dispersively. To read out the resonator state, we perform a heterodyne detection where the readout signal coming out of the resonator travels on the transmission line (i.e., the coaxial cables connecting the inside of the dilution refrigerator to the room-temperature electronics) and is mixed at the room temperature with a strong local oscillator (LO) signal 
\begin{align}
    \hat{V}_{\text{LO}} (t)
    &= \frac{\hat{a}_{\text{LO}}(t) e^{-\ci \phi_{\text{LO}}} + \hat{a}^{\dagger}_{\text{LO}}(t) e^{\ci \phi_{\text{LO}}}}{2}
\nonumber \\ \label{eq:LO_signal_main}
    &\approx 
    V_{\text{LO}} \cos(\omega_{\text{LO}} t - \phi_{\text{LO}}),
\end{align}
whose frequency $\omega_{\text{LO}}$ is different from the readout signal $\omega_{\text{d}}$ by the intermediate frequency (IF) $\omega_{\text{IF}}$, e.g., $\omega_{\text{IF}} = \omega_{\text{LO}} - \omega_{\text{d}}$. In fact, in a typical IQ demodulation stage, the amplified readout signal 
is first divided equally in power and then mixed separately by two LO signals whose phases are $90^{\circ}$-out-of-phase. Subsequently, the analog IF signals, passing through an analog-to-digital converter (ADC), are processed digitally and demodulated finally to DC (zero frequency) as a complex number (so that one can plot the measurement as a point in the phase plane). Unlike a homodyne detection where $\omega_{\text{LO}} = \omega_{\text{d}}$, the heterodyne scheme allows us to measure both quadratures of the field at the same time (but still constrained by the uncertainty principle). In addition, by first moving to an IF frequency ($\omega_{\text{IF}} \sim 100 \text{ MHz}$ in our experiment), the signal experiences less $1/f$ noise.

We will, however, not attempt to model the analog or digital demodulation part of the heterodyne detection using quantum mechanics. Instead, we will work directly with the coherent state coming out of the resonator and assume that we can process the signal in the way described above and retrieve information about the coherent states subject to quantum-mechanical noise and imperfect measurement efficiency. For a fully quantum-mechanical description of the output chain, including filtering and amplification, see \cite{RevModPhys.93.025005}.

Unlike the qubit measurement where the readout frequency is usually set to be $\omega_{\text{r}} + \chi_{\text{qr}} / 2$ so that the detunings seen by the readout probe are exactly $\pm \chi_{\text{qr}} / 2$ and the information can be encoded in only one quadrature, there isn't any symmetry we can utilize to describe a general qudit measured using the heterodyne scheme. To analyze the information encoded in the complex amplitude of a coherent state, we first define (in the rotating frame of $\omega_{\text{d}}$)
\begin{equation}
    \hat{I}_{\phi} = \frac{\hat{a} e^{-\ci \phi} + \hat{a}^{\dagger} e^{\ci \phi}}{2}
\ \ \text{ and } \ \ 
    \hat{Q}_{\phi} = \frac{\hat{a} e^{-\ci \phi} - \hat{a}^{\dagger} e^{\ci \phi}}{2\ci}
\end{equation}
to be the two quadrature operators of the resonator field, where $\phi$ models the net phase coming from the cable delay and any rotation applied during data processing. Similarly, for the coherent states introduced in Section \ref{subsection:analytical_soln_combined_me}, we define
\begin{align}  \label{eq:I_g_def_main}
    \bar{I}_{g}(t) &= \Re(\alpha_g e^{-\ci \phi}),
    \ \ \ \ 
    \bar{Q}_{g}(t) = \Im(\alpha_g e^{-\ci \phi}),
\\  \label{eq:I_e_def_main}
    \bar{I}_{e}(t) &= \Re(\alpha_e e^{-\ci \phi}),
    \ \ \ \ 
    \bar{Q}_{e}(t) = \Im(\alpha_e e^{-\ci \phi}),
\\  \label{eq:I_f_def_main}
    \bar{I}_{f}(t) &= \Re(\alpha_f e^{-\ci \phi}),
    \ \ \ \ 
    \bar{Q}_{f}(t) = \Im(\alpha_f e^{-\ci \phi}).
\end{align}
We, again, first restricted to the qutrit case since the qudit case can be generalized easily.

Note, however, the quadrature fields defined above live in the resonator and what we observe is only the leakage of the resonator to the transmission line. We denote the input and output traveling-wave annihilation operators at port $i=1,2$ as $\hat{a}_{\text{in,out},i}$. Recall that associated with the QLE of the resonator at port 2 is the boundary condition \cite{gardiner2004quantum, PhysRevB.98.045405, PhysRevA.30.1386}
\begin{align} \label{eq:input_output_at_output_port}
    \hat{a}_{\text{out}}(t)
    &\doteq \hat{a}_{\text{out},2}(t)
\nonumber \\
    &=  - \hat{a}_{\text{in},2}(t) 
        + \sqrt{\kappa_{\text{out}}} \hat{a}(t)
    \approx \sqrt{\kappa_\text{out}} \hat{a}(t),
\end{align}
where we have dropped $\hat{a}_{\text{in},2}$ by assuming that the incoming signal at port 2 (i.e., the output port) is isolated by a well-designed circulator/isolator stage and is not amplified at the output stage (i.e., the HEMT is approximately unilateral). Since in the transmission mode, the resonator is a two-port device, there should be another boundary condition for port 1 (i.e., the input port, see Figure \ref{fig:cavity_amplitude_plot_simulations}(a)); in fact, we have been implicitly using it to define the drive $\epsilon = \sqrt{\kappa_{\text{in}}} \bar{a}_{\text{in}}$ entering at port 1 of the resonator. Information leaking out from port 1 will be factored into the quantum efficiency $\eta$ to be discussed.

In the Schr\"{o}dinger picture, the boundary condition at port 2 implies that the transmitted signal is in a coherent state with the complex amplitude
\begin{equation}
    \alpha_{\text{out}}(t) = \sqrt{\kappa_{\text{out}}} \alpha_{a}(t)
\end{equation}
if the resonator is in the coherent state $\ket{\alpha_a(t)}$ for $a \in \{e,g,f\}$. Moreover, since $\hat{a}_{\text{out}}^{\dagger} \hat{a}_{\text{out}}$ is the outgoing photon flux, the average number of photons leaving the resonator from port 2 within an infinitesimally short time $\Delta t$ is given by
\begin{equation}
    \bar{n}(t) 
    = \kappa_{\text{out}} \Delta t |\alpha_{a}(t)|^2.
\end{equation}
In terms of the outgoing quadrature fields on the transmission line, we have 
\begin{gather} \label{eq:I_out_def_main}
    I_{\text{out}}(t)  
    = \sqrt{\kappa_{\text{out}} \Delta t} \bar{I}_a(t),
\\ \label{eq:Q_out_def_main}
    Q_{\text{out}}(t) 
    = \sqrt{\kappa_{\text{out}} \Delta t} \bar{Q}_a(t),
\end{gather}
where $\bar{I}_a$ and $\bar{Q}_a$ are approximately constant over short $\Delta t$. More precisely, it is the signal
\begin{equation}
    \hat{V}_{\text{out}}(t) = \frac{\hat{a}_{\text{out}}(t) e^{- (\ci \omega_{\text{d}} t+\phi)} + \hat{a}_{\text{out}}^{\dagger}(t) e^{ \ci (\omega_{\text{d}} t+\phi)}}{2}
\end{equation}
that is mixed with $\hat{V}_{\text{LO}}(t)$ and its out-of-phase signal $\hat{V}_{\text{LO}}(t - \pi / 2\omega_{\text{LO}})$. Nevertheless, it's not hard to see that Eq.(\ref{eq:I_out_def_main}) and (\ref{eq:Q_out_def_main}), up to a constant scaling due to cable loss and amplification, are effectively what one would get after filtering the higher sideband of the mixer output at $\omega_{\text{d}} + \omega_{\text{LO}}$ and demodulating the filtered signal digitally to DC. For this reason, we will treat $I_{\text{out}}$ and $Q_{\text{out}}$ as the measurement outcome directly and omit calculation involving $\hat{V}_{\text{LO}}$. Any loss due to the nonideality of the mixer will be included in the quantum efficiency.

In reality, however, measurements are not perfect and not all the information encoded in the photon flux can be captured; thus, the effective photon number we can measure is only 
\begin{equation}
    \bar{n}_{\text{eff}}(t) 
    = \eta \kappa \Delta t |\alpha_{a}(t)|^2,
\end{equation}
where $\eta \in [0,1]$ is the measurement efficiency. Since $\eta_{\text{r}} = \kappa_{\text{out}} / \kappa < 1$, the efficiency $\eta$ is naturally lowered by $\eta_{\text{r}}$. Note that $\eta_{\text{r}}$ also contains the effect of $\kappa_{\text{int}}$ since photons lost internally are inaccessible to the detector. In addition, by performing a heterodyne detection, we automatically half the efficiency due to the power division in an IQ mixer. Moreover, note that when $\eta = 0$, we would gain zero information about the system and can only talk about the behavior of the qutrit in an averaged sense; thus, the SME to be constructed should reduce to the unconditioned master equation and we will verify this point later.

Unlike the mean amplitude, the uncertainty/variance associated with a coherent state traveling on the transmission line is fixed (each quadrature has a variance of $1/4$), so the signal-to-noise ratio is proportional to $\Delta t$; in other words, the photon shot noise can be effectively reduced by increasing the measurement time. Concretely, suppose the resonator is in one of the coherent states $\ket{\alpha_a}$ associated with an energy eigenstate $\ket{a}$ of the qutrit. Then, the \textit{conditional} probability density of measuring a particular point $(I,Q)$ in the phase plane with an integration time of $\Delta t$ \textit{given the qutrit state $\ket{a}$} is a two-dimensional Gaussian
\begin{align}
    &f (I,Q | \hat{\rho}_{\mathcal{S}}(t) = \ket{a}\!\bra{a}) 
\nonumber \\
    &\!\! \propto \exp[- \frac{1}{2} \frac{(I - \sqrt{\eta \kappa \Delta t} \, \bar{I}_a)^2 + (Q - \sqrt{\eta \kappa \Delta t} \, \bar{Q}_a)^2}{1/4}]
\end{align}
with a variance $1/4$ in each quadrature. We have also assumed that the two arms of the mixer output have the same conversion loss to use a single $\eta$; in other words, the measurement is balanced.

More generally, if the qutrit is in a superposition state, then the cavity state, after tracing out the qutrit state is given by
\begin{align}
    \hat{\rho}_{\mathcal{R}}(t) 
    &= \Tr_{\mathcal{S}}[\hat{\rho}_{\mathcal{SR}}(t)]
\nonumber \\
    &= \sum_{a \in \{g,e,f\}} \! \!
            \rho_{\mathcal{S}, aa}(t) 
            \ket{\alpha_{a}(t)}\!\bra{\alpha_{a}(t)},
\end{align}
as suggested by Eq.(\ref{eq:general_qutrit_resonator_solution_main}). Hence, the total conditional probability density of measuring $(I,Q)$ in the phase plane is now given by
\begin{widetext}
\begin{equation}
    \label{eq:prob_IQ_general_state_main}
    f(I,Q | \hat{\rho}_{\mathcal{SR}}(t)) 
    \propto 
    \sum_{a \in \{g,e,f\}}
        \rho_{\mathcal{S}, aa}(t)
        \exp[- \frac{1}{2} \frac{(I - \sqrt{\eta \kappa \Delta t} \, \bar{I}_a)^2 + (Q - \sqrt{\eta \kappa \Delta t} \, \bar{Q}_a)^2}{1/4}].
\end{equation}
\end{widetext}
Consequently, the entanglement generated by the dispersive coupling will project the qutrit state to a new state (possibly mixed) based on the measurement outcome $(I,Q)$ after $\Delta t$.

By introducing the integration time $\Delta t$, we have discretized the continuous measurement into time steps $t_0, t_1, t_2,...$ with a step size of $\Delta t$. We formalize measurement and the backaction induced by the measurement output by introducing, at each time step $t_k$, a continuum of POVM \cite{watrous2018theory, breuer2002theory}
\begin{equation}
    \Big\{ 
        \hat{E}_{IQ}(t_k) = \hat{K}_{IQ}^{\dagger}(t_k) \hat{K}_{IQ}(t_k)
        \ | \ 
        (I,Q) \in \mathbf{R}^2 
    \Big \}
\end{equation}
for the qutrit (i.e., the resonator is traced out in this description) with the Kraus operators
\begin{align}
    &\hat{K}_{IQ}(t_k)
\nonumber \\[-1mm]
    &= \mathscr{N}_{k} \! \!
        \sum_{a \in \{g,e,f\}} 
            \! \! \exp \Bigg\{ \! 
                - \Big[ I - \sqrt{\eta \kappa \Delta t} \, \bar{I}_a(t_k) \Big]^2 
\nonumber \\[-1mm]
    & \ \ \ \ \ \ \ \ \ \ \ \ \ \ \ \ \ \ \ \ \ \ \ \ \
                - \Big[ Q - \sqrt{\eta \kappa \Delta t} \, \bar{Q}_a(t_k) \Big]^2 
            \Bigg\}
            \hat{\Pi}_a
\end{align}
for any point $(I,Q)$ in the phase plane. The normalization constant $\mathscr{N}_k$ can be found by imposing
\begin{equation}
    \int_{-\infty}^{\infty} 
        \int_{-\infty}^{\infty} 
            \mathrm{d} I \mathrm{d} Q \,
            \hat{E}_{IQ}(t_k)
    = 1,
\end{equation}
as required by the completeness of the POVM. Using the Kraus operators, the probability density of measuring $(I,Q)$ is
\begin{equation}
    \Tr[\hat{\rho}_{\mathcal{S}}(t_k) \hat{E}_{IQ}(t_k)] 
    = \Tr[\hat{K}_{IQ}(t_k) \hat{\rho}_{\mathcal{S}}(t_k) \hat{K}_{IQ}^{\dagger}(t_k)] ,
\end{equation}
which, of course, must agree with Eq.(\ref{eq:prob_IQ_general_state_main}). Furthermore, the post-measurement state \textit{conditioned} on the outcome $(I,Q)$ is 
\begin{equation} \label{eq:qutrit_state_update_rule_main}
    \hat{\rho}_{\mathcal{S}}(t_{k+1})
    = \hat{\rho}_{\mathcal{S}}(t_k + \Delta t)
    = \frac{\hat{K}_{IQ}(t_k) \hat{\rho}_{\mathcal{S}}(t_k) \hat{K}_{IQ}^{\dagger}(t_k)}{\Tr[\hat{K}_{IQ} (t_k)\hat{\rho}_{\mathcal{S}}(t_k) \hat{K}_{IQ}^{\dagger}(t_k)] }.
\end{equation}
We emphasize that $\hat{\rho}_{\mathcal{S}}(t_{k+1})$ is the conditional reduced density operator and thus is \textit{not} the same as the reduced density operator used in the effective qutrit master equation before. Nevertheless, once averaged over all the possible measurement history, we should reproduce the unconditional density operator. Furthermore, it should be clear from Eq.(\ref{eq:qutrit_state_update_rule_main}) that the series of quantum channels form a Markov chain, making the entire mathematical formalism easier to deal with.

\subsection{Heuristic Derivation of the Qutrit Stochastic Master Equation}
\begin{figure}[h]
    \includegraphics[scale=0.2]{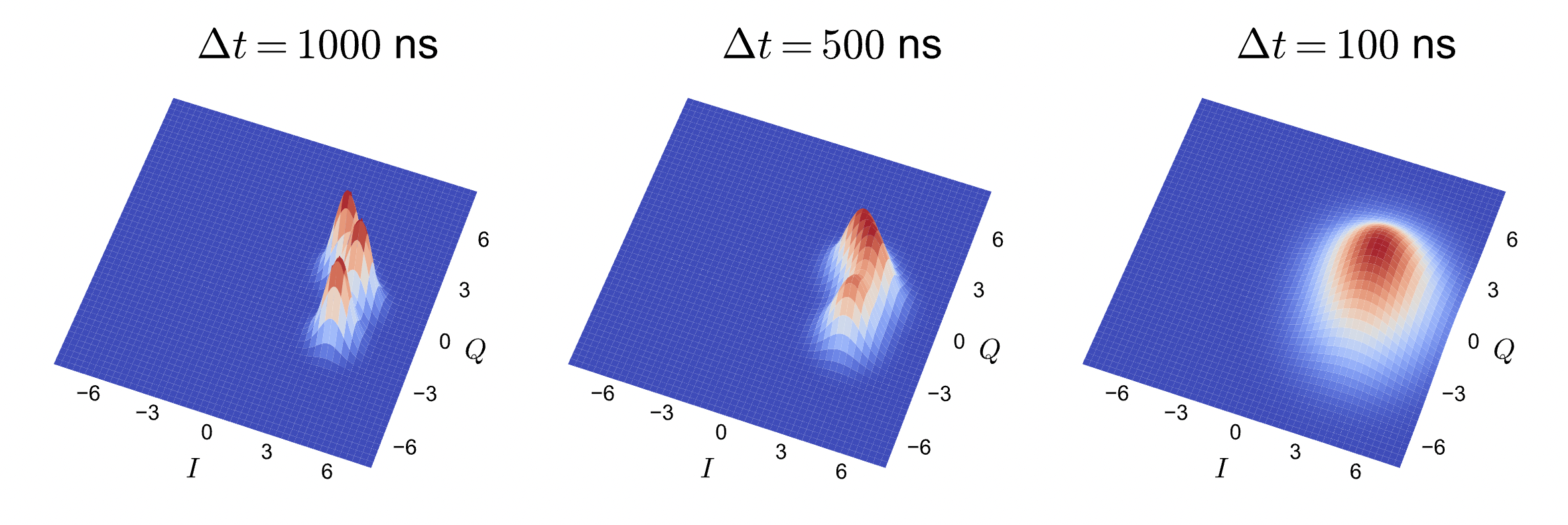}
    \caption{\textbf{Illustration of a weak measurement in the diffusive limit.} As the integration time of each sample reduces, the three Gaussian distributions merge together and can be approximated by a single Gaussian. The mean vector of the approximated distribution is given by the centroid of the three original mean vectors (i.e., $(\Bar{I}_a, \Bar{Q}_a)$) weighted by the probability of the corresponding qutrit state $\ket{a}$.}
    \label{fig:diffusive_limit}
\end{figure}

Based on Eq.(\ref{eq:qutrit_state_update_rule_main}), we look for a stochastic differential equation in the diffusive limit \cite{ROUCHON2022252}, i.e., $\Delta t \rightarrow 0$. As shown in Figure \ref{fig:diffusive_limit}, the three Gaussian clusters in Eq.(\ref{eq:prob_IQ_general_state_main}) merge together to form approximately a new Gaussian distribution as $\Delta t$ becomes sufficiently small \cite{naghiloo2019introduction}. Appendix \ref{Appendix:D} uses this approximation (which is true to the first order in $\Delta t$) to reduce the Kraus operator to 
\begin{align}
    \hat{K}_{IQ}(t_k)
    &\approx 
    \Tilde{\mathscr{N}}_{k} 
        \exp{
            - \Big[
                    I - \sqrt{\eta \kappa \Delta t} 
                    \hat{L}_I(t_k)
                \Big]^2 
        } 
\nonumber \\ \label{eq:Kraus_op_diffusive_main}
    & \ \ \ \ \ \ \ \ 
        \times \exp{
            - \Big[
                    Q - \sqrt{\eta \kappa \Delta t} 
                    \hat{L}_Q(t_k)
                \Big]^2
        },
\end{align}
in the diffusive limit, where the new operators
\begin{align} \label{eq:L_I_def_main}
    \hat{L}_I (t)
    = \bar{I}_g(t) \hat{\Pi}_g
        + \bar{I}_e(t) \hat{\Pi}_e
        + \bar{I}_f(t) \hat{\Pi}_f,
\\  \label{eq:L_Q_def_main}
    \hat{L}_Q(t) 
    = \bar{Q}_g(t) \hat{\Pi}_g
        + \bar{Q}_e(t) \hat{\Pi}_e
        + \bar{Q}_f(t) \hat{\Pi}_f.
\end{align}
act as the Lindblad operators for measuring the $I$ and $Q$ quadrature, respectively.

By expanding Eq.(\ref{eq:qutrit_state_update_rule_main}) using Eq.(\ref{eq:Kraus_op_diffusive_main}), one can derive (see Appendix \ref{Appendix:D}) the effective SME 
\begin{align}
    \mathrm{d} \hat{\rho}
    &= \bigg(
        - \frac{\ci}{\hbar}
            \Big[ 
                \hat{H}_{\text{q,eff}}, 
                \hat{\rho}
            \Big] 
        + \gamma_{1,ge} 
            \mathcalboondox{D}
                \Bigsl[
                    \hat{\sigma}_{ge}
                \Bigsr] 
            \hat{\rho} 
        + \gamma_{1,gf} 
            \mathcalboondox{D}
                \Bigsl[
                    \hat{\sigma}_{gf}
                \Bigsr] 
            \hat{\rho} 
\nonumber \\
    & \ \ \ \ \ \ \ \ 
        + \gamma_{1,ef} 
            \mathcalboondox{D}
                \Bigsl[
                    \hat{\sigma}_{ef}
                \Bigsr] 
            \hat{\rho} 
        + \frac{\gamma_{\phi,ge}}{2} 
            \mathcalboondox{D}
                \Bigsl[
                    \hat{\sigma}_{z,ge}
                \Bigsr]
            \hat{\rho}
\nonumber \\
    & \ \ \ \ \ \ \ \ 
        + \frac{\gamma_{\phi,gf}}{2} 
            \mathcalboondox{D}
                \Bigsl[
                    \hat{\sigma}_{z,gf}
                \Bigsr]
            \hat{\rho}
        + \frac{\gamma_{\phi,ef}}{2} 
            \mathcalboondox{D}
                \Bigsl[
                    \hat{\sigma}_{z,ef}
                \Bigsr]
            \hat{\rho}
        \bigg)
        \mathrm{d} t
\nonumber \\
    &  \ \ \ \ 
        + \left(
            \kappa 
                \mathcalboondox{D}
                    \Bigsl[
                        \hat{L}_{I}
                    \Bigsr]
                \hat{\rho}
            + \kappa 
                \mathcalboondox{D}
                    \Bigsl[
                        \hat{L}_{Q}
                    \Bigsr]
                \hat{\rho}
        \right)
        \mathrm{d} t
\nonumber\\ \label{eq:qutrit_SME_main}
        &  \ \ \ \
            + \sqrt{\eta \kappa} 
            \mathcalboondox{M}
                    \Bigsl[
                        \hat{L}_{I}
                    \Bigsr]
                \hat{\rho} \,
                \mathrm{d} W_{I}
        + \sqrt{\eta \kappa} 
            \mathcalboondox{M}
                    \Bigsl[
                        \hat{L}_{Q}
                    \Bigsr]
                \hat{\rho} \,
                \mathrm{d} W_{Q}
\end{align}
for the conditional reduced density operator of the qutrit, with the heterodyne measurement outcomes (i.e., the complex signal demodulated to DC) encoded in two classical stochastic differential equations
\begin{gather} \label{eq:voltage_I_def_main}
    V_{I}(t)
    = \sqrt{\eta \kappa} 
            \Bigsl \langle
                2 \hat{L}_{I} (t)
            \Bigsr \rangle
        + \xi_{I}(t),
\\ \label{eq:voltage_Q_def_main}
    V_{Q}(t)
    = \sqrt{\eta \kappa} 
            \Bigsl \langle
                2 \hat{L}_{Q} (t)
            \Bigsr \rangle
        + \xi_{Q}(t),
\end{gather}
where $\langle \hat{c} \rangle = \Tr(\hat{\rho} \hat{c})$. The outcomes $V_{I,Q}(t)$ are proportional to $I$ and $Q$ signals measured by the ADC in the real experiment, but are rescaled to remove $\sqrt{\Delta t}$ in Eq.(\ref{eq:I_out_def_main}) and (\ref{eq:Q_out_def_main}). Note that the subscript $\mathcal{S}$ is dropped with the understanding that $\hat{\rho}$ represents the conditional density operator of the qutrit only.

In the SME, the measurement superoperator $\mathcalboondox{M}\Bigsl[\hat{L}\Bigsr]$ associated with an operator $\hat{L}$ is defined via \cite{wiseman2010quantum, ROUCHON2022252}
\begin{equation}
    \mathcalboondox{M}
        \Bigsl[
            \hat{L}
        \Bigsr]
    \hat{\rho}
    = \hat{L}
            \hat{\rho}
        + \hat{\rho}
            \hat{L}^{\dagger}
        - \Bigsl \langle
                \hat{L}
                + \hat{L}^{\dagger}
            \Bigsr \rangle
            \hat{\rho}.
\end{equation}
In addition, $W_{I,Q}$ are two independent classical Wiener processes and $\xi_{I,Q}(t) = \dot{W}_{I,Q}(t)$ are classical white-noise signals satisfying
\begin{align}
    &\mathbb{E}[\xi_{I}(t)] 
    = \mathbb{E}[\xi_{Q}(t)] 
    = \mathbb{E}[\xi_{I}(t)\xi_{Q}(t')] 
    = 0,
\\
    &\mathbb{E}[\xi_{I}(t)\xi_{I}(t')]
    = \mathbb{E}[\xi_{Q}(t)\xi_{Q}(t')]
    = \delta(t-t').
\end{align}
Formally, we should use It\^{o}'s rule, i.e., $\mathrm{d} W_{I}^2 = \mathrm{d} W_{Q}^2 = \mathrm{d} t$ and $\mathrm{d} W_{I} \mathrm{d} W_{Q} = 0$ almost surely \cite{evans2012introduction}.

\subsection{Measurement Rates v.s. Measurement-Induced Dephasing Rates}
At first glance, the measurement-induced dephasing rates defined in Eq.(\ref{eq:measurement_induced_dephasing_ge_main})-(\ref{eq:measurement_induced_dephasing_ge_main}) seem to be missed by the SME shown in Eq.(\ref{eq:qutrit_SME_main}). In fact, $\kappa \mathcalboondox{D} \Bigsl[ \hat{L}_{I} \Bigsr] \hat{\rho} + \kappa \mathcalboondox{D} \Bigsl[ \hat{L}_{Q} \Bigsr] \hat{\rho}$ did not show up in the any of the unconditioned master equations introduced before, at least not obvious in its current form. Nevertheless, one can show that
\begin{align}
    & \kappa 
        \mathcalboondox{D}
                \Bigsl[
                    \hat{L}_{I}
                \Bigsr]
            \hat{\rho}
        + \kappa 
        \mathcalboondox{D} 
                \Bigsl[
                    \hat{L}_{Q}
                \Bigsr]
            \hat{\rho}
\nonumber \\
    &= \frac{\Gamma_{\text{m}, ge}}{4} 
                \mathcalboondox{D}
                    \Bigsl[
                        \hat{\sigma}_{z,ge}
                    \Bigsr]
                \hat{\rho}
            + \frac{\Gamma_{\text{m}, gf}}{4} 
                \mathcalboondox{D}
                    \Bigsl[
                        \hat{\sigma}_{z,gf}
                    \Bigsr]
                \hat{\rho}
\nonumber \\ \label{eq:equivalence_L_IQ_and_sigma_z_ab}
    & \ \ \ \ \ \ \ \ 
            + \frac{\Gamma_{\text{m}, ef}}{4} 
                \mathcalboondox{D}
                    \Bigsl[
                        \hat{\sigma}_{z,ef}
                    \Bigsr]
                \hat{\rho}
\end{align}
by a simple rearrangement of the dissipation superoperators. In Eq.(\ref{eq:equivalence_L_IQ_and_sigma_z_ab}), 
\begin{equation} \label{eq:measurement_rate_def_main}
    \Gamma_{\text{m}, ab}(t) = \kappa |\beta_{ab}(t)|^2
\end{equation}
is the measurement rate associated with the operator $\hat{\sigma}_{z,ab}$ and $\beta_{ab}(t) = \alpha_a(t) - \alpha_b(t)$ represents the vector connecting the two resonator coherent states in the phase plane. Intuitively, the greater the separation between any two coherent states or the larger the resonator decay rate, the easier the state classification becomes and, thus, the higher the measurement rate.

Moreover, it turns out that the same rates $\Gamma_{\text{m}, ab}$ also appear as the dephasing rates in the unconditioned master equation of the combined system in the \textit{displaced} frame (see Appendix \ref{Appendix:B}); thus, Eq.(\ref{eq:qutrit_SME_main}) can be related to the unconditioned master equation in a heuristic sense. 

\begin{figure}
\includegraphics[scale=0.22]{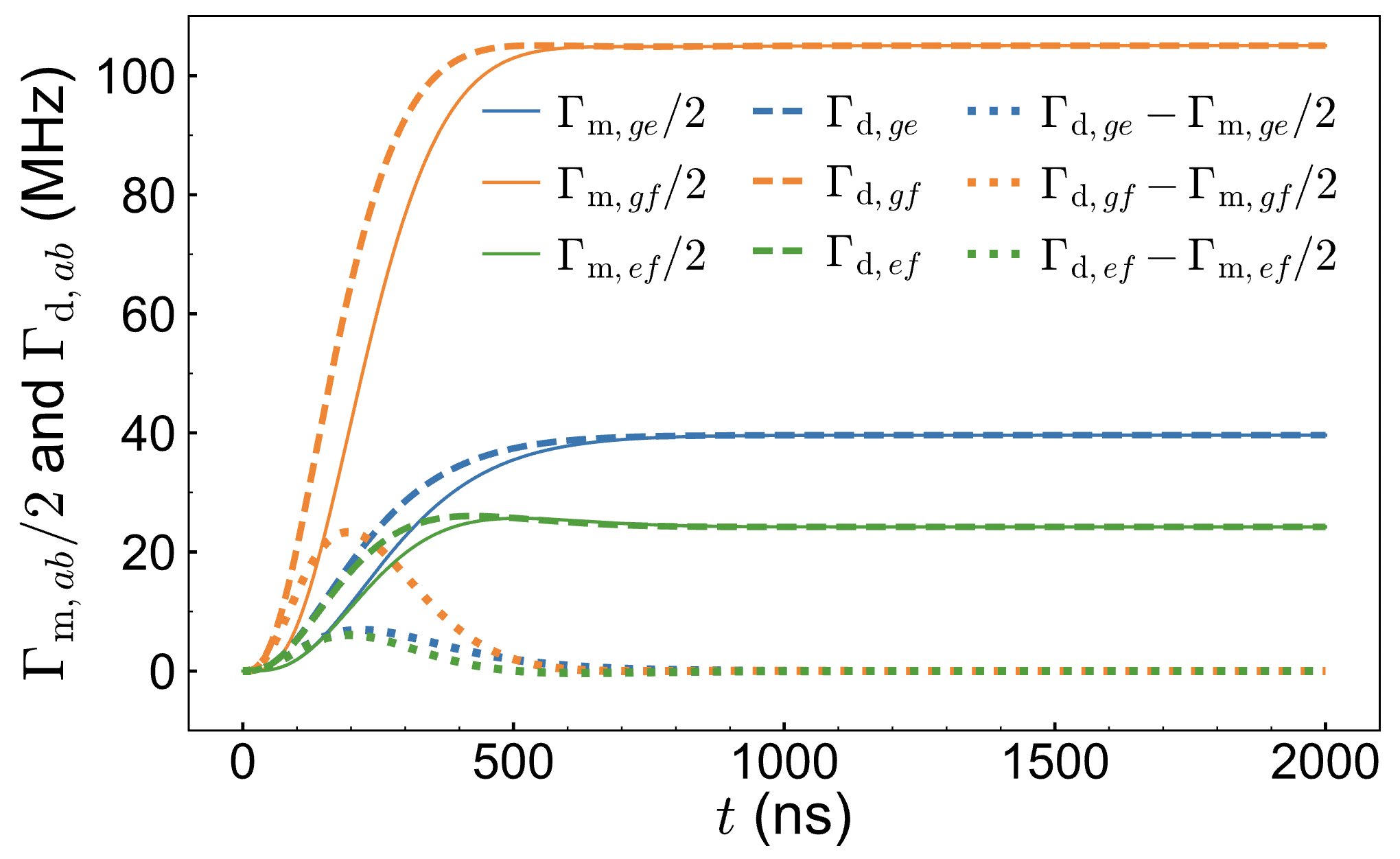}
\caption{\textbf{A comparison of the measurement dephasing rates derived in the laboratory frame and the measurement rates found in the displaced frame.} Since we have assumed a Markovian system in the derivation of the SME, the measurement rate and measurement-induced dephasing rates are not exactly the same in the transient parts. However, when the system reaches its steady state, the two rates are equivalent.}
\label{fig:dephasing_rate_comparison}
\end{figure}

Even though we have reproduced something that appeared in the displaced frame, we have still not derived the exact measurement-induced dephasing rates $\Gamma_{\text{d}, ab}$ that appear in the unconditioned master equation of the qutrit. However, if we choose to use the steady-state values of $\alpha_{g,e,f}$ listed in Eq.(\ref{eq:steady_state_cavity_amplitude_alpha_g_main})-(\ref{eq:steady_state_cavity_amplitude_alpha_f_main}), we can show analytically that 
\begin{equation} \label{eq:equivalence_Gamma_d_and_m}
    \Gamma_{\text{m}, ab} (+\infty) = 2 \Gamma_{\text{d}, ab}(+\infty)
\end{equation}
for $a \neq b$ and, thus, the heterodyne measurement indeed induces a dephasing at rate $\Gamma_{\text{d}, ab}$ between two energy levels of the qutrit at steady state. The steady-state behavior was observed in the qubit case \cite{PhysRevA.77.012112}, but our results imply the generality of such a relationship for a qutrit (and a qudit). Figure \ref{fig:dephasing_rate_comparison} provides a detailed comparison between $\Gamma_{\text{m}, ge}$ and $2\Gamma_{\text{d}, ge}$ before reaching the steady state. Such a discrepancy can exist since we have been assuming a Markovian system; from the derivation of the effective qutrit master equation, however, we know that the qutrit alone is not really Markovian and the information loss in the transient of the resonator evolution is not fully gained by the heterodyne detection. 

To capture the full measurement-induced dephasing using the quantum trajectory approach, one must include the resonator as part of the SME with the measurement operators acting on the resonator directly. Appendix \ref{subsection:full_SME_derivation} provides a derivation of the effective qutrit SME from the SME of the combined system, showing that we can simply replace $\Gamma_{\text{m}, ab}(t)$ in the decoherence terms of Eq.(\ref{eq:qutrit_SME_main}) with $2\Gamma_{\text{d}, ab}(t)$ as expected. 
% The form of $\hat{L}_{I}$ and $\hat{L}_{Q}$ changes to 
% \begin{align}
%     \hat{L}_I' (t)
%     &= \Big[ 
%             \alpha_g(t) \hat{\Pi}_g
%             + \alpha_e(t) \hat{\Pi}_e
%             + \alpha_f(t) \hat{\Pi}_f 
%         \Big] 
%         e^{-\ci \phi},
% \\
%     \hat{L}_Q' (t) 
%     &= - \Big[ 
%             \ci \alpha_g(t) \hat{\Pi}_g
%             + \ci \alpha_e(t) \hat{\Pi}_e
%             + \ci \alpha_f(t) \hat{\Pi}_f
%         \Big] 
%         e^{-\ci \phi}.
% \end{align}
% Compared with Eq.(\ref{eq:L_I_def_main}) and (\ref{eq:L_Q_def_main}), the new measurement operators affect the qutrit state via a different set of phases, i.e., $(\Re(\alpha_a), \Im(\alpha_a)) \rightarrow(\alpha_a, -\ci \alpha_a)$; nevertheless, the trajectories simulated with $\Bigsl(\hat{L}_I,\hat{L}_Q\Bigsr)$ and with $\Bigsl(\hat{L}_I',\hat{L}_Q'\Bigsr)$ exhibit no clear difference and the heuristic argument is still valuable given the intuition it provides. Using $\Bigsl(\hat{L}_I',\hat{L}_Q'\Bigsr)$, 
Consequently, we arrive at the corrected effective qutrit stochastic master equation
\begin{align}
    \mathrm{d} \hat{\rho}
    &= \bigg(
        - \frac{\ci}{\hbar}
            \Big[ 
                \hat{H}_{\text{q,eff}}, 
                \hat{\rho}
            \Big] 
        + \gamma_{1,ge} 
            \mathcalboondox{D}
                \Bigsl[
                    \hat{\sigma}_{ge}
                \Bigsr] 
            \hat{\rho} 
\nonumber \\
    &  \ \ \ \ \ \ \ \ 
        + \gamma_{1,gf} 
            \mathcalboondox{D}
                \Bigsl[
                    \hat{\sigma}_{gf}
                \Bigsr] 
            \hat{\rho} 
        + \gamma_{1,ef} 
            \mathcalboondox{D}
                \Bigsl[
                    \hat{\sigma}_{ef}
                \Bigsr] 
            \hat{\rho} 
\nonumber \\
    &  \ \ \ \ \ \ \ \ 
        + \frac{\gamma_{\phi,ge} + \Gamma_{\text{d},ge}}{2} 
            \mathcalboondox{D}
                \Bigsl[
                    \hat{\sigma}_{z,ge}
                \Bigsr]
            \hat{\rho}
\nonumber \\
    &  \ \ \ \ \ \ \ \ 
        + \frac{\gamma_{\phi,gf}+ \Gamma_{\text{d},gf}}{2} 
            \mathcalboondox{D}
                \Bigsl[
                    \hat{\sigma}_{z,gf}
                \Bigsr]
            \hat{\rho}
\nonumber \\
    &  \ \ \ \ \ \ \ \ 
        + \frac{\gamma_{\phi,ef}+ \Gamma_{\text{d},ef}}{2} 
            \mathcalboondox{D}
                \Bigsl[
                    \hat{\sigma}_{z,ef}
                \Bigsr]
            \hat{\rho}
        \bigg)
        \mathrm{d} t
\nonumber\\ \label{eq:sme_final_form}
    &  \ \ \ \ 
        + \sqrt{\eta \kappa} 
            \mathcalboondox{M}
                    \Bigsl[
                        \hat{L}_{I}
                    \Bigsr]
                \hat{\rho} \,
                \mathrm{d} W_{I}
        + \sqrt{\eta \kappa} 
            \mathcalboondox{M}
                    \Bigsl[
                        \hat{L}_{Q}
                    \Bigsr]
                \hat{\rho} \,
                \mathrm{d} W_{Q},
\end{align}
where $\eta < 1/2$ as a consequence of the heterodyne detection. Moreover, since $\mathbb{E}(\mathrm{d}W_{I,Q}) = 0$, we indeed reproduce the effective qutrit master equation described in Section \ref{subsection:effective_qutrit_me} (at steady state) by taking the ensemble average of the SME.

\subsection{Generalization to the Measurement of a Qudit}
The diffusive SME for the qutrit can be generalized in the qudit case. Besides adding more decoherence channels, we need to redefine the Lindblad operators for measurement:
\begin{equation} \label{eq:def_L_IQ_for_qudit}
    \hat{L}_I
    = \sum_{j=0}^{D-1} 
        \bar{I}_{j}
        \hat{\Pi}_j, 
    \ \ \text{ and } \ \ 
     \hat{L}_Q
    = \sum_{j=0}^{D-1} 
        \bar{Q}_{j}
        \hat{\Pi}_j,
\end{equation}
where the quadrature amplitudes of the $j$th coherent state are given by
\begin{equation}
    \bar{I}_{j} = \Re(\alpha_j e^{-\ci \phi})
    \ \ \text{ and } \ \ 
    \bar{Q}_{j} = \Im(\alpha_j e^{-\ci \phi})
\end{equation}
for $j=0,...,D-1$. 
Then, the effective qudit SME is given by
\begin{align}
    \mathrm{d} \hat{\rho}
    &= - \frac{\ci}{\hbar}
            \Big[ 
                \hat{H}_{\text{q,eff}}, 
                \hat{\rho}
            \Big] \mathrm{d} t
        + \sum_{j=0}^{D-1} \sum_{k>j} 
            \gamma_{1,jk} 
            \mathcalboondox{D}
                \Bigsl[
                    \hat{\sigma}_{jk}
                \Bigsr] 
            \hat{\rho} \, \mathrm{d} t
\nonumber \\
    & \ \ \ \  \ \ \ 
        + \sum_{j=0}^{D-1} \sum_{k>j} 
            \frac{\gamma_{\phi,jk}+ \Gamma_{\text{d},jk}}{2} 
            \mathcalboondox{D}
                \Bigsl[
                    \hat{\sigma}_{z,jk}
                \Bigsr]
            \hat{\rho} \, \mathrm{d} t
\nonumber\\ \label{eq:qudit_sme_final_form}
    &  \ \ \ \  \ \ \
        + \sqrt{\eta \kappa} 
            \mathcalboondox{M}
                    \Bigsl[
                        \hat{L}_{I}
                    \Bigsr]
                \hat{\rho} \,
                \mathrm{d} W_{I}
        + \sqrt{\eta \kappa} 
            \mathcalboondox{M}
                    \Bigsl[
                        \hat{L}_{Q}
                    \Bigsr]
                \hat{\rho} \,
                \mathrm{d} W_{Q},
\end{align}
where $\hat{\rho}$ represents the density operator of the qudit \textit{conditioned} on the measurement history. The measurement outcomes are still governed by Eq.(\ref{eq:voltage_I_def_main}) and (\ref{eq:voltage_Q_def_main}) but $\hat{L}_{I}$ and $\hat{L}_{Q}$ are now defined by Eq.(\ref{eq:def_L_IQ_for_qudit}). Finally, taking the ensemble average of Eq.(\ref{eq:qudit_sme_final_form}) gives the unconditioned master equation stated in Eq.(\ref{eq:effective_qudit_me_main}).

\begin{figure*}
    \includegraphics[scale=0.36]{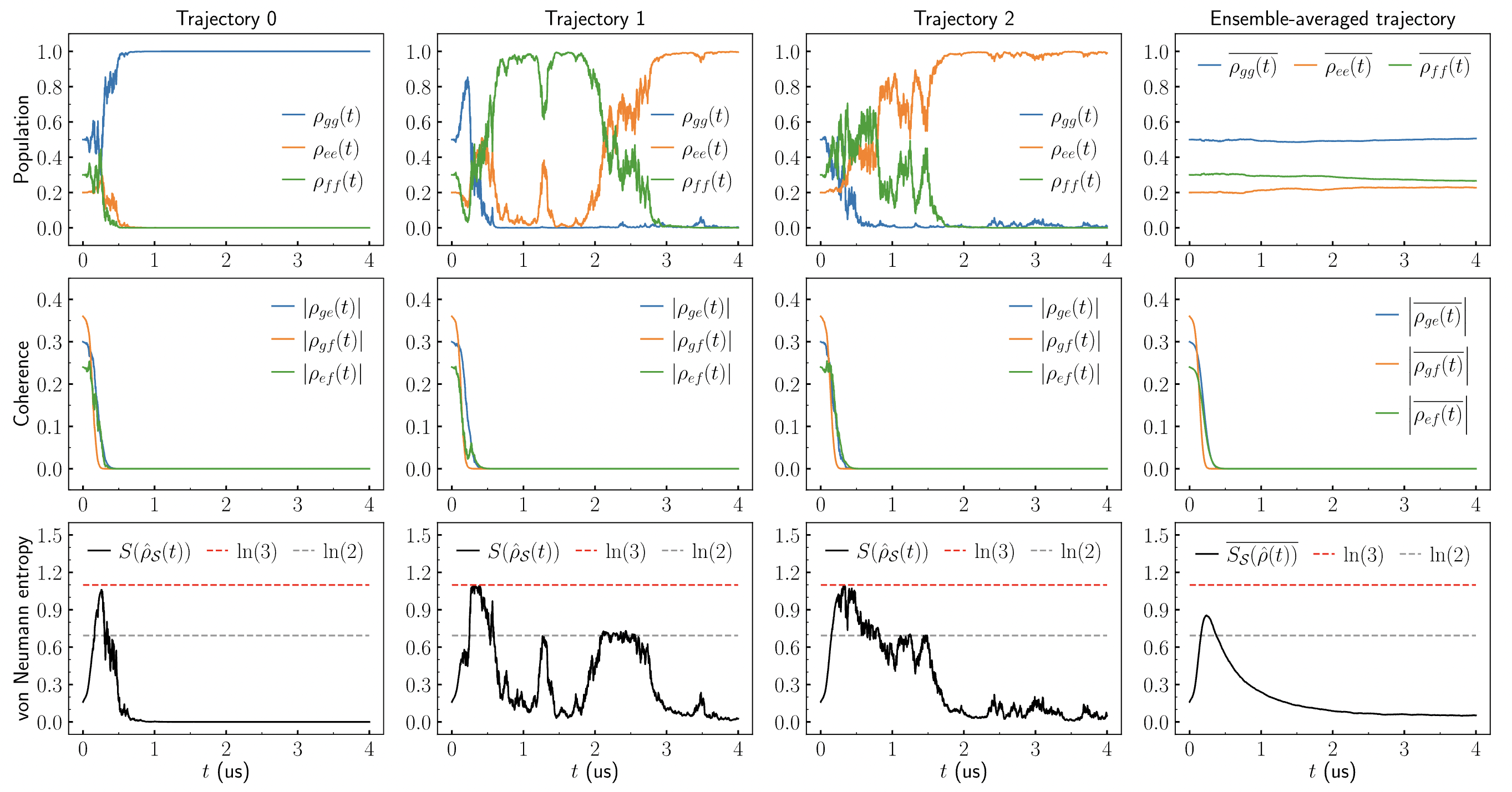}
    \caption{\textbf{Monte Carlo simulation of the qutrit stochastic master equation with an initial state given by Eq.(\ref{eq:sme_sim_initial_state}).} Three sample trajectories are shown in the first three columns. Then, a thousand trajectories are averaged to produce the last column. The first two rows plot the matrix elements of the qutrit density operator. The last row shows the von Neumann entropy $S(\hat{\rho}) = - \Tr(\hat{\rho} \ln \hat{\rho})$, which peaks when the coherence drops to zero. This is because the measurement-induced dephasing will first erase the phase information in a superposition state; however, since the state will be steered towards one of the energy eigenstates, the uncertainty drops to zero eventually.}
    \label{fig:sme_MC_sim_plot}
\end{figure*}

\subsection{Unraveling of
the SME}
In Eq.(\ref{eq:sme_final_form}) and (\ref{eq:qudit_sme_final_form}), the heterodyne measurement is modeled using stochastic processes while the other decoherence terms are analyzed by taking the ensemble average. Instead of having a mixture of quantum trajectories and averaged time evolution in the same equation, one can unravel the SME to write down a stochastic Schr\"{o}dinger equation for the wave function. Since a mixed state is not allowed in a Schr\"{o}dinger equation, all the decoherence channels must be described stochastically, i.e., by some random processes, resulting in infinitely many realizations of the wave function. The unravelings of an SME are not unique, but they must reproduce the decoherence terms in the SME after averaging over all the realizations of the random processes associated with the decoherence channels. 

One possible unraveling uses the quantum state diffusion mode, which works naturally with the diffusive limit used in describing the continuous measurement. Given the set of pairs 
$\Bigsl \{ \Bigsl(\hat{L}_i, \gamma_i\Bigsr) \Bigsr \}_i$ with Lindblad operators $\hat{L}_i$ and the associated rates $\gamma_i$, the quantum state diffusion equation is given by
\begin{align}
    \mathrm{d} \ket{\Psi} 
    &= - \frac{\ci}{\hbar}           
            \hat{H}_{\text{eff}} 
            \ket{\Psi} 
            \mathrm{d} t
        + \sum_{i} 
            \sqrt{\gamma_i} \Big( 
                \hat{L}_i 
                - \Bigsl\langle \hat{L}_i \Bigsr\rangle 
            \Big)
            \mathrm{d} W_i
\nonumber \\ \label{eq:quantum_state_diffusion_eqn_general}
    & 
        - \frac{1}{2} 
            \sum_{i} 
                \gamma_i \left( 
                    \hat{L}_i^{\dagger} 
                        \hat{L}_i
                    + \Bigsl\langle 
                            \hat{L}_i 
                        \Bigsr\rangle 
                        \Bigsl\langle 
                            \hat{L}_i^{\dagger}
                        \Bigsr\rangle 
                    - 2 \Bigsl\langle 
                            \hat{L}_i^{\dagger} 
                        \Bigsr\rangle 
                        \hat{L}_i 
                \right)
                \ket{\Psi}
                \mathrm{d} t,
\end{align}
where $\Bigsl \langle \hat{L}_i^{\dagger} \Bigsr\rangle = \bra{\Psi} \hat{L}_i^{\dagger} 
\ket{\Psi}$ is the expectation value computed for a single realization of $\ket{\Psi}$ and $\{W_i\}_i$ are independent Wiener processes satisfying $\mathbb{E}[\mathrm{d} W_i(t)] = \mathbb{E}[\mathrm{d} W_i(t) \mathrm{d} W_j(t)] = 0$ for $i \neq j$ and $\mathbb{E}[\mathrm{d} W_i^2(t)] = \mathrm{d} t$. By substituting
\begin{equation} \label{eq:list_of_Lindblad_ops}
    \Bigsl \{ \Bigsl(\hat{L}_i, \gamma_i\Bigsr) \Bigsr \}_i
    = \Big \{   
        \Bigsl(\hat{\sigma}_{jk}, \gamma_{1,jk}\Bigsr),
        \Bigsl(\hat{\sigma}_{z, jk}, \gamma_{\phi,jk}/2 \Bigsr)
    \Big \}_{k>j}
\end{equation}
into Eq.(\ref{eq:quantum_state_diffusion_eqn_general}) and using the properties of the Wiener processes, one can easily compute $\mathbb{E}[\mathrm{d}\hat{\rho}]$ and obtain the decoherence terms in Eq.(\ref{eq:qudit_sme_final_form}). In other words, the conditioning of the wave function on the realizations of the Wiener processes is marginalized to create a mixed state of the qudit.

Moreover, the random processes $W_I$ and $W_Q$ can be treated as a part of Eq.(\ref{eq:quantum_state_diffusion_eqn_general}) by adding $\Bigsl(\hat{L}_I, \sqrt{\eta \kappa}\Bigsr)$, $\Bigsl(\hat{L}_I, \sqrt{(1-\eta) \kappa}\Bigsr)$, $\Bigsl(\hat{L}_Q, \sqrt{\eta \kappa}\Bigsr)$ and $\Bigsl(\hat{L}_Q, \sqrt{(1-\eta) \kappa}\Bigsr)$ into Eq.(\ref{eq:list_of_Lindblad_ops}) with independent Wiener processes $W_{I}$, $W_{I}'$, $W_{Q}$, and $W_{Q}'$. Then, the full SME (in the steady state) can be found by only taking the expectation with respect to $W_{I}'$ and $W_{Q}'$, thus producing the correct measurement efficiency (see Appendix \ref{Appendix:D} for modeling an imperfect measurement).

\section{\label{sec:5}Simulations and Experiments}
\subsection{Quantum Trajectories and Ensemble Averages}
Eq.(\ref{eq:sme_final_form}) can be simulated using finite difference just like a normal differential equation. However, since the measurement is random, each finite-difference simulation of Eq.(\ref{eq:sme_final_form}) must be accompanied by a specific realization of $W_{I}$ and $W_{Q}$. In other words, Eq.(\ref{eq:sme_final_form}) can only be simulated in the Monte-Carlo sense. In addition, since a finite-difference simulation discretizes the time into steps of size $\Delta t$, one need to draw $\Delta W_I$ and $\Delta W_Q$ from two independent Gaussian distributions both with mean zero and variance $\Delta t$.

For example, consider an arbitrary qutrit state  
\begin{equation} \label{eq:sme_sim_initial_state}
    \hat{\rho}_{\mathcal{S}}(0)
    = \begin{pmatrix}
        0.5  & 0.3 & 0.36 \\ 
        0.3  & 0.2 & 0.24 \\
        0.36 &0.24 & 0.3
    \end{pmatrix}.
\end{equation}
If we repeat the simulation with this initial state a thousand times, we will obtain a thousand distinct quantum trajectories of the qutrit. Three sample trajectories are shown in Figure \ref{fig:sme_MC_sim_plot}, along with the corresponding von Neumann entropy. The last column of the figure gives the sample-averaged state and entropy. Indeed, the sample-averaged population and coherence agree with the ensemble-averaged ones shown in Figure \ref{fig:cavity_amplitude_plot_simulations}(e)-(f), verifying that the unconditioned master equation is the expectation of the SME.

What is not obvious from the unconditioned master equation is the convergence of the qutrit state to one of the energy eigenstates as shown in the first row of Figure \ref{fig:sme_MC_sim_plot}. Although the qutrit is measured continuously and weakly, each infinitesimal measurement can push the qutrit to a new state; consequently, due to the nature of the dispersive coupling, the qutrit will slowly converge to the pointer states $\ket{g}$, $\ket{e}$, and $\ket{f}$. For the qubit SME in the long-$T_1$ limit, one can construct a Lyapunov function to show the convergence of the qubit state to either $\ket{g}$ and $\ket{e}$ when the readout probe is sent at $\omega_{\text{r}} + \chi_{\text{qr}} / 2$ \cite{ROUCHON2022252}. However, for a qutrit or a qudit, a Lyapunov function is not known to the best of the author's knowledge. Despite the fact that each trajectory converges to an energy eigenstate, the sample-averaged population is still flat as shown in the last column of \ref{fig:sme_MC_sim_plot}, which is a manifestation of the QND nature of dispersive measurement. Since the dispersive measurements on average do not modify the populations of each qutrit levels, we can conclude that the measurement outcomes on average are a faithful reproduction of the underlying qutrit state.

\begin{figure}[h]
    \centering
    \includegraphics[scale=0.2]{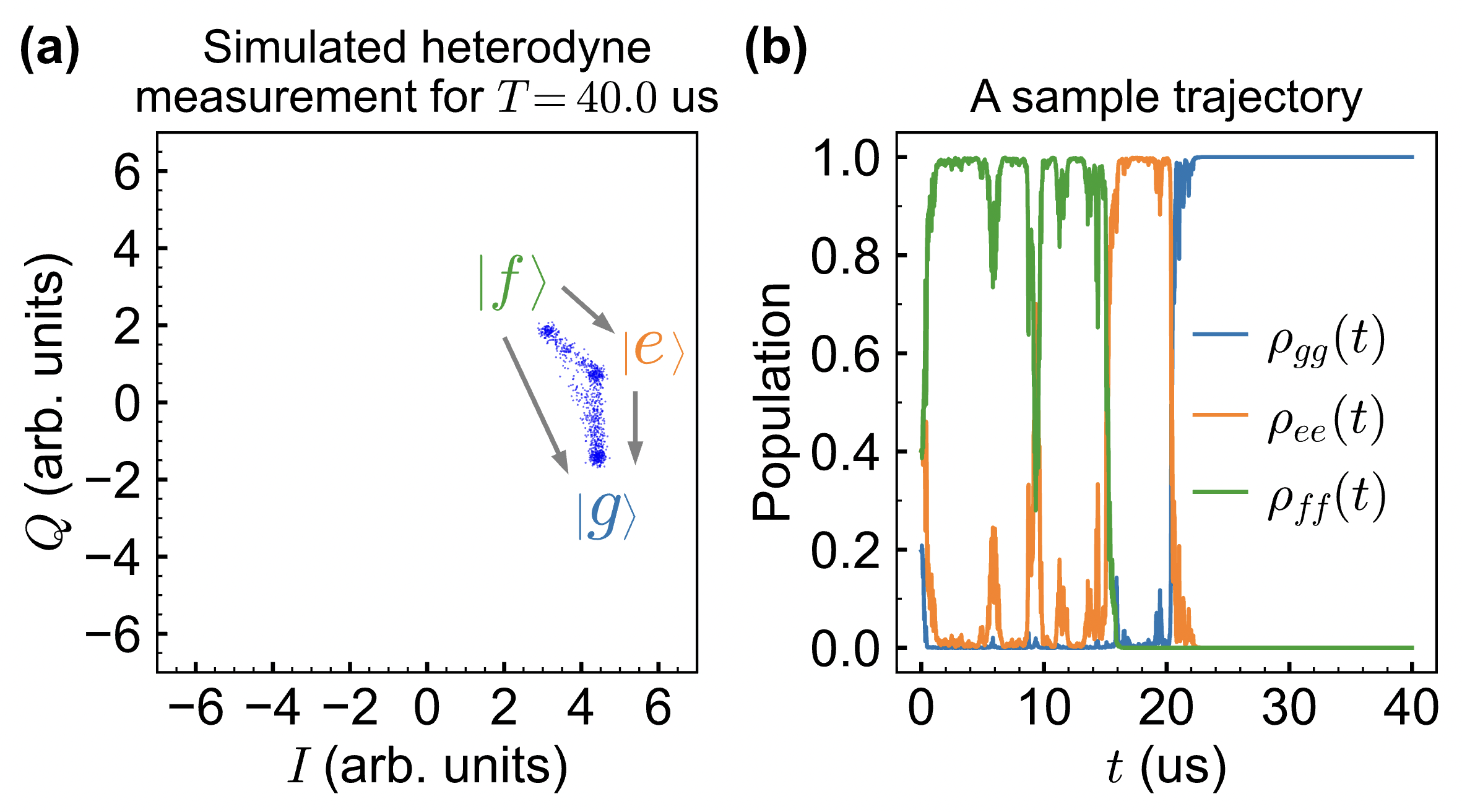}
    \caption{\textbf{Effect of qutrit decay in a long heterodyne measurement}. A thousand random quantum trajectories have been simulated with a measurement time $T=40 \text{ us}$ and with the qutrit parameters $1/\gamma_{1,ge} =1/\gamma_{1,ef} = 35 \text{ us}$ and $1/\gamma_{1,gf} = 1 \text{ ms}$. \textbf{a}. For each quantum trajectory, the heterodyne measurement outcomes $V_I(t)$ and $V_Q(t)$ are averaged over the measurement time $T$ and are plotted as a point in the $I-Q$ plane. Besides the three Gaussian clusters expected from a QND measurement, we also observe streams of points connecting the three clusters, which represent the leakage of population from the excited states to the ground states. \textbf{b}. When the measurement time is on the order of the decay time, the random process exhibits a jumping characteristic on the large time scale.}
    \label{fig:long_measurement_time_effect}
\end{figure}

However, when $\gamma_{1,ab}$ are finite, the convergence of the qutrit is not exactly true even if the quantum trajectories appear to have converged to an eigenstate in a short amount of time. We can gain more insight by making the simulation longer, which effectively increases the measurement time. As shown in Figure \ref{fig:long_measurement_time_effect}, if we set the measurement time to be longer than $1/\gamma_{1,ab}$, we would observe the shift of populations in the phase plane. In other words, the sample-averaged populations, i.e., $\rho_{gg}$, $\rho_{ee}$, and $\rho_{ff}$, are no longer a constant as appeared in Figure \ref{fig:sme_MC_sim_plot} if we extend the simulation time. Hence, in practice, the QND nature of the measurement is limited by the decay rates and we cannot simply improve the signal-to-noise ratio by increasing the measurement time arbitrarily. 

Furthermore, by looking at the time evolution of the qutrit populations, we observe jumps when the measurement time is sufficiently long. Similar behavior of the quantum trajectories was discussed for the qubit case in \cite{PhysRevA.77.012112}; for a general qudit, as long as $\gamma_{1,jk}$ is nonzero, similar jumps will happen so that the qudit state can eventually reach the ground state as required by the unconditioned master equation. 

% \subsection{Testing the Quantum Regression Theorem}
\subsection{Measurement Shot Noise}
\begin{figure*}
    % \includesvg[scale=0.09]{fig_folder/readout_time_comparison}
    \includegraphics[scale=0.09]{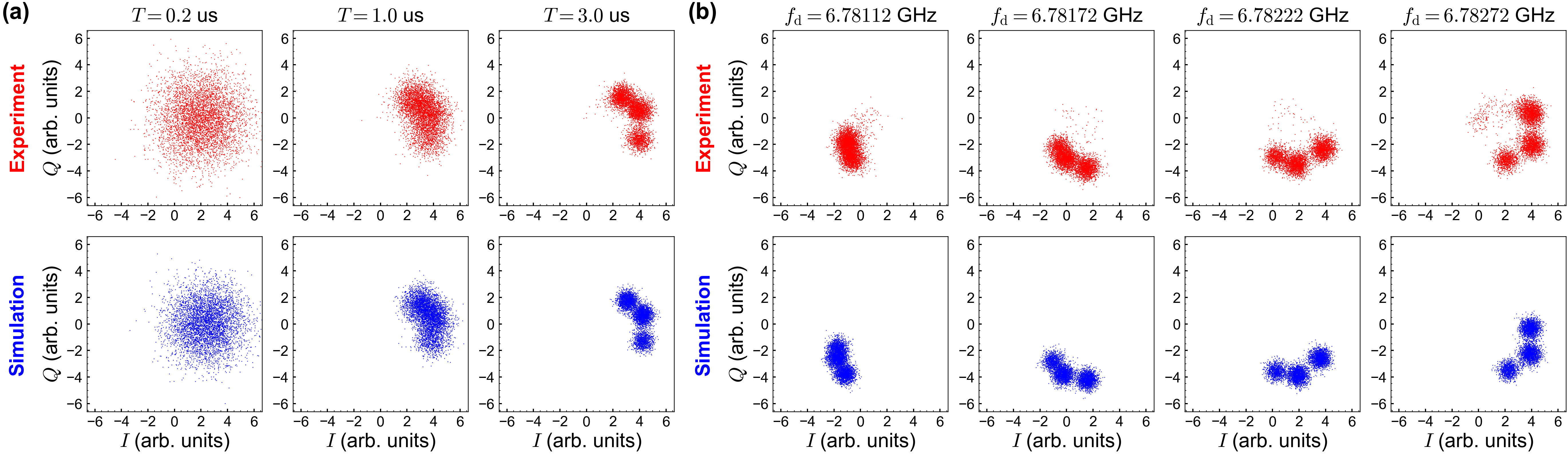}
    \caption{\textbf{Comparison between the simulation and experiment.} The total cavity decay rate is measured to be $\kappa/2\pi = 2.7$ MHz and the dispersive shift is $\chi_{\text{qr}}/2\pi = 0.6$ MHz. The same parameters are used in the simulation with the measurement efficiency set to $\eta = 0.04$ (the experiment is performed without a quantum-limited amplifier). The quadrature signals are collected by AlazarTech digitizer with a sampling frequency of $1$ GHz, which can be treated at $1/\Delta t$ in the derivation of the diffusive SME. \textbf{a.} Changing the measurement time $T$. \textbf{b.} Sweeping the readout frequency $f_{\text{d}}$ (i.e., the cavity drive frequency).}
    \label{fig:sme_sim_exp_comparison}
\end{figure*}

To validate the qutrit SME, real dispersive measurement is performed on a transmon qutrit with $\tilde{\omega}_{\text{q}}/2\pi = 4.48$ GHz and $|\alpha_{\text{q}}|/2\pi \approx 280$ GHz coupled to a 3D aluminum cavity with $\omega_{\text{r}}/2\pi = 6.7835$ GHz when the qutrit is in the ground state. The qutrit decay rates have been characterized to be larger than $30$ us for $T_{1,ge}$, $T_{1,gf}$, and $T_{1,ef}$ and the maximum readout time used in the experiment is much shorter than the decay time to ensure the validity of the QND measurement. In addition, the shortest dephasing time is measured to be $T_{2,gf} = 1 / \gamma_{2,gf} = 3$ us, which does not play a significant role since the measurement-induced dephasing happens at a much higher rate as can be seen in the second row of Figure \ref{fig:sme_MC_sim_plot}.

In all the subsequent experiments, the qutrit is excited to an equal superposition state by two consecutive $\pi$-pulses to create three equally weighted clusters in the phase plane. We include two types of comparison between the experiment and theory. The first one is on the measurement time, showing the reduction of the measurement uncertainty as more information leaks out of the readout cavity. The second comparison tests how accurately the theory can predict the steady-state amplitudes of the coherent states as a function of the readout detuning.

In a real dispersive measurement, what we have access to are the transmitted signals $V_I(t)$ and $V_Q(t)$. By calculating the time averages of $V_I(t)$ and $V_Q(t)$ (with possibly weighting functions to account for the transient behavior), denoted by $\bar{V}_I$ and $\bar{V}_Q$, we obtain a complex number $\bar{V}_I + \ci \bar{V}_Q$, which can be plotted on a phase plane. Repeating the experiment a large number of times generates a scattering plot, whose distribution reflects the qutrit state. One can then compare the simulated $\bar{V}_I + \ci \bar{V}_Q$ with the one measured in experiments to verify the correctness of the SME. In Figure \ref{fig:sme_sim_exp_comparison}(a), we vary the measurement time up to $3$ us to verify the effect of the shot noise due to the quadrature of the coherent states and the broadening due to the measurement inefficiency. As suggested by Eq.(\ref{eq:voltage_I_def_main}) and (\ref{eq:voltage_Q_def_main}), by increasing the measurement time and computing a more accurate time average of $V_{I,Q}$, one will have access to the value of $\Bigsl \langle \hat{L}_{I,Q} (t) \Bigsr \rangle$ up to small variations in the transient part of $\hat{L}_{I,Q} (t)$. Since $\hat{L}_{I,Q} (t)$ is a superposition of the projection operators weighted by the quadrature amplitudes of the resonator coherent state, the qutrit state (which determines the probability amplitude of each energy level) directly affects the chance that one of the projection operator $\hat{\Pi}_a$ is picked in the frequentist's point of view. Hence, $\Bigsl \langle \hat{L}_{I,Q} (t) \Bigsr \rangle$ encodes the information about the qutrit state and a longer measurement time effectively increases signal-to-noise ratio for measuring $\Bigsl \langle \hat{L}_{I,Q} (t) \Bigsr \rangle$ as illustrated in Figure \ref{fig:sme_sim_exp_comparison}(a).

\subsection{Readout Frequency}
With the readout time fixed at $3$ us, we sweep the readout frequency and observe the change in the amplitude and phase of the (time-averaged) transmitted signal. A comparison between the simulated and measured phase planes is shown in Figure \ref{fig:sme_sim_exp_comparison}(b). As predicted by Figure \ref{fig:cavity_amplitude_plot_simulations}(b), the center of each Gaussian cluster lies on a circle that goes through the center. 

When the readout frequency is far away from any of the state-dependent resonator frequencies (i.e., $\omega_{\text{r}} + j\chi_{\text{qr}}$ for $j = 0,1,2$), the three Gaussian clusters overlap on one another, making the state classification impossible. However, this also means that the qutrit states are well protected (both in terms of the population and coherence) since the rate at which information can leak out of the readout cavity at the steady state is proportional to the separation between the clusters (see Eq.(\ref{eq:measurement_rate_def_main}) and (\ref{eq:equivalence_Gamma_d_and_m})). 

For a transmon-type qutrit where the dispersive shifts are $0$, $\chi_{\text{qr}}$ and $2\chi_{\text{qr}}$, we can set the readout frequency to be $\omega_{\text{r}} + \chi_{\text{qr}}$ so that the three Gaussian clusters will be symmetrically placed in the phase plane as shown in the last column of \ref{fig:sme_sim_exp_comparison}(b). However, for a general qudit with unequally-spaced dispersive shifts, there isn't any symmetry one can utilize, which makes the SME simulation of great usefulness. 

Given that the simulation and experiment can be well matched, one can now find an optimal set of design and experimental parameters that corresponds to a large separation between the clusters and thus gives the best state classification. In addition, other parameters such as measurement efficiency can be determined for each experimental setup; subsequently, the simulation used to design the qudits can adopt the same measurement efficiency to predict the outcomes of the heterodyne detection.

\section{\label{sec:7}Conclusion}
In this work, we have introduced and solved the unconditioned master equation describing the dispersive coupling between a qudit and a resonator. Consequently, the concept of measurement-induced dephasing and frequency shifts are extended to a qutrit and a general qudit. Two approaches -- the positive $P$-representation and the displaced frame -- employed in the qubit analysis have been adopted to solve the unconditioned master equation for the qudit for the first time. Unlike the qubit case where the shift of the transition frequency is unconstrained, we observe a non-Markovian nature of the dispersive measurement for qudit with $D \geq 3$. In the presence of negligible frequency shifts due to the readout probe, we arrive at an effective (unconditioned) master equation for a qudit. 

Given the analysis of the unconditioned qudit state, we then study the conditioned qudit state, where the information leaving the combined system (i.e., the transmitted readout signal) is retrieved and processed in both the analog and digital domain. In particular, we focus on heterodyne detection, where both quadratures of the coherent signals traveling along the transmission line can be measured with the efficiency halved. In the diffusive limit where each measurement has an infinitesimal integration time, we can model the continuous monitoring of the qudit by a stochastic master equation (SME). In addition, the measurement outcomes of the two quadratures are described by two classical stochastic differential equations with the same Wiener processes that appeared in the SME. In this way, the quantum dynamics of the qudit are related to the classical information about the coherent signal, allowing us to depict the stochastic nature of quantum measurement. 

Finally, we compare the simulation of the SME with real experiments on a transmon qutrit coupled to a readout cavity. The accuracy of the model is tested with a sweep of the measurement time and of the readout frequency. Both comparisons demonstrate the preciseness of the SME in predicting the formation of clusters in the phase plane. Since our model does not rely on the type of qudit, it can be applied to the measurement of other new superconducting qudit topologies, such as a multimon. In practice, the analytical solution to the unconditioned master SMEs can be critical guidance to the qudit design and readout, enabling rapid prediction of the location of the clusters in the phase plane and providing a potential theoretical foundation for understanding the role of weak measurement in quantum feedback control.

\begin{acknowledgments}
All simulations of the density operators were performed using QuTiP and NumPy. The work at the University of California, Los Angeles (UCLA) is supported by the National Science Foundation (NSF) Graduate Research Fellowship Program under grant number DGE-2034835, the NSF Graduate Research Traineeship on Quantum Science and Engineering at UCLA under grant number DGE-2125924, the Office of Naval Research, and the Army Research Office. Any opinions, findings, and conclusions or recommendations expressed in this material are those of the author(s) and do not necessarily reflect the views of the NSF, the Office of Naval Research, and the Army Research Office. The work at Lawrence Livermore National Laboratory is performed under Contract No. DE-AC52-07NA27344 by the U.S. Department of Energy.
\end{acknowledgments}

\appendix

\section{\label{Appendix:A}Solving the Master Equation of the Combined System (Qutrit + Resonator)}

\subsection{\label{A.1}Zero Temperature}
To solve the master equation, we project the density operator of the composite system onto the energy eigenbasis of the qutrit and thus introduce the operators
% \cite{PhysRevA.74.042318}
\begin{equation}
    \hat{\rho}_{ab}(t) 
    = \bra{a}
            \hat{\rho}_{\mathcal{SR}}(t)
        \ket{b}
\end{equation}
for $a, b \in \{ g, e, f\}$; in other words, the reduced density operator can be decomposed into
\begin{align}
    \hat{\rho}_{\mathcal{SR}}(t)
    &= \hat{\rho}_{gg}
        \ket{g} \! \bra{g} 
        + \hat{\rho}_{ge}
        \ket{g} \! \bra{e} 
        + \hat{\rho}_{gf}
        \ket{g} \! \bra{f} 
\nonumber \\
    & \ \ \ \ 
        + \hat{\rho}_{eg}
        \ket{e} \! \bra{g} 
        + \hat{\rho}_{ee}
        \ket{e} \! \bra{e} 
        + \hat{\rho}_{ef}
        \ket{e} \! \bra{f} 
\nonumber \\
    & \ \ \ \ 
        + \hat{\rho}_{fg}
        \ket{f} \! \bra{g} 
        + \hat{\rho}_{fe}
        \ket{f} \! \bra{e} 
        + \hat{\rho}_{ff}
        \ket{f} \! \bra{f},
\end{align}
and, after expanding the master equation using the nine operators, we obtain nine coupled \textit{operator} differential equations
\begin{align} \label{eq:rho_qutrit_and_cavity_gg}
    \dot{\hat{\rho}}_{gg}
    &= - \ci \Delta_{\text{rd}} \left[ 
            \hat{a}^{\dagger}\hat{a}, \hat{\rho}_{gg}
        \right]
\nonumber \\
    & \ \ \ \ 
        + \ci 
            \left[ 
                \epsilon \hat{a}^{\dagger}
                    + \epsilon^* \hat{a},
                \hat{\rho}_{gg} 
            \right]
            + \kappa \mathcalboondox{D}[\hat{a}]
                \hat{\rho}_{gg},
\end{align}
\begin{align} \label{eq:rho_qutrit_and_cavity_ee}
    \dot{\hat{\rho}}_{ee}
    &= - \ci (\chi_{\text{qr}} + \Delta_{\text{rd}}) \left[ 
            \hat{a}^{\dagger}\hat{a}, \hat{\rho}_{ee}
        \right]
\nonumber \\
    & \ \ \ \ 
        + \ci \left[ 
                \epsilon \hat{a}^{\dagger}
                    + \epsilon^* \hat{a},
                \hat{\rho}_{ee} 
            \right]
            + \kappa \mathcalboondox{D}[\hat{a}]
                \hat{\rho}_{ee},
\end{align}
\begin{align} \label{eq:rho_qutrit_and_cavity_ff}
    \dot{\hat{\rho}}_{ff}
    &= - \ci (2\chi_{\text{qr}} + \Delta_{\text{rd}}) \left[ 
            \hat{a}^{\dagger}\hat{a}, \hat{\rho}_{ff}
        \right]
\nonumber \\
    & \ \ \ \ 
        + \ci \left[ 
                \epsilon \hat{a}^{\dagger}
                    + \epsilon^* \hat{a},
                \hat{\rho}_{ff} 
            \right]
            + \kappa \mathcalboondox{D}[\hat{a}]
                \hat{\rho}_{ff},
\end{align}
\begin{align} \label{eq:rho_qutrit_and_cavity_ge}
    \dot{\hat{\rho}}_{ge}
    &= \ci \tilde{\omega}_{\text{q}} \hat{\rho}_{ge} 
        - \ci \Delta_{\text{rd}} \left[ 
            \hat{a}^{\dagger}\hat{a}, \hat{\rho}_{ge}
        \right] \!
        + \ci\chi_{\text{qr}} \hat{\rho}_{ge}
            \hat{a}^{\dagger}\hat{a}
\nonumber \\
    & \ \ \ \
        + \ci \left[ 
                \epsilon \hat{a}^{\dagger}
                    + \epsilon^* \hat{a},
                \hat{\rho}_{ge} 
            \right] \!
        + \kappa \mathcalboondox{D}[\hat{a}]
                \hat{\rho}_{ge}
        - \gamma_{2,ge} \hat{\rho}_{ge},
\end{align}
\begin{align} \label{eq:rho_qutrit_and_cavity_eg}
    \dot{\hat{\rho}}_{eg}
    &= - \ci \tilde{\omega}_{\text{q}} \hat{\rho}_{eg} 
        - \ci \Delta_{\text{rd}} \left[ 
            \hat{a}^{\dagger}\hat{a}, \hat{\rho}_{eg}
        \right] \!
        - \ci\chi_{\text{qr}} 
            \hat{a}^{\dagger}\hat{a}
            \hat{\rho}_{eg}
\nonumber \\
    & \ \ \ \ 
        + \ci \left[ 
                \epsilon \hat{a}^{\dagger}
                    + \epsilon^* \hat{a},
                \hat{\rho}_{eg} 
            \right] \!
        + \kappa \mathcalboondox{D}[\hat{a}]
                \hat{\rho}_{eg}
        - \gamma_{2,eg} \hat{\rho}_{eg},
\end{align}
\begin{align} \label{eq:rho_qutrit_and_cavity_gf}
    \dot{\hat{\rho}}_{gf}
    &= \ci (2\tilde{\omega}_{\text{q}} + \alpha_{\text{q}})\hat{\rho}_{gf} 
        - \ci \Delta_{\text{rd}} \left[ 
            \hat{a}^{\dagger}\hat{a}, \hat{\rho}_{gf}
        \right] \!
\nonumber \\
    & \ \ \ \ 
        + \ci 2 \chi_{\text{qr}} \hat{\rho}_{gf}
            \hat{a}^{\dagger}\hat{a}
        + \ci \left[ 
                \epsilon \hat{a}^{\dagger}
                    + \epsilon^* \hat{a},
                \hat{\rho}_{gf} 
            \right] \!
\nonumber \\
    & \ \ \ \ 
        + \kappa 
            \mathcalboondox{D} 
            [\hat{a}]
                \hat{\rho}_{gf}
        - \gamma_{2,gf} \hat{\rho}_{gf},
\end{align}
\begin{align} \label{eq:rho_qutrit_and_cavity_fg}
    \dot{\hat{\rho}}_{fg}
    &= -\ci (
                2\tilde{\omega}_{\text{q}} + \alpha_{\text{q}}
            )
            \hat{\rho}_{fg} 
        - \ci \Delta_{\text{rd}} 
            \left[ 
                \hat{a}^{\dagger}\hat{a}, \hat{\rho}_{fg}
            \right] \!
\nonumber \\
    & \ \ \ \ 
        - \ci 2 \chi_{\text{qr}} 
            \hat{a}^{\dagger} \hat{a}
            \hat{\rho}_{fg}
        + \ci \left[ 
                \epsilon 
                        \hat{a}^{\dagger}
                    + \epsilon^* 
                        \hat{a},
                \hat{\rho}_{fg} 
            \right] \!
\nonumber \\
    & \ \ \ \ 
        + \kappa \mathcalboondox{D}[\hat{a}]
                \hat{\rho}_{fg}
        - \gamma_{2,fg} \hat{\rho}_{fg},
\end{align}
\begin{align} \label{eq:rho_qutrit_and_cavity_ef}
    \dot{\hat{\rho}}_{ef}
    &= \ci (\tilde{\omega}_{\text{q}} + \alpha_{\text{q}}) \hat{\rho}_{ef} 
        - \ci \Delta_{\text{rd}} \left[ 
            \hat{a}^{\dagger}\hat{a}, \hat{\rho}_{ef}
        \right] \!
\nonumber \\
    & \ \ \ \ 
        + \ci \chi_{\text{qr}} (2\hat{\rho}_{ef}
            \hat{a}^{\dagger}\hat{a}
            - \hat{a}^{\dagger}\hat{a} \hat{\rho}_{ef})
\nonumber \\
    & \ \ \ \ 
        + \ci \left[ 
                \epsilon \hat{a}^{\dagger}
                    + \epsilon^* \hat{a},
                \hat{\rho}_{ef} 
            \right] \!
        + \kappa \mathcalboondox{D}[\hat{a}]
                \hat{\rho}_{ef}
        - \gamma_{2,ef} \hat{\rho}_{ef},
\end{align}
\begin{align} \label{eq:rho_qutrit_and_cavity_fe}
    \dot{\hat{\rho}}_{fe}
    &= - \ci (\tilde{\omega}_{\text{q}} + \alpha_{\text{q}}) \hat{\rho}_{fe} 
        - \ci \Delta_{\text{rd}} \left[ 
            \hat{a}^{\dagger}\hat{a}, \hat{\rho}_{fe}
        \right] \!
\nonumber \\
    & \ \ \ \ 
        - \ci \chi_{\text{qr}}  
            (
                2 \hat{a}^{\dagger}
                    \hat{a}
                    \hat{\rho}_{fe}
                - \hat{\rho}_{fe}
                    \hat{a}^{\dagger}
                    \hat{a} 
            )
\nonumber \\
    & \ \ \ \ 
        + \ci \left[ 
                \epsilon \hat{a}^{\dagger}
                    + \epsilon^* \hat{a},
                \hat{\rho}_{fe} 
            \right] \!
        + \kappa \mathcalboondox{D}[\hat{a}]
                \hat{\rho}_{fe}
        - \gamma_{2,fe} \hat{\rho}_{fe}.
\end{align}
The total decay rates of the coherence are defined in the main text and are reproduced below:
\begin{align}
    \gamma_{2,ge} 
    = \tilde{\gamma}_{2,ge} 
        + (\tilde{\gamma}_{2,gf} 
        + \tilde{\gamma}_{2,ef}) / 4,
\\
    \gamma_{2,gf} 
    = \tilde{\gamma}_{2,gf} 
        + (\tilde{\gamma}_{2,ge} 
        + \tilde{\gamma}_{2,ef}) / 4,
\\
    \gamma_{2,ef} 
    = \tilde{\gamma}_{2,ef} 
        + (\tilde{\gamma}_{2,ge}
        + \tilde{\gamma}_{2,gf}) / 4.
\end{align}

Note that each operator $\hat{\rho}_{ab}$ lives in an infinite dimensional space since $\mathscr{H}_{\mathcal{R}}$ is a Fock space. Nevertheless, it's possible to find a closed-form solution by invoking the positive $P$-representation
\begin{equation}
    \hat{\rho}_{ab}(t) 
    = \int \mathrm{d}^2 \alpha
        \int \mathrm{d}^2 \beta
            \frac{\ket{\alpha}\!\bra{\beta^*}}{\braket{\beta^*}{\alpha}} P_{ab}(\alpha, \beta, t).
\end{equation}
The reader can verify that the action of the creation and annihilation operators in the operator space can be translated to some simple operations in the positive $P$-representation:
\begin{align} \label{eq:operator_correspondance_1}
    \hat{a} \hat{\rho} (t)  
    &\ \ \longrightarrow \ \ 
    \alpha 
    P(\alpha, \beta, t),
\\ \label{eq:operator_correspondance_2}
    \hat{a}^{\dagger} \hat{\rho} (t)  
    &\ \ \longrightarrow \ \ 
    \left(
        \beta - \frac{\partial}{\partial \alpha} 
    \right)
    P(\alpha, \beta, t),
\\ \label{eq:operator_correspondance_3}
    \hat{\rho} (t) \hat{a}^{\dagger}
    &\ \ \longrightarrow \ \ 
    \beta P(\alpha, \beta, t),
\\ \label{eq:operator_correspondance_4}
    \hat{\rho} (t) \hat{a}
    &\ \ \longrightarrow \ \ 
    \left(
        \alpha - \frac{\partial}{\partial \beta} 
    \right) P(\alpha, \beta, t).
\end{align}
As an example, let us use Eq.(\ref{eq:operator_correspondance_1})-(\ref{eq:operator_correspondance_4}) to transform Eq.(\ref{eq:rho_qutrit_and_cavity_gg}) into a scalar equation:
\begin{align}
    \dot{P}_{gg}
    &= - \ci \Delta_{\text{rd}}
            \left [
                \left(
                    \beta - \frac{\partial}{\partial \alpha} 
                \right)
                (\alpha P_{gg})
                - \left(
                    \alpha - \frac{\partial}{\partial \beta} 
                \right)
                (\beta P_{gg})
            \right]
\nonumber \\
    & \ \ \ \ 
        + \ci 
            \bigg[
                \epsilon 
                    \left(
                        \beta 
                        - \frac{\partial}{\partial \alpha} 
                    \right) P_{gg}
                + \epsilon^* \alpha P_{gg}
\nonumber \\
    & \ \ \ \ \ \ \ \ \ \ \ \ 
                - \epsilon \beta P_{gg}
                - \epsilon^* 
                    \left(
                        \alpha 
                        - \frac{\partial}{\partial \beta} 
                    \right) P_{gg}
            \bigg]
\nonumber \\
    & \ \ \ \ 
        + \kappa 
            \bigg[
                \alpha \beta P_{gg}
                - \frac{1}{2} 
                \left(
                    \beta - \frac{\partial}{\partial \alpha} 
                \right)
                (\alpha P_{gg})
\nonumber \\
    & \ \ \ \ \ \ \ \ \ \ \ \ 
                - \frac{1}{2} \left(
                    \alpha - \frac{\partial}{\partial \beta} 
                \right)
                (\beta P_{gg})
            \bigg]
\nonumber \\
    &= \frac{\partial}{\partial \alpha}
        \left[
            \left(
                -\ci \epsilon 
                + \ci \Delta_{\text{rd}} \alpha 
                + \frac{\kappa \alpha}{2} 
            \right) 
            P_{gg}
        \right]
\nonumber \\
    & \ \ \ \ 
        + \frac{\partial}{\partial \beta}
        \left[
            \left(
                \ci \epsilon^{*} 
                - \ci \Delta_{\text{rd}} \beta
                + \frac{\kappa \beta}{2}
            \right) 
            P_{gg}
        \right],
\end{align}
where $\partial_t P_{gg} (\alpha, \beta, t)$ is shorthanded as $\dot{P}_{gg}$. By applying the same procedure to the other eight operators, we get, in total, nine coupled \textit{scalar} differential equations
\begin{align} \label{eq:complex_P_gg}
    \!\!\!\! \dot{P}_{gg}
    &= \frac{\partial}{\partial \alpha}
        \left[
            (-\ci \epsilon 
            + \ci \Delta_{\text{rd}} \alpha 
            + \kappa \alpha/2) 
            P_{gg}
        \right]
\nonumber \\
    & \ \ \ \ 
        + \frac{\partial}{\partial \beta}
        \left[
            (\ci \epsilon^{*}  
            - \ci \Delta_{\text{rd}} \beta
            + \kappa \beta/2) 
            P_{gg}
        \right],
\end{align}
\begin{align}  \label{eq:complex_P_ee}
    \!\!\!\! \dot{P}_{ee}
    &= \frac{\partial}{\partial \alpha}
        \left[
            (-\ci \epsilon 
            + \ci \chi_{\text{qr}} \alpha 
            + \ci \Delta_{\text{rd}} \alpha 
            + \kappa \alpha/2) 
            P_{ee}
        \right]
\nonumber \\
    & \ \ \ \ 
        + \frac{\partial}{\partial \beta}
        \left[
            (\ci \epsilon^{*} 
            - \ci \chi_{\text{qr}} \beta
            - \ci \Delta_{\text{rd}} \beta
            + \kappa \beta/2) 
            P_{ee}
        \right],
\end{align}
\begin{align}  \label{eq:complex_P_ff}
    \!\!\!\! \dot{P}_{ff}
    &= \frac{\partial}{\partial \alpha}
        \left[
            (-\ci \epsilon 
            + \ci 2\chi_{\text{qr}} \alpha 
            + \ci \Delta_{\text{rd}} \alpha 
            + \kappa \alpha/2) 
            P_{ff}
        \right]
\nonumber \\
    & \ \ \ \ 
        + \frac{\partial}{\partial \beta}
        \left[
            (\ci \epsilon^{*} 
            - \ci 2\chi_{\text{qr}} \beta
            - \ci \Delta_{\text{rd}} \beta
            + \kappa \beta/2) 
            P_{ff}
        \right],
\end{align}
\begin{align}
    \!\!\!\! \dot{P}_{ge}
    &= \frac{\partial}{\partial \alpha}
        \left[
            (- \ci \epsilon 
            + \ci \Delta_{\text{rd}} \alpha 
            + \kappa \alpha/2) 
            P_{ge}
        \right]
\nonumber \\
    & \ \ \ \ 
        + \frac{\partial}{\partial \beta}
        \left[
            (\ci \epsilon^{*}  
            - \ci \chi_{\text{qr}} \beta
            - \ci \Delta_{\text{rd}} \beta
            + \kappa \beta/2) 
            P_{ge}
        \right]
\nonumber \\ \label{eq:complex_P_ge}
    & \ \ \ \
        + \ci \chi_{\text{qr}} \alpha \beta P_{ge}
        + \ci \tilde{\omega}_{\text{q}} P_{ge} - \gamma_{2,ge} P_{ge},
\end{align}
\begin{align} 
    \!\!\!\! \dot{P}_{eg}
    &= \frac{\partial}{\partial \alpha}
        \left[
            (- \ci \epsilon 
            + \ci \chi_{\text{qr}} \alpha
            + \ci \Delta_{\text{rd}} \alpha 
            + \kappa \alpha/2) 
            P_{eg}
        \right]
\nonumber \\
    & \ \ \ \ 
        + \frac{\partial}{\partial \beta}
        \left[
            (\ci \epsilon^{*}  
            - \ci \Delta_{\text{rd}} \beta
            + \kappa \beta/2) 
            P_{eg}
        \right]
\nonumber \\ \label{eq:complex_P_eg}
    & \ \ \ \ 
        - \ci \chi_{\text{qr}} \alpha \beta P_{eg}
        - \ci \tilde{\omega}_{\text{q}} P_{eg} - \gamma_{2,ge} P_{eg},
\end{align}
\begin{align}
    \!\!\!\! \dot{P}_{gf}
    &= \frac{\partial}{\partial \alpha}
        \left[
            (- \ci \epsilon 
            + \ci \Delta_{\text{rd}} \alpha 
            + \kappa \alpha/2) 
            P_{gf}
        \right]
\nonumber \\
    & \ \ \ \ 
        + \frac{\partial}{\partial \beta}
        \left[
            (\ci \epsilon^{*}  
            - \ci 2\chi_{\text{qr}} \beta
            - \ci \Delta_{\text{rd}} \beta
            + \kappa \beta/2) 
            P_{gf}
        \right]
\nonumber \\ \label{eq:complex_P_gf}
    & \ \ \ \ 
        + \ci 2\chi_{\text{qr}} \alpha \beta P_{gf}
        + \ci (2\tilde{\omega}_{\text{q}} + \alpha_{\text{q}}) P_{gf} - \gamma_{2,gf} P_{gf},
\end{align}
\begin{align}
    \dot{P}_{fg}
    &= \frac{\partial}{\partial \alpha}
        \left[
            (- \ci \epsilon 
            + \ci 2\chi_{\text{qr}} \alpha 
            + \ci \Delta_{\text{rd}} \alpha 
            + \kappa \alpha/2) 
            P_{fg}
        \right]
\nonumber \\
    & \ \ \ \ 
        + \frac{\partial}{\partial \beta}
        \left[
            (\ci \epsilon^{*}  
            - \ci \Delta_{\text{rd}} \beta
            + \kappa \beta/2) 
            P_{fg}
        \right]
\nonumber \\ \label{eq:complex_P_fg}
    & \ \ \ \
        - \ci 2\chi_{\text{qr}} \alpha \beta P_{fg}
        - \ci (2\tilde{\omega}_{\text{q}} + \alpha_{\text{q}}) P_{fg} - \gamma_{2,gf} P_{fg},
\end{align}
\begin{align}
    \dot{P}_{ef}
    &= \frac{\partial}{\partial \alpha}
        \left[
            (- \ci \epsilon 
            + \ci \chi_{\text{qr}} \alpha 
            + \ci \Delta_{\text{rd}} \alpha 
            + \kappa \alpha/2) 
            P_{ef}
        \right]
\nonumber \\
    & \ \ \ \ 
        + \frac{\partial}{\partial \beta}
        \left[
            (\ci \epsilon^{*}  
            - \ci 2\chi_{\text{qr}} \beta
            - \ci \Delta_{\text{rd}} \beta
            + \kappa \beta/2) 
            P_{ef}
        \right]
\nonumber \\ \label{eq:complex_P_ef}
    & \ \ \ \ 
        + \ci \chi_{\text{qr}} \alpha \beta P_{ef}
        + \ci (\tilde{\omega}_{\text{q}} + \alpha_{\text{q}}) P_{ef} - \gamma_{2,ef} P_{ef},
\end{align}
\begin{align}
    \dot{P}_{fe}
    &= \frac{\partial}{\partial \alpha}
        \left[
            (- \ci \epsilon 
            + \ci 2\chi_{\text{qr}} \alpha 
            + \ci \Delta_{\text{rd}} \alpha 
            + \kappa \alpha/2) 
            P_{fe}
        \right]
\nonumber \\
    & \ \ \ \ 
        + \frac{\partial}{\partial \beta}
        \left[
            (\ci \epsilon^{*}  
            - \ci \chi_{\text{qr}} \beta
            - \ci \Delta_{\text{rd}} \beta
            + \kappa \beta/2) 
            P_{fe}
        \right]
\nonumber \\ \label{eq:complex_P_fe}
    & \ \ \ \ 
        - \ci \chi_{\text{qr}} \alpha \beta P_{fe}
        - \ci (\tilde{\omega}_{\text{q}} + \alpha_{\text{q}}) P_{fe} - \gamma_{2,ef} P_{fe}.
\end{align}
It should be noted that the differential equations of $P_{ab}$ are usually of the type of a Fokker-Planck equation, which also includes the diffusive terms (i.e., the second partial derivatives with respect to $\alpha$ and $\beta$). However, since we have assumed that $\Bar{N}(\omega_{\text{r}}) = 0$, there is no terms of the form
\begin{equation}
    \hat{a}^{\dagger} \hat{\rho}_{ab} \hat{a} 
    \ \longrightarrow \ \ 
    \left(
        \beta - \frac{\partial}{\partial \alpha} 
    \right)
    \left(
        \alpha - \frac{\partial}{\partial \beta} 
    \right)
    P(\alpha, \beta, t).
\end{equation}
Even in the case where $\Bar{N} > 0$, the method of the positive $P$-representation will still work but a sharp coherent state (see below) inside the cavity will broaden itself diffusively in the phase plane.

Although looking complicated, the nine coupled equations admit simple trajectories in the complex planes of $\alpha$ and $\beta$. We use the ansatze 
% \cite{PhysRevA.74.042318}
\begin{align}
    P_{gg}(\alpha, \beta, t)
    &= \delta^{(2)}(\alpha - \alpha_g(t)) 
        \delta^{(2)}(\beta - \alpha_g^*(t)),
\\
    P_{ee}(\alpha, \beta, t)
    &= \delta^{(2)}(\alpha - \alpha_e(t)) 
        \delta^{(2)}(\beta - \alpha_e^*(t)),
\\ 
    P_{ff}(\alpha, \beta, t)
    &= \delta^{(2)}(\alpha - \alpha_f(t)) 
        \delta^{(2)}(\beta - \alpha_f^*(t))
\end{align}
for the diagonal terms and 
\begin{align}
    \!\!\!\! P_{ge}(\alpha, \beta, t)
    &= c_{ge}(t) \delta^{(2)}(\alpha - \alpha_g(t)) 
        \delta^{(2)}(\beta - \alpha_e^*(t)),
\\
    \!\!\!\! P_{eg}(\alpha, \beta, t)
    &= c_{eg}(t) \delta^{(2)}(\alpha - \alpha_e(t)) 
        \delta^{(2)}(\beta - \alpha_g^*(t)),
\\ 
    \!\!\!\! P_{gf}(\alpha, \beta, t)
    &= c_{gf}(t) \delta^{(2)}(\alpha - \alpha_g(t)) 
        \delta^{(2)}(\beta - \alpha_f^*(t)),
\\ 
    \!\!\!\! P_{fg}(\alpha, \beta, t)
    &= c_{fg}(t) \delta^{(2)}(\alpha - \alpha_f(t)) 
        \delta^{(2)}(\beta - \alpha_g^*(t)),
\\ 
    \!\!\!\! P_{ef}(\alpha, \beta, t)
    &= c_{ef}(t) \delta^{(2)}(\alpha - \alpha_e(t)) 
        \delta^{(2)}(\beta - \alpha_f^*(t)),
\\
    \!\!\!\! P_{fe}(\alpha, \beta, t)
    &= c_{fe}(t)\delta^{(2)}(\alpha - \alpha_f(t)) 
        \delta^{(2)}(\beta - \alpha_e^*(t))
\end{align}
for off-diagonal terms. Each diagonal term $P_{aa}$ represents a single coherent state whose amplitudes are specified by the two delta functions. Plugging the ansatze into Eq.(\ref{eq:complex_P_gg})-(\ref{eq:complex_P_ff}), we obtain the time evolution of the coherent states 
\begin{align} \label{eq:differential_eqn_alpha_g}
    \dot{\alpha}_g
    &= - \ci (\Delta_{\text{rd}} - \ci \kappa/2) \alpha_g
        + \ci \epsilon,
\\ \label{eq:differential_eqn_alpha_e}
    \dot{\alpha}_e
    &=  - \ci (\Delta_{\text{rd}} + \chi_{\text{qr}} - \ci \kappa/2) \alpha_e
        + \ci \epsilon,
\\ \label{eq:differential_eqn_alpha_f}
    \dot{\alpha}_f
    &= - \ci (\Delta_{\text{rd}} + 2 \chi_{\text{qr}} - \ci \kappa/2) \alpha_f
        + \ci \epsilon.
\end{align}
Besides the time evolution brought by the coherent states, each non-diagonal term $P_{ab}$ ($a \neq b$) is modulated by an envelope function $c_{ab}$. By substituting the ansatze together with Eq.(\ref{eq:differential_eqn_alpha_g})-(\ref{eq:differential_eqn_alpha_f}) into Eq.(\ref{eq:complex_P_ge})-(\ref{eq:complex_P_fe}), we deduce that
\begin{align} \label{eq:c_ge_diff_eqn}
    \dot{c}_{ge}
    &= \ci(\tilde{\omega}_{\text{q}} + \ci \gamma_{2,ge}) c_{ge} 
        + \ci \chi_{\text{qr}} \alpha_g \alpha_e^* c_{ge},
\\ \label{eq:c_eg_diff_eqn}
    c_{eg}
    &= c_{ge}^{*},
\\ \label{eq:c_gf_diff_eqn}
    \dot{c}_{gf}
    &= \ci(2\tilde{\omega}_{\text{q}} + \alpha_{\text{q}} + \ci \gamma_{2,gf}) c_{gf} 
        + \ci 2\chi_{\text{qr}} \alpha_g \alpha_f^* c_{gf},
\\ \label{eq:c_fg_diff_eqn}
    c_{fg}
    &= c_{gf}^{*},
\\ \label{eq:c_ef_diff_eqn}
    \dot{c}_{ef}
    &= \ci(\tilde{\omega}_{\text{q}} + \alpha_{\text{q}} + \ci \gamma_{2,ef}) c_{ef} 
        + \ci \chi_{\text{qr}} \alpha_e \alpha_f^* c_{ef},
\\ \label{eq:c_fe_diff_eqn}
    c_{fe}
    &= c_{ef}^{*}.
\end{align}
Given the arbitrary initial conditions, the detailed time evolution of $\alpha_{a}$ and $c_{ab}$ can be solved numerically. After integrating the delta functions in the complex plane of $\alpha$ and $\beta$, we arrive at the general solution (in the rotating frame)
\begin{align}
    &\hat{\rho}_{\mathcal{SR}}(t) 
    = \sum_{a \in \{g,e,f\}} p_a(0) \ket{a} \! \bra{a} \otimes  \ket{\alpha_a(t)} \! \bra{\alpha_a(t)}
\nonumber \\ \label{eq:general_qutrit_resonator_solution}
    &+ \sum_{a,b \in \{g,e,f\}} \frac{c_{ab}(t)}{\bra{\alpha_{b}(t)}\ket{\alpha_{a}(t)}} \ket{a} \! \bra{b} \otimes  \ket{\alpha_a(t)} \! \bra{\alpha_b(t)},
\end{align}
where $p_{g,e,f}(0)$ are the initial populations in each energy eigenstate of the qutrit. The steady-state (complex) amplitudes of the cavity coherent states are given by
\begin{align} \label{eq:steady_state_coherent_amp_g}
    \alpha_g (+\infty)
    &= \frac{\sqrt{\kappa_{\text{in}}} \bar{a}_{\text{in}}}{\Delta_{\text{rd}} - \ci \kappa/2},
\\  \label{eq:steady_state_coherent_amp_e}
    \alpha_e (+\infty)
    &= \frac{\sqrt{\kappa_{\text{in}}} \bar{a}_{\text{in}}}{\Delta_{\text{rd}} + \chi_{\text{qr}} - \ci \kappa/2},
\\  \label{eq:steady_state_coherent_amp_f}
    \alpha_f (+\infty)
    &= \frac{\sqrt{\kappa_{\text{in}}} \bar{a}_{\text{in}}}{\Delta_{\text{rd}} + 2\chi_{\text{qr}} - \ci \kappa/2}.
\end{align}

In the Appendix \ref{Appendix:B} and \ref{Appendix:C}, we will approach the same problem with a different technique; nevertheless, the result will be identical, except that we will include $\gamma_{1,ab}$ for completeness.

\subsection{\label{A.2}Nonzero Temperature}
When $\Bar{N} > 0$, the master equation takes the form
\begin{align}
    \dot{\hat{\rho}}_{\mathcal{SR}}
    &= - \frac{\ci}{\hbar}
            \left[ 
                \hat{H}_{\text{eff}}, 
                \hat{\rho}_{\mathcal{SR}}
            \right]
        + \kappa \big(\bar{N} + 1 \big)
            \mathcalboondox{D}[\hat{a}] 
            \hat{\rho}_{\mathcal{SR}}
\nonumber \\
    & \ \ \ \ 
        + \kappa \bar{N} 
            \mathcalboondox{D}\Bigsl[\hat{a}^{\dagger}\Bigsr] 
            \hat{\rho}_{\mathcal{SR}}
        + \frac{\gamma_{2,ge}}{2} 
            \mathcalboondox{D}\big[\hat{\sigma}_{z,ge}\big]
            \hat{\rho}_{\mathcal{SR}}
\nonumber \\ \label{eq:general_qutrit_cavity_master_equation_finite_temp}
    & \ \ \ \ 
        + \frac{\gamma_{2,gf}}{2} 
            \mathcalboondox{D}\big[\hat{\sigma}_{z,gf}\big]
            \hat{\rho}_{\mathcal{SR}}
        + \frac{\gamma_{2,ef}}{2} 
            \mathcalboondox{D}\big[\hat{\sigma}_{z,ef}\big]
            \hat{\rho}_{\mathcal{SR}}
\end{align}
in the long-$T_1$ limit. The operator differential equations of $\hat{\rho}_{ab}$ are almost the same as before except that we replace $\kappa \mathcalboondox{D}[\hat{a}] \hat{\rho}_{ab}$ with 
\begin{equation}
    \kappa \big(\bar{N} + 1 \big)
        \mathcalboondox{D}[\hat{a}] \hat{\rho}_{ab}
    + \kappa \bar{N} 
            \mathcalboondox{D}\Bigsl[\hat{a}^{\dagger}\Bigsr] \hat{\rho}_{ab}.
\end{equation}
Consequently, the scalar differential equations for the positive $P$-representations $P_{ab}$ acquire the second partial derivatives mention before, i.e.,
\begin{equation} \label{eq:fokker_planck_P_ab}
    \dot{P}_{ab}
    = \big( \text{terms from the case $\bar{N} = 0$} \big) 
        + \kappa \bar{N} 
            \frac{\partial^2}{\partial \alpha \partial \beta} P_{ab}.
\end{equation}

Since Eq.(\ref{eq:fokker_planck_P_ab}) with $a=b$ has the same form as the classical Fokker-Planck equation, we use Gaussian distributions now as the new ansatze
% \cite{PhysRevA.74.042318}
\begin{align}
    &P_{gg}(\alpha, \beta, t)
\nonumber \\
    &= \frac{1}{\pi N(t)} 
        \exp
            \left\{
                - \frac{1}{N(t)} 
                \big[ \alpha - \alpha_g(t) \big]
                \big[ \beta - \alpha_g^*(t) \big]
            \right\},
\\
    &P_{ee}(\alpha, \beta, t)
\nonumber \\
    &= \frac{1}{\pi N(t)} 
        \exp
            \left\{
                - \frac{1}{N(t)} 
                \big[ \alpha - \alpha_e(t) \big]
                \big[ \beta - \alpha_e^*(t) \big]
            \right\},
\\ 
    &P_{ff}(\alpha, \beta, t)
\nonumber \\
    &= \frac{1}{\pi N(t)} 
        \exp
            \left\{
                - \frac{1}{N(t)} 
                \big[ \alpha - \alpha_f(t) \big]
                \big[ \beta - \alpha_f^*(t) \big]
            \right\},
\end{align}
where we require that $\alpha_{g,e,f}$ still satisfy Eq.(\ref{eq:differential_eqn_alpha_g}), (\ref{eq:differential_eqn_alpha_e}), and (\ref{eq:differential_eqn_alpha_f}), respectively. Substituting the Gaussian distributions into the Fokker-Planck equation, we obtain the differential equation of the variance $N$ (more precisely, the variance of the two-dimensional Gaussian is $N/2$)
\begin{equation}
    \dot{N}(t) 
    = - \kappa 
        \big[ 
            N(t) - \bar{N}
        \big].
\end{equation}

Suppose the composite system was in thermal equilibrium with the bath before receiving the drive $\varepsilon_{\text{d}}(t)$, then we will simply use $N(+\infty) = \Bar{N}$ in $P_{gg}$, $P_{ee}$, and $P_{ff}$. In other words, instead of building up a coherent state in the resonator, the external drive will excite a Gaussian state with a quadrature uncertainty broadened by the thermal bath. This also means that the resonator is a continuum combination of coherent states with amplitudes near $\alpha_g$, $\alpha_e$, and $\alpha_f$. In contrast, if the bath is in a vacuum state, $N(+\infty) = 0$ and a coherent state initially excited in the resonator will remain coherent.

Unlike the case where $\bar{N} = 0$, the coherence $c_{ab}$ also depends on $\alpha$ and $\beta$ now. This is because $c_{ab}$ are affected by an infinite collection of coherent states around $\alpha_g$, $\alpha_e$, and $\alpha_f$. Nevertheless, we expect $c_{ab}$ to vanish on similar timescales set by Eq.(\ref{eq:c_ge_diff_eqn})-(\ref{eq:c_fe_diff_eqn}).

\section{\label{Appendix:B}Deriving the Master Equation of the Combined System (Qutrit + Resonator) in the Displaced Frame}
\subsection{\label{B.1}Qutrit-State-Dependent Displacement}
Due to dispersive coupling, each eigenstate of the qutrit is entangled with a coherent state of the resonator. In the last section, we have found three differential equations for the complex amplitudes $\alpha_g$, $\alpha_e$, and $\alpha_f$ of the coherent states. To make this entanglement explicit, we define a unitary operator
\begin{equation} \label{eq:dispaced_frame_transformation}
    \hat{\mathsf{P}}(t)
    = \hat{\Pi}_g \hat{D}(\alpha_g(t))
        + \hat{\Pi}_e \hat{D}(\alpha_e(t))
        + \hat{\Pi}_f \hat{D}(\alpha_f(t)),
\end{equation}
where $\hat{\Pi}_a = \ket{a} \! \bra{a}$ are the projection operator onto the energy eigenstate $\ket{a}$ of the qutrit and $\hat{D}(\alpha_a(t))$ are the displacement operators (see the Eq.(\ref{eq:displacement_ops_def_main}) for the definition). Intuitively, $\hat{\mathsf{P}}$ entangles each projection $\hat{\Pi}_a$ of the qutrit with a displacement operator of the resonator such that if the qutrit is in an energy eigenstate, the resonator coherent state will be displaced to the vacuum state. For the subsequent derivation, we follow the notation used in \cite{PhysRevA.77.012112} and use
\begin{equation}
    \hat{O}^{\mathsf{P}}
    = \mathsf{P}^{\dagger} 
        \hat{O} 
        \mathsf{P}
\end{equation}
to denote any operator $\hat{O}$ in the displaced frame.

In the new frame, the density operator of the composite system is given by
\begin{equation}
    \hat{\rho}^{\mathsf{P}}(t) 
    = \mathsf{P}^{\dagger} \hat{\rho}(t) \mathsf{P},
\end{equation}
where, to simplify the notation, we will start to use $\hat{\rho} = \hat{\rho}_{\mathcal{SR}}$.
In addition, if we define
\begin{equation}
    \hat{\rho}^{\mathsf{P}}_{nmab} (t)
    = \bra{n, a} 
        \hat{\rho}^{\mathsf{P}} (t)
            \ket{m, b} 
\end{equation}
to be the matrix element of $\hat{\rho}^{\mathsf{P}}$ in the energy basis of the qutrit and the number basis of the resonator, then
\begin{equation}
    \hat{\rho}^{\mathsf{P}}
    = \sum_{n,m = 0}^{\infty}
        \sum_{a,b \in \{g,e,f\}}
            \hat{\rho}^{\mathsf{P}}_{nmab} 
                \ket{n, a} \! \bra{m, b} .
\end{equation}

Our goal is to find the time evolution of the qutrit reduced density operator, i.e.,
\begin{equation}
    \hat{\rho}_{\mathcal{S}}(t)
    = \Tr_{\mathcal{R}}
        \Bigsl[
            \hat{\rho}(t) 
        \Bigsr]
    = \Tr_{\mathcal{R}}
        \Bigsl[
            \mathsf{P} \hat{\rho}^{\mathsf{P}}(t) \mathsf{P}^{\dagger}
        \Bigsr].
\end{equation}
By using Eq.(\ref{eq:dispaced_frame_transformation}), we obtain
\begin{align}
    \hat{\rho}_{\mathcal{S}}(t)
    &= \sum_{n} 
            \Big(
                \rho^{\mathsf{P}}_{nngg}
                    \ket{g} \! \bra{g}
                + \rho^{\mathsf{P}}_{nnee}
                    \ket{e} \! \bra{e}
                + \rho^{\mathsf{P}}_{nnff}
                    \ket{f} \! \bra{f}
            \Big)
\nonumber \\
    & \ \ \ \ 
        + \sum_{n,m} 
            \Big(
                \lambda_{nmmn}^{ge}
                    \ket{g} \! \bra{e}
                + \lambda_{mnnm}^{ge *}
                    \ket{e} \! \bra{g}
            \Big)
\nonumber \\
    & \ \ \ \ 
        + \sum_{n,m} 
            \Big(
                \lambda_{nmmn}^{gf}
                    \ket{g} \! \bra{f}
                + \lambda_{mnnm}^{gf *}
                    \ket{f} \! \bra{g}
            \Big)
\nonumber \\ \label{eq:reduced_rho_in_terms_of_displaced_rho}
    & \ \ \ \ 
        + \sum_{n,m} 
            \Big(
                \lambda_{nmmn}^{ef}
                    \ket{e} \! \bra{f}
                + \lambda_{mnnm}^{ef *}
                    \ket{f} \! \bra{e}
            \Big),
\end{align}
where
\begin{align} \label{eq:lambda_nmpq_ge_def}
    \lambda_{nmpq}^{ge}(t)
    &= \rho^{\mathsf{P}}_{nmge} e^{-\ci \Im(\alpha_e \alpha_g^*)} d_{pq},
\\  \label{eq:lambda_nmpq_gf_def}
    \lambda_{nmpq}^{gf}(t)
    &= \rho^{\mathsf{P}}_{nmgf} e^{-\ci \Im(\alpha_f \alpha_g^*)} d_{pq},
\\  \label{eq:lambda_nmpq_ef_def}
    \lambda_{nmpq}^{ef}(t)
    &= \rho^{\mathsf{P}}_{nmef} e^{-\ci \Im(\alpha_f \alpha_e^*)} d_{pq},
\end{align}
with $d_{pq}(t) = \bra{p} \hat{D}(\beta_{ge}) \ket{q}$. To arrive at Eq.(\ref{eq:reduced_rho_in_terms_of_displaced_rho}), we have used the fact that
\begin{equation}
    \sum_{p} 
        \bra{m} \hat{D}^{\dagger}(\alpha) \ket{p}
        \bra{p} \hat{D}(\alpha) \ket{n}
    = \delta_{mn}
\end{equation}
since $\hat{D}(\alpha)$ is unitary.
Once the matrix elements of $\rho^{\mathsf{P}}$ (and thus $\lambda_{nmpq}^{ab}$) are known, the matrix elements of the qutrit reduced density operator can be computed trivially. For example, 
\begin{equation}
    \hat{\rho}_{\mathcal{S},gg}
    = \bra{g} \hat{\rho}_{\mathcal{S}} \ket{g}
    = \sum_{n} 
            \rho^{\mathsf{P}}_{nngg}
\end{equation}
and 
\begin{equation} \label{eq:rho_s_ge_and_lambda_nmmn_ge}
    \hat{\rho}_{\mathcal{S},ge}
    = \bra{g} \hat{\rho}_{\mathcal{S}} \ket{e}
    = \sum_{n,m} 
        \lambda_{nmmn}^{ge}.
\end{equation}

\subsection{\label{B.2}Master Equation in the Displaced Frame}
The density operator in the displaced frame satisfies the master equation
\begin{align}
    \dot{\hat{\rho}}^{\mathsf{P}}
    &= - \frac{\ci}{\hbar}
            \Big[ 
                \hat{H}_{\text{eff}}^{\mathsf{P}}, 
                \hat{\rho}^{\mathsf{P}}
            \Big]
        - \hat{\mathsf{P}}^{\dagger} 
            \dot{\hat{\mathsf{P}}}
            \hat{\rho}^{\mathsf{P}} 
        - \hat{\rho}^{\mathsf{P}}   
            \dot{\hat{\mathsf{P}}}^{\dagger}
            \hat{\mathsf{P}}
        + \kappa 
            \mathcalboondox{D}
                \Bigsl[
                    \hat{a}^{\mathsf{P}}
                \Bigsr] 
            \hat{\rho}^{\mathsf{P}}
\nonumber \\
    & \ \ \ \ 
        + \gamma_{1,ge} 
            \mathcalboondox{D}
                \Bigsl[
                    \hat{\sigma}_{ge}^{\mathsf{P}}
                \Bigsr] 
            \hat{\rho}^{\mathsf{P}}
        + \gamma_{1,gf} 
            \mathcalboondox{D}
                \Bigsl[
                    \hat{\sigma}_{gf}^{\mathsf{P}}
                \Bigsr] 
            \hat{\rho}^{\mathsf{P}}
\nonumber \\
    & \ \ \ \ 
        + \gamma_{1,ef} 
            \mathcalboondox{D}
                \Bigsl[
                    \hat{\sigma}_{ef}^{\mathsf{P}}
                \Bigsr] 
            \hat{\rho}^{\mathsf{P}}
        + \frac{\gamma_{\phi,ge}}{2} 
            \mathcalboondox{D}
                \Bigsl[
                    \hat{\sigma}_{z,ge}^{\mathsf{P}}
                \Bigsr]
            \hat{\rho}^{\mathsf{P}}
\nonumber \\
    & \ \ \ \ 
        + \frac{\gamma_{\phi,gf}}{2} 
            \mathcalboondox{D}
                \Bigsl[
                    \hat{\sigma}_{z,gf}^{\mathsf{P}}
                \Bigsr]
            \hat{\rho}^{\mathsf{P}}
        + \frac{\gamma_{\phi,ef}}{2} 
            \mathcalboondox{D}
                \Bigsl[
                    \hat{\sigma}_{z,ef}^{\mathsf{P}}
                \Bigsr]
            \hat{\rho}^{\mathsf{P}}.
\end{align}
As for any time-dependent unitary transformation, the extra terms $- \hat{\mathsf{P}}^{\dagger} \dot{\hat{\mathsf{P}}} \hat{\rho}^{\mathsf{P}} - \hat{\rho}^{\mathsf{P}} \dot{\hat{\mathsf{P}}}^{\dagger} \hat{\mathsf{P}}$ appear in the new master equation to eliminate the readout drive terms in the original master equation. Moreover, the Hamiltonian of the combined system (qutrit + cavity) still takes the form
\begin{align}
    \hat{H}_{\text{eff}} / \hbar
    &= \hat{H}_{\mathcal{SR},\text{rot}}^{\text{disp}} / \hbar
\nonumber \\
    &= \tilde{\omega}_{\text{q}} 
            \hat{\Pi}_e
        + (
                2 \tilde{\omega}_{\text{q}} 
                + \alpha_{\text{q}}
            ) 
            \hat{\Pi}_f
        + \Delta_{\text{rd}} 
            \hat{a}^{\dagger} \hat{a} 
\nonumber \\
    & \ \ \ \ 
        + \chi_{\text{qr}} 
                \Bigsl(
                    \hat{\Pi}_e 
                    + 2 \hat{\Pi}_f
                \Bigsr)
            \hat{a}^{\dagger} \hat{a} 
        - \left( 
            \epsilon \hat{a}^{\dagger}
            + \epsilon^* \hat{a}
        \right).
\end{align}
Note, however, we have kept the qutrit decay (i.e., $\gamma_{1,ab}$) in the master equation for full generality.

To begin simplifying each term, we will need various operators rewritten in the displaced frame. For the cavity operators, we have
\begin{equation}
\label{eq:displaced_a_op}
    \hat{a}^{\mathsf{P}}
    = \hat{a}
        + \left(
            \alpha_g \hat{\Pi}_g 
            + \alpha_e \hat{\Pi}_e
            + \alpha_f \hat{\Pi}_f
            \right)
    = \hat{a} + \hat{\Pi}_{\alpha},
\end{equation}
\begin{align}
    \Bigsl( 
        \hat{a}^{\dagger}\hat{a} 
    \Bigsr)^{\mathsf{P}}
    &= \hat{a}^{\dagger}\hat{a} 
        + \hat{a}^{\dagger} \hat{\Pi}_{\alpha}
        + \hat{a} \hat{\Pi}_{\alpha}^{\dagger}
        + \hat{\Pi}_{\alpha}^{\dagger} 
            \hat{\Pi}_{\alpha}
\nonumber \\
    &= \hat{a}^{\dagger}\hat{a} 
        + \hat{a}^{\dagger} \hat{\Pi}_{\alpha}
        + \hat{a} \hat{\Pi}_{\alpha}^{\dagger}
\nonumber \\
    & \ \ \ \ 
        + |\alpha_g|^2 \hat{\Pi}_{g}
        + |\alpha_e|^2 \hat{\Pi}_{e}
        + |\alpha_f|^2 \hat{\Pi}_{f},
\end{align}
where we denote
\begin{equation}
    \hat{\Pi}_{\alpha}(t) 
    = \alpha_g(t) \hat{\Pi}_g 
        + \alpha_e(t) \hat{\Pi}_e
        + \alpha_f(t) \hat{\Pi}_f.
\end{equation}
Similarly, the operators associated with the qutrit subspace in the displaced frame are given by
\begin{equation} \label{eq:displaced_sigma_minus_op}
    \hat{\sigma}_{z,ab}^{\mathsf{P}}
    = \hat{\sigma}_{z,ab}
\ \ \text{ and } \ \ 
    \hat{\sigma}_{ab}^{\mathsf{P}}
    = \hat{\sigma}_{ab} 
        \hat{D}^{\dagger}(\alpha_a)
        \hat{D}(\alpha_b).
\end{equation} 

First, we start with the transformed Hamiltonian. By using Eq.(\ref{eq:displaced_a_op})-(\ref{eq:displaced_sigma_minus_op}), we obtain
\begin{align}
    \frac{\hat{H}_{\text{eff}}^{\mathsf{P}}}{\hbar}
    &= \tilde{\omega}_{\text{q}} 
            \ket{e}\!\bra{e}
        + (
                2\tilde{\omega}_{\text{q}} 
                + \alpha_{\text{q}}
            ) 
            \ket{f}\!\bra{f}
\nonumber \\
    & \ \ \ \
        - \left[ 
                \epsilon    
                    \Big(
                        \hat{a}^{\dagger} + \hat{\Pi}_{\alpha}^{\dagger} 
                    \Big)
                + \epsilon^*
                    \Big(
                        \hat{a} + \hat{\Pi}_{\alpha}
                    \Big)
        \right]
\nonumber \\
    & \ \ \ \
        + \Big[ 
                \Delta_{\text{rd}} 
                + \chi_{\text{qr}} 
                    \Bigsl(
                        \hat{\Pi}_e 
                        + 2 \hat{\Pi}_f
                    \Bigsr)
            \Big]
\nonumber \\
    & \ \ \ \ \ \ \ \ 
            \times \Big(
                \hat{a}^{\dagger}\hat{a} 
                + \hat{a}^{\dagger} \hat{\Pi}_{\alpha}
                + \hat{a} \hat{\Pi}_{\alpha}^{\dagger}
\nonumber \\
    & \ \ \ \ \ \ \ \  \ \ \ \ 
                + |\alpha_g|^2 \hat{\Pi}_{g}
                + |\alpha_e|^2 \hat{\Pi}_{e}
                + |\alpha_f|^2 \hat{\Pi}_{f}
            \Big)
\nonumber \\
    &= \tilde{\omega}_{\text{q}} 
            \ket{e}\!\bra{e}
        + (
                2\tilde{\omega}_{\text{q}} 
                + \alpha_{\text{q}}
            ) 
            \ket{f}\!\bra{f}
\nonumber \\
    & \ \ \ \
        - \Big(
            \epsilon \hat{a}^{\dagger}
            + \epsilon^{*} \hat{a}
            \Big)
        - \Big(
            \epsilon \hat{\Pi}_{\alpha}^{\dagger}
            + \epsilon^{*} \hat{\Pi}_{\alpha}
            \Big)
\nonumber \\
    & \ \ \ \
        + \Big[ 
                \Delta_{\text{rd}} 
                + \chi_{\text{qr}} 
                    \Bigsl(
                        \hat{\Pi}_e 
                        + 2 \hat{\Pi}_f
                    \Bigsr)
            \Big]
            \hat{a}^{\dagger}\hat{a} 
\nonumber \\
    & \ \ \ \
        + \Big[ 
                \Delta_{\text{rd}} 
                + \chi_{\text{qr}} 
                (
                \ket{e}\!\bra{e} + 2 \ket{f}\!\bra{f}
                )
            \Big]
            \Big( 
                \hat{a}^{\dagger} \hat{\Pi}_{\alpha}
                + \hat{a} \hat{\Pi}_{\alpha}^{\dagger}
            \Big)
\nonumber \\
    & \ \ \ \
        + \Big[ 
                \Delta_{\text{rd}} 
                + \chi_{\text{qr}} 
                    \Bigsl(
                        \hat{\Pi}_e 
                        + 2 \hat{\Pi}_f
                    \Bigsr)
            \Big]
\nonumber \\
    & \ \ \ \  \ \ \ \
            \times
            \Big(
                |\alpha_g|^2 \hat{\Pi}_{g}
                + |\alpha_e|^2 \hat{\Pi}_{e}
                + |\alpha_f|^2 \hat{\Pi}_{f}
            \Big).
\end{align}

Next, to simplify $- \hat{\mathsf{P}}^{\dagger} \dot{\hat{\mathsf{P}}} \hat{\rho}^{\mathsf{P}} - \hat{\rho}^{\mathsf{P}} \dot{\hat{\mathsf{P}}}^{\dagger} \hat{\mathsf{P}}$, we invoke the identity
\begin{align}
    &\frac{\mathrm{d}}{\mathrm{d} t} 
            \hat{D}(\alpha(t))
\nonumber \\
    &= \dot{\alpha} \hat{a}^{\dagger} \hat{D}(\alpha(t))
        - \hat{D}(\alpha(t)) \dot{\alpha}^* \hat{a}
        - \frac{\alpha^* \dot{\alpha} + \dot{\alpha}^* \alpha}{2} \hat{D}(\alpha(t))
\nonumber \\ \label{eq:time_derivative_displacement_op}
    &= \left[ 
            \dot{\alpha} \hat{a}^{\dagger}
            - \dot{\alpha}^* \hat{a}
            - \frac{\alpha^* \dot{\alpha} - \dot{\alpha}^* \alpha}{2}
        \right]
        \hat{D}(\alpha(t))
\end{align}
for a time-dependent displacement operator. Note that $\partial_t e^{\hat{A}(t)}  = \dot{\hat{A}} (t) e^{\hat{A}(t)}$ is only true if $\Bigsl[\hat{A}(t), \hat{A}(s)\Bigsr] = 0$ for any $t$ and $s$, which does not apply to the case of the displacement operator. Consequently, using Eq.(\ref{eq:time_derivative_displacement_op}) results in
\begin{align}
    \dot{\hat{\mathsf{P}}}
    &= \hat{\Pi}_g
            \left[ 
                \left(
                    \dot{\alpha}_g \hat{a}^{\dagger} 
                    - \dot{\alpha}_g^* \hat{a}
                \right) 
                + \left(
                    \dot{\alpha}_g^* \alpha_g
                    - \alpha_g^* \dot{\alpha}_g
                \right)/2 
            \right] \hat{D}(\alpha_g)
\nonumber \\
    & \ \ \ \ 
        + \hat{\Pi}_e
            \left[
                \left(
                    \dot{\alpha}_e \hat{a}^{\dagger} 
                    - \dot{\alpha}_e^* \hat{a}
                \right) 
                + \left(
                    \dot{\alpha}_e^* \alpha_e
                    - \alpha_e^* \dot{\alpha}_e
                \right) /2
            \right] \hat{D}(\alpha_e)
\nonumber \\
    & \ \ \ \ 
        + \hat{\Pi}_f 
            \left[
                \left(
                    \dot{\alpha}_f \hat{a}^{\dagger} 
                    - \dot{\alpha}_f^* \hat{a}
                \right) 
                + \left(
                    \dot{\alpha}_f^* \alpha_f
                    - \alpha_f^* \dot{\alpha}_f
                \right) /2
            \right] \hat{D}(\alpha_f)
\end{align}
and
\begin{align}
    &\hat{\mathsf{P}}^{\dagger} \dot{\hat{\mathsf{P}}}
\nonumber \\
    &= \hat{\Pi}_g 
            \hat{D}^{\dagger}(\alpha_g)
            \big[ 
                \left(
                    \dot{\alpha}_g \hat{a}^{\dagger} 
                    - \dot{\alpha}_g^* \hat{a}
                \right) 
\nonumber \\
    & \ \ \ \ \ \ \ \ \ \ \ \ \ \ \ \ \ \ \ \
                + \left(
                    \dot{\alpha}_g^* \alpha_g
                    - \alpha_g^* \dot{\alpha}_g
                \right) /2
            \big] 
            \hat{D}(\alpha_g)
\nonumber \\
    & \ \ \ \ 
        + \hat{\Pi}_e 
            \hat{D}^{\dagger}(\alpha_e)
            \big[
                \left(
                    \dot{\alpha}_e \hat{a}^{\dagger} 
                    - \dot{\alpha}_e^* \hat{a}
                \right) 
\nonumber \\
    & \ \ \ \ \ \ \ \ \ \ \ \ \ \ \ \ \ \ \ \ \ \ \ \ 
                + \left(
                    \dot{\alpha}_e^* \alpha_e
                    - \alpha_e^* \dot{\alpha}_e
                \right) /2
            \big] 
            \hat{D}(\alpha_e)
\nonumber \\
    & \ \ \ \ 
        + \hat{\Pi}_f 
            \hat{D}^{\dagger} (\alpha_f)
            \big[
                \left(
                    \dot{\alpha}_f \hat{a}^{\dagger} 
                    - \dot{\alpha}_f^* \hat{a}
                \right) 
\nonumber \\
    & \ \ \ \ \ \ \ \ \ \ \ \ \ \ \ \ \ \ \ \ \ \ \ \ 
                + \left(
                    \dot{\alpha}_f^* \alpha_f
                    - \alpha_f^* \dot{\alpha}_f
                \right) /2
            \big] 
            \hat{D}(\alpha_f)
\nonumber \\
    &= \hat{\Pi}_g 
            \left[ 
                \left(
                    \dot{\alpha}_g \hat{a}^{\dagger} 
                    - \dot{\alpha}_g^* \hat{a}
                \right) 
                - \left(
                    \dot{\alpha}_g^* \alpha_g
                    - \alpha_g^* \dot{\alpha}_g
                \right) /2
            \right]
\nonumber \\
    & \ \ \ \ 
        + \hat{\Pi}_e 
            \left[
                \left(
                    \dot{\alpha}_e \hat{a}^{\dagger} 
                    - \dot{\alpha}_e^* \hat{a}
                \right) 
                - \left(
                    \dot{\alpha}_e^* \alpha_e
                    - \alpha_e^* \dot{\alpha}_e
                \right) /2
            \right] 
% \nonumber \\
%     & \ \ \ \
%         + \hat{\Pi}_f 
%             \left[
%                 \left(
%                     \dot{\alpha}_f \hat{a}^{\dagger} 
%                     - \dot{\alpha}_f^* \hat{a}
%                 \right) 
%                 - \left(
%                     \dot{\alpha}_f^* \alpha_f
%                     - \alpha_f^* \dot{\alpha}_f
%                 \right) /2
%             \right] 
\nonumber \\
    &= \dot{\hat{\Pi}}_{\alpha} \hat{a}^{\dagger} 
        - \dot{\hat{\Pi}}_{\alpha}^{\dagger}  \hat{a}
        + \ci \Im(\alpha_g^* \dot{\alpha}_g) \hat{\Pi}_g
\nonumber \\
    & \ \ \ \ 
        + \ci \Im(\alpha_e^* \dot{\alpha}_e) \hat{\Pi}_e
        + \ci \Im(\alpha_f^* \dot{\alpha}_f) \hat{\Pi}_f.
\end{align}
Since $\dot{\hat{\mathsf{P}}}^{\dagger} \hat{\mathsf{P}} = - \hat{\mathsf{P}}^{\dagger} \dot{\hat{\mathsf{P}}}$, we have
\begin{align}
    &- \hat{\mathsf{P}}^{\dagger} 
            \dot{\hat{\mathsf{P}}}
            \hat{\rho}^{\mathsf{P}} 
        - \hat{\rho}^{\mathsf{P}}   
            \dot{\hat{\mathsf{P}}}^{\dagger}
            \hat{\mathsf{P}}
\nonumber \\ \label{eq:displaced_frame_time_derivative_terms}
    &= - \left[
                \dot{\hat{\Pi}}_{\alpha} 
                    \hat{a}^{\dagger}
                - \dot{\hat{\Pi}}_{\alpha}^{\dagger} 
                    \hat{a},
                \hat{\rho}^{\mathsf{P}} 
            \right]
\nonumber \\
    & \ \ \ \ 
        - \ci 
            \left[
                \Im(\alpha_g^* \dot{\alpha}_g) \hat{\Pi}_g
                + \Im(\alpha_e^* \dot{\alpha}_e) \hat{\Pi}_e
                + \Im(\alpha_f^* \dot{\alpha}_f) \hat{\Pi}_f,
                \hat{\rho}^{\mathsf{P}} 
            \right].
\end{align}
To proceed further, we substitute Eq.(\ref{eq:differential_eqn_alpha_g})-(\ref{eq:differential_eqn_alpha_f}) found for the combined system into Eq.(\ref{eq:displaced_frame_time_derivative_terms}). In particular, 
\begin{align}
    \dot{\hat{\Pi}}_{\alpha} 
    &= \dot{\alpha}_g \hat{\Pi}_g
        + \dot{\alpha}_e \hat{\Pi}_e
        + \dot{\alpha}_f \hat{\Pi}_f
\nonumber \\
    &= \Big[
             - \ci (\Delta_{\text{rd}} - \ci \kappa/2) \alpha_g
            + \ci \epsilon
        \Big] \hat{\Pi}_g
\nonumber \\
    & \ \ \ \ 
        + \Big[
             - \ci (\Delta_{\text{rd}} + \chi_{\text{qr}} - \ci \kappa/2) \alpha_e
            + \ci \epsilon
        \Big] \hat{\Pi}_e
\nonumber \\
    & \ \ \ \ 
        + \Big[
             - \ci (\Delta_{\text{rd}} + 2 \chi_{\text{qr}} - \ci \kappa/2) \alpha_f
             + \ci \epsilon
        \Big] \hat{\Pi}_f
\nonumber \\
    &= \ci \epsilon
        - \ci 
            \Big[ 
                \Delta_{\text{rd}} 
                + \chi_{\text{qr}} 
                    \Bigsl(
                        \hat{\Pi}_e 
                        + 2 \hat{\Pi}_f
                    \Bigsr)
            \Big]
            \hat{\Pi}_{\alpha}
        - \frac{\kappa}{2} \hat{\Pi}_{\alpha}
\end{align}
and the first term in Eq.(\ref{eq:displaced_frame_time_derivative_terms}) becomes
\begin{align}
    &\dot{\hat{\Pi}}_{\alpha} \hat{a}^{\dagger}
        - \dot{\hat{\Pi}}_{\alpha}^{\dagger} \hat{a}
\nonumber \\
    &= \ci \epsilon \hat{a}^{\dagger}
        - \ci 
            \Big[ 
                \Delta_{\text{rd}} 
                + \chi_{\text{qr}} 
                    \Bigsl(
                        \hat{\Pi}_e 
                        + 2 \hat{\Pi}_f
                    \Bigsr)
            \Big]
            \hat{\Pi}_{\alpha} 
            \hat{a}^{\dagger}
        - \frac{\kappa}{2} \hat{\Pi}_{\alpha}
            \hat{a}^{\dagger}
\nonumber \\
    & \ \ \ \ 
        + \ci \epsilon^* \hat{a}
        - \ci 
            \Big[ 
                \Delta_{\text{rd}} 
                + \chi_{\text{qr}} 
                    \Bigsl(
                        \hat{\Pi}_e 
                        + 2 \hat{\Pi}_f
                    \Bigsr)
            \Big]
            \hat{\Pi}_{\alpha}^{\dagger}
            \hat{a}
        + \frac{\kappa}{2} \hat{\Pi}_{\alpha}^{\dagger}
            \hat{a}
\nonumber \\
    &= \ci \Big(
            \epsilon \hat{a}^{\dagger}
            + \epsilon^* \hat{a}
        \Big)
\nonumber \\
    & \ \ \ \ 
        - \ci 
            \Big[ 
                \Delta_{\text{rd}} 
                + \chi_{\text{qr}} 
                    \Bigsl(
                        \hat{\Pi}_e 
                        + 2 \hat{\Pi}_f
                    \Bigsr)
            \Big]
            \Big(
                \hat{\Pi}_{\alpha} 
                    \hat{a}^{\dagger}
                + \hat{\Pi}_{\alpha}^{\dagger}
                    \hat{a}
            \Big)
\nonumber \\
    & \ \ \ \ 
        - \frac{\kappa}{2} 
            \Big(
                \hat{\Pi}_{\alpha} 
                    \hat{a}^{\dagger}
                - \hat{\Pi}_{\alpha}^{\dagger}
                    \hat{a}
            \Big).
\end{align}
With a similar manipulation, the second term in Eq.(\ref{eq:displaced_frame_time_derivative_terms}) reduces to 
\begin{align}
    &
    - \ci \Im(\alpha_g^* \dot{\alpha}_g) \hat{\Pi}_g
        - \ci \Im(\alpha_e^* \dot{\alpha}_e) \hat{\Pi}_e
        - \ci \Im(\alpha_f^* \dot{\alpha}_f) \hat{\Pi}_f
\nonumber \\
    &= -\ci \Big[
                \Im(\ci \alpha_g^* \epsilon ) 
                - \Delta_{\text{rd}} |\alpha_g|^2
            \Big]
            \hat{\Pi}_g
\nonumber \\
    & \ \ \ \
        -\ci \Big[ 
                \Im( \ci \alpha_e^* \epsilon ) 
                - (\Delta_{\text{rd}} + \chi_{\text{qr}}) |\alpha_e|^2
            \Big]
            \hat{\Pi}_e
\nonumber \\
    & \ \ \ \ 
        -\ci \Big[ 
                \Im( \ci \alpha_f^* \epsilon )
                - (\Delta_{\text{rd}} + 2\chi_{\text{qr}}) |\alpha_e|^2
            \Big] \hat{\Pi}_f
\nonumber \\
    &= - \ci 
            \Big[ 
                \Im(\ci \alpha_g^* \epsilon ) 
                    \hat{\Pi}_g
                + \Im(\ci \alpha_e^* \epsilon ) 
                    \hat{\Pi}_e
                + \Im(\ci \alpha_f^* \epsilon ) 
                    \hat{\Pi}_f
            \Big]
\nonumber \\ \label{eq:expression_with_identity_in_it}
    & \ \ \ \
        + \ci \Big[ 
                \Delta_{\text{rd}} 
                + \chi_{\text{qr}} 
                (
                \ket{e}\!\bra{e} + 2 \ket{f}\!\bra{f}
                )
            \Big]
\nonumber \\
    & \ \ \ \ \ \ \ \
            \times \Big( 
                |\alpha_g|^2 \hat{\Pi}_g
                + |\alpha_e|^2 \hat{\Pi}_e
                + |\alpha_f|^2 \hat{\Pi}_f
            \Big).
\end{align}
Since $\Bigsl[\hat{1}, \hat{\rho}^{\mathsf{P}}\Bigsr] = 0$, we can remove a multiple of the identity operator from the first term of Eq.(\ref{eq:expression_with_identity_in_it}); specifically, we can write
\begin{align}
    &- \ci 
        \Big[ 
            \Im(\ci \alpha_g^* \epsilon ) 
                \hat{\Pi}_g
            + \Im(\ci \alpha_e^* \epsilon ) 
                \hat{\Pi}_e
            + \Im(\ci \alpha_f^* \epsilon ) 
                \hat{\Pi}_f
        \Big]
\nonumber \\
    &= - \ci \frac{\alpha_g^* \epsilon + \alpha_g \epsilon^*}{2} 
        \hat{\Pi}_g
        - \ci \frac{\alpha_e^* \epsilon + \alpha_e \epsilon^*}{2} 
        \hat{\Pi}_e
        - \ci \frac{\alpha_f^* \epsilon + \alpha_f \epsilon^*}{2} 
        \hat{\Pi}_f
\nonumber \\
    &= - \frac{\ci}{2} \frac{\epsilon^*(\alpha_g+\alpha_e+\alpha_f)+\epsilon(\alpha_g+\alpha_e+\alpha_f)^*}{3} \hat{1} 
\nonumber \\
    & \ \ \ \
        - \frac{\ci}{2} \frac{\epsilon^* \beta_{ge} + \epsilon \beta_{ge}^*}{3} 
        \hat{\sigma}_{z,ge}
        - \frac{\ci}{2} \frac{\epsilon^* \beta_{gf} + \epsilon \beta_{gf}^*}{3} 
        \hat{\sigma}_{z,gf}
\nonumber \\
    & \ \ \ \ 
        - \frac{\ci}{2} \frac{\epsilon^* \beta_{ef} + \epsilon \beta_{ef}^*}{3} 
        \hat{\sigma}_{z,ef}
\\[2mm]
    &= - \ci C_1 \hat{1} 
        + \ci \Delta_{g,1}
            \hat{\Pi}_{g}
        + \ci \Delta_{e,1}
            \hat{\Pi}_{e}
        + \ci \Delta_{f,1}
            \hat{\Pi}_{f},
\end{align}
where we have defined $C_1 = [\epsilon^*(\alpha_g+\alpha_e+\alpha_f)+\epsilon(\alpha_g+\alpha_e+\alpha_f)^*]/6$, 
\begin{align}
    \beta_{ge}(t) &= \alpha_g(t) - \alpha_e(t),
\\
    \beta_{gf}(t) &= \alpha_g(t) - \alpha_f(t),
\\
    \beta_{ef}(t) &= \alpha_e(t) - \alpha_f(t),
\end{align}
and 
\begin{align}
    \Delta_{g,1}(t) 
    &= \frac{1}{6} \Big[
            - \Big(
                \epsilon^* \beta_{ge} 
                +\epsilon \beta_{ge}^*
            \Big)
            - \Big(
                \epsilon^* \beta_{gf} 
                +\epsilon \beta_{gf}^*
            \Big)
        \Big],
\\
    \Delta_{e,1}(t)
    &= \frac{1}{6} \Big[
            + \Big(
                \epsilon^* \beta_{ge} 
                +\epsilon \beta_{ge}^*
            \Big)
            - \Big(
                \epsilon^* \beta_{ef} 
                +\epsilon \beta_{ef}^*
            \Big)
        \Big],
\\
    \Delta_{f,1}(t)
    &= \frac{1}{6} \Big[
            + \Big(
                \epsilon^* \beta_{gf} 
                +\epsilon \beta_{gf}^*
            \Big)
            + \Big(
                \epsilon^* \beta_{ef} 
                +\epsilon \beta_{ef}^*
            \Big)
        \Big].
\end{align}
In addition, the same argument can be applied to terms in $\hat{H}_{\text{eff}}^{\mathsf{P}}$, i.e.,
\begin{align}
    &\epsilon \hat{\Pi}_{\alpha}^{\dagger}
        + \epsilon^{*} \hat{\Pi}_{\alpha}
\nonumber \\
    &= 2C_1 \hat{1} 
        - 2 \Delta_{g,1}
            \hat{\Pi}_{g}
        - 2 \Delta_{e,1}
            \hat{\Pi}_{e}
        - 2 \Delta_{f,1}
            \hat{\Pi}_{f},
\end{align}
and the net effect is that
\begin{align}
    & -\frac{\ci}{\hbar}    
            \Bigsl[
                \hat{H}_{\text{eff}}^{\mathsf{P}},
                \hat{\rho}^{\mathsf{P}}   
            \Bigsr]
        - \hat{\mathsf{P}}^{\dagger} 
            \dot{\hat{\mathsf{P}}}
            \hat{\rho}^{\mathsf{P}} 
        - \hat{\rho}^{\mathsf{P}}   
            \dot{\hat{\mathsf{P}}}^{\dagger}
            \hat{\mathsf{P}}
\nonumber\\
    &= - \ci
            \Big[ 
                    \Delta_{g,1} 
                    \hat{\Pi}_g
                + (
                    \tilde{\omega}_{\text{q}} 
                    + \Delta_{e,1}
                    )
                    \hat{\Pi}_e
\nonumber \\
    & \ \ \ \  \ \ \ \  \ \ \ \  \ \ \ \  \ \ \ \
                + (
                    2\tilde{\omega}_{\text{q}} 
                    + \Delta_{f,1})
                    + \alpha_{\text{q}}
                    ) 
                    \hat{\Pi}_f,
                \hat{\rho}^{\mathsf{P}}   
            \Big]
\nonumber \\ \label{eq:H_eff_and_displaced_frame_derivative_terms}
    & \ \ \ \
        - \ci 
            \Big[ 
                \Bigsl[ 
                    \Delta_{\text{rd}} 
                    + \chi_{\text{qr}} 
                    \Bigsl(
                        \hat{\Pi}_e 
                        + 2 \hat{\Pi}_f
                    \Bigsr)
                \Bigsr]
                \hat{a}^{\dagger}\hat{a}, 
                \hat{\rho}^{\mathsf{P}}   
            \Big]
\nonumber \\
    & \ \ \ \ 
        + \frac{\kappa}{2} 
            \Big[
                \hat{\Pi}_{\alpha} 
                    \hat{a}^{\dagger}
                - \hat{\Pi}_{\alpha}^{\dagger}
                    \hat{a},
                \hat{\rho}^{\mathsf{P}}   
            \Big].
\end{align}

Now, we focus our attention on the cavity decay term
\begin{align}
    &\mathcalboondox{D}
        \Bigsl[
            \hat{a}^{\mathsf{P}}
        \Bigsr] 
        \hat{\rho}^{\mathsf{P}}
\nonumber \\
    &= \Big( 
                \hat{a} + \hat{\Pi}_{\alpha}
            \Big)
            \hat{\rho}^{\mathsf{P}}
            \Big( 
                \hat{a}^{\dagger} + \hat{\Pi}_{\alpha}^{\dagger}
            \Big)
\nonumber \\
    & \ \ \ \ 
        - \frac{1}{2}
            \hat{\rho}^{\mathsf{P}}
            \Big(
                \hat{a}^{\dagger}\hat{a} 
                + \hat{a}^{\dagger} 
                    \hat{\Pi}_{\alpha}
                + \hat{a} 
                    \hat{\Pi}_{\alpha}^{\dagger}
                + \hat{\Pi}_{\alpha}^{\dagger}
                    \hat{\Pi}_{\alpha}
            \Big)
\nonumber \\
    & \ \ \ \ 
        - \frac{1}{2}
            \Big(
                \hat{a}^{\dagger}\hat{a} 
                + \hat{a}^{\dagger} 
                    \hat{\Pi}_{\alpha}
                + \hat{a} 
                    \hat{\Pi}_{\alpha}^{\dagger}
                + \hat{\Pi}_{\alpha}^{\dagger}
                    \hat{\Pi}_{\alpha}
            \Big)
            \hat{\rho}^{\mathsf{P}}
\nonumber \\
    &= \mathcalboondox{D}
            \Bigsl[
                \hat{a}
            \Bigsr] 
            \hat{\rho}^{\mathsf{P}}
        + \mathcalboondox{D}
            \Bigsl[
                \hat{\Pi}_{\alpha} 
            \Bigsr] 
            \hat{\rho}^{\mathsf{P}}
        + \hat{a} 
            \hat{\rho}^{\mathsf{P}}  
            \hat{\Pi}^{\dagger}
        + \hat{a}^{\dagger}
            \hat{\rho}^{\mathsf{P}}  
            \hat{\Pi}
\nonumber \\
    & \ \ \ \ 
        - \frac{1}{2}
            \hat{\rho}^{\mathsf{P}}
            \hat{a}^{\dagger} 
            \hat{\Pi}_{\alpha}
        - \frac{1}{2}
            \hat{\rho}^{\mathsf{P}}
            \hat{a} 
            \hat{\Pi}_{\alpha}^{\dagger}
        - \frac{1}{2}
            \hat{a}^{\dagger} 
            \hat{\Pi}_{\alpha}
            \hat{\rho}^{\mathsf{P}}
        - \frac{1}{2} \hat{a} 
            \hat{\Pi}_{\alpha}^{\dagger}
            \hat{\rho}^{\mathsf{P}}.
\end{align}
The second term $\mathcalboondox{D} \Bigsl[\hat{\Pi}_{\alpha} \Bigsr] \hat{\rho}^{\mathsf{P}}$ contains both frequency shifts and dephasing. To separate the two effects, we can simply expand the expression in the energy eigenbasis of the qutrit. For example,
\begin{align}
    &\bra{g} \mathcalboondox{D} \Bigsl[\hat{\Pi}_{\alpha} \Bigsr] \hat{\rho}^{\mathsf{P}} \ket{e}
\nonumber \\
    &= \Big( 
            \alpha_g \alpha_e^* 
            - \frac{1}{2} |\alpha_g|^2 
            - \frac{1}{2} |\alpha_e|^2
        \Big)
        \bra{g} \hat{\rho}^{\mathsf{P}} \ket{e}
\nonumber \\
    &= \Big( 
            - \frac{1}{2} |\beta_{ge}|^2 
            -\ci \Im(\alpha_e \alpha_g^*)
        \Big) \bra{g} \hat{\rho}^{\mathsf{P}} \ket{e}.
\end{align}
By applying the same calculation to the other off-diagonal terms and noting that the diagonal terms vanish in the chosen basis, we find that
\begin{align}
    &\mathcalboondox{D} \Bigsl[\hat{\Pi}_{\alpha} \Bigsr] \hat{\rho}^{\mathsf{P}}
\nonumber \\
    &= \frac{\Gamma_{\text{m},ge}}{4\kappa} 
        \mathcalboondox{D} 
            \Bigsl[
                \hat{\sigma}_{z,ge}
            \Bigsr] 
            \hat{\rho}^{\mathsf{P}}
        + \frac{\Gamma_{\text{m},gf}}{4\kappa} |\beta_{gf}|^2 
        \mathcalboondox{D} 
            \Bigsl[
                \hat{\sigma}_{z,gf}
            \Bigsr] 
            \hat{\rho}^{\mathsf{P}}
\nonumber \\
    & \ \ \ \ 
        + \frac{\Gamma_{\text{m},ef}}{4\kappa} |\beta_{ef}|^2 
        \mathcalboondox{D} 
            \Bigsl[
                \hat{\sigma}_{z,ef}
            \Bigsr] 
            \hat{\rho}^{\mathsf{P}}
\nonumber \\ \label{eq:dephasing_terms_in_displaced_frame}
    & \ \ \ \ 
        - \frac{\ci}{2}
            \Bigsl[
                \Im(\alpha_e \alpha_g^*) 
                    \hat{\sigma}_{z,ge}
                + \Im(\alpha_f \alpha_g^*) 
                    \hat{\sigma}_{z,gf}
\nonumber \\
    & \ \ \ \  \ \ \ \  \ \ \ \  \ \ \ \  \ \ \ \  \ \ \ \  \ \ \ \  \ \ \ \ 
                + \Im(\alpha_f \alpha_e^*) 
                    \hat{\sigma}_{z,ef},
                \hat{\rho}^{\mathsf{P}}   
            \Bigsr],
\end{align}
where we have introduced three dephasing rates
\begin{align}
    \Gamma_{\text{m},ge} 
    &= \kappa |\beta_{ge}|^2, 
\\
    \Gamma_{\text{m},gf}
    &= \kappa |\beta_{gf}|^2,
\\
    \Gamma_{\text{m},ef} 
    &= \kappa |\beta_{ef}|^2.
\end{align}
With the help of Eq.(\ref{eq:dephasing_terms_in_displaced_frame}) and the observation that
\begin{align}
    \hat{\Pi}_{\alpha}
    &= (\alpha_g + \alpha_e + \alpha_f) \hat{1}
\nonumber \\
    & \ \ \ \ 
        + \frac{\beta_{ge}}{3} \hat{\sigma}_{z,ge}
        + \frac{\beta_{gf}}{3} \hat{\sigma}_{z,gf}
        + \frac{\beta_{ef}}{3} \hat{\sigma}_{z,ef},
\end{align}
we find
\begin{align}
    &\mathcalboondox{D}
        \Bigsl[
            \hat{a}^{\mathsf{P}}
        \Bigsr] 
        \hat{\rho}^{\mathsf{P}}
\nonumber \\
    &= \mathcalboondox{D}
            \Bigsl[
                \hat{a}
            \Bigsr] 
            \hat{\rho}^{\mathsf{P}}
        - \frac{1}{2}
            \Big[
                \hat{\Pi}_{\alpha} 
                    \hat{a}^{\dagger}
                - \hat{\Pi}_{\alpha}^{\dagger}
                    \hat{a},
                \hat{\rho}^{\mathsf{P}}
            \Big]
\nonumber \\
    & \ \ \ \ 
        + \frac{\beta_{ge}^*}{3} 
            \hat{a}
            \Big[
                \hat{\rho}^{\mathsf{P}},
                \hat{\sigma}_{z,ge}
            \Big]
        + \frac{\beta_{ge}}{3} 
            \Big[
                \hat{\sigma}_{z,ge}, 
                \hat{\rho}^{\mathsf{P}}
            \Big]
            \hat{a}^{\dagger}
\nonumber \\
    & \ \ \ \ 
        + \frac{\beta_{gf}^*}{3} 
            \hat{a}
            \Big[
                \hat{\rho}^{\mathsf{P}},
                \hat{\sigma}_{z,gf}
            \Big]
        + \frac{\beta_{gf}}{3} 
            \Big[
                \hat{\sigma}_{z,gf}, 
                \hat{\rho}^{\mathsf{P}}
            \Big]
            \hat{a}^{\dagger}
\nonumber \\
    & \ \ \ \ 
        + \frac{\beta_{ef}^*}{3} 
            \hat{a}
            \Big[
                \hat{\rho}^{\mathsf{P}},
                \hat{\sigma}_{z,ef}
            \Big]
        + \frac{\beta_{ef}}{3} 
            \Big[
                \hat{\sigma}_{z,ef}, 
                \hat{\rho}^{\mathsf{P}}
            \Big]
            \hat{a}^{\dagger}
\nonumber \\
    & \ \ \ \ 
        +  \frac{\Gamma_{\text{m},ge}}{4\kappa} 
        \mathcalboondox{D} 
            \Bigsl[
                \hat{\sigma}_{z,ge}
            \Bigsr] 
            \hat{\rho}^{\mathsf{P}}
        + \frac{\Gamma_{\text{m},gf}}{4\kappa} |\beta_{gf}|^2 
        \mathcalboondox{D} 
            \Bigsl[
                \hat{\sigma}_{z,gf}
            \Bigsr] 
            \hat{\rho}^{\mathsf{P}}
\nonumber \\ \label{eq:cavity_decay_terms_displaced_frame}
    & \ \ \ \ 
        + \frac{\Gamma_{\text{m},ef}}{4\kappa} |\beta_{ef}|^2 
        \mathcalboondox{D} 
            \Bigsl[
                \hat{\sigma}_{z,ef}
            \Bigsr] 
            \hat{\rho}^{\mathsf{P}}
        - \frac{\ci}{2}
            \Bigsl[
                \Im(\alpha_e \alpha_g^*) 
                    \hat{\sigma}_{z,ge}
\nonumber \\
    & \ \ \ \  \ \ \ \ \ \ \
                + \Im(\alpha_f \alpha_g^*) 
                    \hat{\sigma}_{z,gf}
                + \Im(\alpha_f \alpha_e^*) 
                    \hat{\sigma}_{z,ef},
                \hat{\rho}^{\mathsf{P}}   
    \Bigsr].
\end{align}

Finally, by combining Eq.(\ref{eq:displaced_sigma_minus_op}), (\ref{eq:H_eff_and_displaced_frame_derivative_terms}), and (\ref{eq:cavity_decay_terms_displaced_frame}), we arrive at the master equation of the composite system in the displaced frame
\begin{align}
    &\dot{\rho}^{\mathsf{P}}
\nonumber\\
    &= - \frac{\ci}{\hbar} \Big[ 
                \hat{H}'_{\text{q,eff}},
                \hat{\rho}^{\mathsf{P}}
            \Big]
        - \ci \Big[  
                \Bigsl[ 
                    \Delta_{\text{rd}} 
                    + \chi_{\text{qr}} 
                    \Bigsl(
                        \hat{\Pi}_e 
                        + 2 \hat{\Pi}_f
                    \Bigsr)
                \Bigsr]
                \hat{a}^{\dagger}\hat{a},
                \hat{\rho}^{\mathsf{P}}
            \Big]
\nonumber \\
    & \ \ \ \
        + \kappa \mathcalboondox{D}
            \Bigsl[
                \hat{a}
            \Bigsr]
            \hat{\rho}^{\mathsf{P}}
        + \frac{\kappa \beta_{ge}^*}{3} 
            \hat{a}
            \Big[
                \hat{\rho}^{\mathsf{P}},
                \hat{\sigma}_{z,ge}
            \Big]
        + \frac{\kappa \beta_{gf}^*}{3} 
            \hat{a}
            \Big[
                \hat{\rho}^{\mathsf{P}},
                \hat{\sigma}_{z,gf}
            \Big]
\nonumber \\
    & \ \ \ \
        + \frac{\kappa \beta_{ef}^*}{3} 
            \hat{a}
            \Big[
                \hat{\rho}^{\mathsf{P}},
                \hat{\sigma}_{z,ef}
            \Big]
        + \frac{\kappa \beta_{ge}}{3} 
            \Big[
                \hat{\sigma}_{z,ge}, 
                \hat{\rho}^{\mathsf{P}}
            \Big]
            \hat{a}^{\dagger}
\nonumber \\
    & \ \ \ \
        + \frac{\kappa \beta_{gf}}{3} 
            \Big[
                \hat{\sigma}_{z,gf}, 
                \hat{\rho}^{\mathsf{P}}
            \Big]
            \hat{a}^{\dagger}
        + \frac{\kappa \beta_{ef}}{3} 
            \Big[
                \hat{\sigma}_{z,ef}, 
                \hat{\rho}^{\mathsf{P}}
            \Big]
            \hat{a}^{\dagger}
\nonumber \\
    & \ \ \ \
        + \gamma_{1,ge} 
            \mathcalboondox{D}
                \Bigsl[
                    \hat{\sigma}_{ge} 
                    \hat{D}^{\dagger}(\alpha_g)
                    \hat{D}(\alpha_e)
                \Bigsr] 
            \hat{\rho}^{\mathsf{P}}
\nonumber \\
    & \ \ \ \ 
        + \gamma_{1,gf} 
            \mathcalboondox{D}
                \Bigsl[
                    \hat{\sigma}_{gf} 
                    \hat{D}^{\dagger}(\alpha_g)
                    \hat{D}(\alpha_f)
                \Bigsr] 
            \hat{\rho}^{\mathsf{P}}
\nonumber \\
    & \ \ \ \
        + \gamma_{1,ef} 
            \mathcalboondox{D}
                \Bigsl[
                    \hat{\sigma}_{ef} 
                    \hat{D}^{\dagger}(\alpha_e)
                    \hat{D}(\alpha_f)
                \Bigsr] 
            \hat{\rho}^{\mathsf{P}}
\nonumber \\
    & \ \ \ \ 
        + \frac{\gamma_{\phi,ge}}{2} 
            \mathcalboondox{D}
                \Bigsl[
                    \hat{\sigma}_{z,ge}
                \Bigsr]
            \hat{\rho}^{\mathsf{P}}
        + \frac{\gamma_{\phi,gf}}{2} 
            \mathcalboondox{D}
                \Bigsl[
                    \hat{\sigma}_{z,gf}
                \Bigsr]
            \hat{\rho}^{\mathsf{P}}
\nonumber \\
    & \ \ \ \
        + \frac{\gamma_{\phi,ef}}{2} 
            \mathcalboondox{D}
                \Bigsl[
                    \hat{\sigma}_{z,ef}
                \Bigsr]
            \hat{\rho}^{\mathsf{P}}
        + \frac{\Gamma_{\text{m},ge}}{4} 
            \mathcalboondox{D} 
            \Bigsl[
                \hat{\sigma}_{z,ge}
            \Bigsr] 
            \hat{\rho}^{\mathsf{P}}
\nonumber \\ \label{eq:displaced_frame_me_full}
    & \ \ \ \ 
        + \frac{\Gamma_{\text{m},gf}}{4} |\beta_{gf}|^2 
        \mathcalboondox{D} 
            \Bigsl[
                \hat{\sigma}_{z,gf}
            \Bigsr] 
            \hat{\rho}^{\mathsf{P}}
        + \frac{\Gamma_{\text{m},ef}}{4} |\beta_{ef}|^2 
        \mathcalboondox{D} 
            \Bigsl[
                \hat{\sigma}_{z,ef}
            \Bigsr] 
            \hat{\rho}^{\mathsf{P}},
\end{align}
where we have defined
\begin{align}
    &\hat{H}'_{\text{q,eff}} / \hbar 
\nonumber \\
    &= \Delta_{g,1}
            \hat{\Pi}_g
        + (
                \Tilde{\omega}_{\text{q}} 
                + \Delta_{e,1} 
            )
            \hat{\Pi}_e
        + (
                2\Tilde{\omega}_{\text{q}} 
                + \Delta_{f,1}
                + \alpha_{\text{q}}
            ) 
            \hat{\Pi}_f
\nonumber \\
    & \ \ \ \ 
        + \frac{\kappa }{2}
            \Bigsl[
                \Im(\alpha_e \alpha_g^*) 
                    \hat{\sigma}_{z,ge}
\nonumber \\
    & \ \ \ \  \ \ \ \  \ \ \ \ 
                + \Im(\alpha_f \alpha_g^*) 
                    \hat{\sigma}_{z,gf} 
                + \Im(\alpha_f \alpha_e^*) 
                    \hat{\sigma}_{z,ef}
            \Bigsr]
\nonumber \\
    &\doteq 
        \Tilde{\omega}_g' \hat{\Pi}_g
        + \Tilde{\omega}_e' \hat{\Pi}_e
        + \Tilde{\omega}_f' \hat{\Pi}_f
\end{align}
to be the effective qubit Hamiltonian in the displaced frame. Note that $\hat{H}'_{\text{q,eff}}$ and $\omega_{ba}' = \omega_{b}' - \omega_{a}'$ are not the final effective Hamiltonian and transition frequencies of the qutrit since we are still in the displaced frame; transforming back to the laboratory frame will cancel some of the shifts seen in the displaced frame.

\section{\label{Appendix:C}Derivation of the Effective Qutrit Master Equation}
In the last section, we have established the connection between the matrix elements of the density operator in the displaced frame to the qutrit density operator in the laboratory frame. To find the effective master equation of the qutrit, we first rewrite the master equation in the displaced frame in terms of the matrix elements of $\hat{\rho}^{\mathsf{P}}$:
\begin{align} 
    \dot{\rho}^{\mathsf{P}}_{nmgg}
    &= \big[
            -\ci \Delta_{\text{rd}} (n-m) 
            - \kappa (n + m) / 2
        \big]
        \rho^{\mathsf{P}}_{nmgg}
\nonumber \\
    & \ \ \ \ 
        + \gamma_{1, ge} 
            \sum_{p,q} 
                d_{pn}^{*} 
                d_{qm}
                \rho^{\mathsf{P}}_{pqee} 
\nonumber \\
    & \ \ \ \ 
        + \gamma_{1, gf} 
            \sum_{p,q} 
                d_{pn}^{*} 
                d_{qm}
                \rho^{\mathsf{P}}_{pqff}  
\nonumber \\  \label{eq:dispaced_frame_rho_nmgg}
    & \ \ \ \ 
        + \kappa \sqrt{(n+1)(m+1)} \rho^{\mathsf{P}}_{(n+1)(m+1)gg},
\end{align}
\begin{align}
    \dot{\rho}^{\mathsf{P}}_{nmee}
    &= \big[
            -\ci (\Delta_{\text{rd}} + \chi_{\text{qr}}) (n-m) 
            - \gamma_{1, ge}
\nonumber \\
    & \ \ \ \ \ \ \ \ 
            - \kappa (n + m) / 2
        \big]
        \rho^{\mathsf{P}}_{nmee}
\nonumber \\ 
    & \ \ \ \ 
        + \gamma_{1, ef} 
            \sum_{p,q} 
                d_{pn}^{*} 
                d_{qm}
                \rho^{\mathsf{P}}_{pqff}
\nonumber \\ \label{eq:dispaced_frame_rho_nmee}
    & \ \ \ \ 
        + \kappa \sqrt{(n+1)(m+1)} \rho^{\mathsf{P}}_{(n+1)(m+1)ee},
\end{align}
\begin{align}
    \dot{\rho}^{\mathsf{P}}_{nmff}
    &= \big[
            -\ci (\Delta_{\text{rd}} + 2\chi_{\text{qr}}) (n-m) 
            - (\gamma_{1, gf} + \gamma_{1, ef})
\nonumber \\
    & \ \ \ \ \ \ \ \ 
            - \kappa (n + m) / 2
        \big]
        \rho^{\mathsf{P}}_{nmff}
\nonumber \\[1mm] \label{eq:dispaced_frame_rho_nmff}
    & \ \ \ \
        + \kappa \sqrt{(n+1)(m+1)} \rho^{\mathsf{P}}_{(n+1)(m+1)ff},
\end{align}
\begin{align}
    \dot{\rho}^{\mathsf{P}}_{nmge}
    &= \big[
            \ci \Tilde{\omega}_{eg}'
            -\ci \Delta_{\text{rd}} (n-m) 
            + \ci \chi_{\text{qr}} m
\nonumber \\
    & \ \ \ \ \ \ \ \ 
            - \gamma_{2,ge} 
            - \kappa (n + m) / 2
            - \Gamma_{m,ge}
        \big]
        \rho^{\mathsf{P}}_{nmge}
\nonumber \\
    & \ \ \ \ 
        + \kappa \sqrt{(n+1)(m+1)}
            \rho^{\mathsf{P}}_{(n+1)(m+1)ge}
\nonumber \\
    & \ \ \ \ 
        - \frac{2\kappa \beta_{ge}^*}{3} 
            \sqrt{n+1}
            \rho^{\mathsf{P}}_{(n+1)mge}
\nonumber \\ \label{eq:dispaced_frame_rho_nmge}
    & \ \ \ \ 
        + \frac{2\kappa \beta_{ge}}{3} 
            \sqrt{m+1}
            \rho^{\mathsf{P}}_{n(m+1)ge},
\end{align}
\begin{align}
    \dot{\rho}^{\mathsf{P}}_{nmgf}
    &= \big[
            \ci \Tilde{\omega}_{fg}'
            -\ci \Delta_{\text{rd}} (n-m) 
            + \ci \chi_{\text{qr}} 2 m
\nonumber \\
    & \ \ \ \ \ \ \ \ 
            - \gamma_{2,gf} 
            - \kappa (n + m) / 2
            - \Gamma_{m,gf}
        \big]
        \rho^{\mathsf{P}}_{nmgf}
\nonumber \\
    & \ \ \ \ 
        + \kappa \sqrt{(n+1)(m+1)}
            \rho^{\mathsf{P}}_{(n+1)(m+1)gf}
\nonumber \\
    & \ \ \ \ 
        - \frac{2\kappa \beta_{gf}^*}{3} 
            \sqrt{n+1}
            \rho^{\mathsf{P}}_{(n+1)mgf}
\nonumber \\ \label{eq:dispaced_frame_rho_nmgf}
    & \ \ \ \ 
        + \frac{2\kappa \beta_{gf}}{3} 
            \sqrt{m+1}
            \rho^{\mathsf{P}}_{n(m+1)gf},
\end{align}
\begin{align}
    \dot{\rho}^{\mathsf{P}}_{nmef}
    &= \big[
            \ci \Tilde{\omega}_{fe}'
            -\ci \Delta_{\text{rd}} (n-m) 
            + \ci \chi_{\text{qr}} (2m-n)
\nonumber \\
    & \ \ \ \ \ \ \ \ 
            - \gamma_{2,ef} 
            - \kappa (n + m) / 2
            - \Gamma_{m,ef}
        \big]
        \rho^{\mathsf{P}}_{nmef}
\nonumber \\
    & \ \ \ \ 
        + \kappa \sqrt{(n+1)(m+1)}
            \rho^{\mathsf{P}}_{(n+1)(m+1)ef}
\nonumber \\
    & \ \ \ \ 
        - \frac{2\kappa \beta_{ef}^*}{3} 
            \sqrt{n+1}
            \rho^{\mathsf{P}}_{(n+1)mef}
\nonumber \\ \label{eq:dispaced_frame_rho_nmef}
    & \ \ \ \ 
        + \frac{2\kappa \beta_{ef}}{3} 
            \sqrt{m+1}
            \rho^{\mathsf{P}}_{n(m+1)ef}.
\end{align}
There are three other differential equations, but they are simply the complex conjugates of Eq.(\ref{eq:dispaced_frame_rho_nmge})-(\ref{eq:dispaced_frame_rho_nmef}). 
Now, given Eq.(\ref{eq:dispaced_frame_rho_nmgg})-(\ref{eq:dispaced_frame_rho_nmff}), the differential equations governing the time evolution of the diagonal matrix elements of $\hat{\rho}_{\mathcal{S}}$ (i.e., the populations of the qutrit eigenstates) are found to be
\begin{align}
    \dot{\rho}_{\mathcal{S},gg}
    &= \sum_{n}
        \dot{\rho}^{\mathsf{P}}_{nngg}
\nonumber \\
    &= \gamma_{1, ge} 
            \sum_{p,q} 
                \rho^{\mathsf{P}}_{pqee} 
                \sum_{n}
                    d_{pn}^{*} 
                    d_{qn}
\nonumber \\
    & \ \ \ \ 
        + \gamma_{1, gf} 
            \sum_{p,q} 
                \rho^{\mathsf{P}}_{pqff}  
                \sum_{n} 
                    d_{pn}^{*} 
                    d_{qn}
\nonumber \\\label{eq:qutrit_population_gg}
    &= \gamma_{1, ge} \rho_{\mathcal{S},ee} 
        + \gamma_{1, gf} \rho_{\mathcal{S},ff} ,
\end{align}
\begin{align}
    \dot{\rho}_{\mathcal{S},ee}
    &= \sum_{n}
        \dot{\rho}^{\mathsf{P}}_{nnee}
\nonumber \\
    &= - \gamma_{1, ge} 
            \sum_{n} \rho^{\mathsf{P}}_{nnee}
\nonumber \\
    & \ \ \ \ 
        + \gamma_{1, ef} 
            \sum_{p,q}
                \rho^{\mathsf{P}}_{pqff} 
                \sum_{n}
                    d_{pn}^{*} 
                    d_{qn}
\nonumber \\ \label{eq:qutrit_population_ee}
    &= - \gamma_{1, ge} \rho_{\mathcal{S},ee} 
        + \gamma_{1, ef} \rho_{\mathcal{S},ff} ,
\end{align}
\begin{align}
    \dot{\rho}_{\mathcal{S},ff}
    &= \sum_{n}
        \dot{\rho}^{\mathsf{P}}_{nnff}
\nonumber \\
    &= - (\gamma_{1, gf} + \gamma_{1, ef})
        \sum_{n} \rho^{\mathsf{P}}_{nnff}
\nonumber \\ \label{eq:qutrit_population_ff}
    &= - (\gamma_{1, gf} + \gamma_{1, ef}) \rho_{\mathcal{S},ff}.
\end{align}

Next, according to Eq.(\ref{eq:reduced_rho_in_terms_of_displaced_rho}), computing the off-diagonal terms of $\hat{\rho}_{\mathcal{S}}$ also requires us to know all $\lambda_{nmpq}$, whose time derivatives follow (see Eq.(\ref{eq:lambda_nmpq_ge_def})-(\ref{eq:lambda_nmpq_ef_def}))
\begin{align}  \label{eq:lambda_nmpq_ge}
    \dot{\lambda}_{nmpq}^{ge}
    &= \dot{\rho}_{nmge}^{\mathsf{P}}
            d_{p,q} 
            e^{-\ci \Im(\alpha_e \alpha_g^*)}
        - \ci \frac{\mathrm{d} \Im(\alpha_e \alpha_g^*)}{\mathrm{d} t} 
            \lambda_{nmpq}^{ge}
\nonumber \\
    & \ \ \ \ 
        + \sqrt{p} \dot{\beta}_{ge} \lambda_{nm(p-1)q}^{ge}
        - \sqrt{q} \dot{\beta}_{ge}^* \lambda_{nmp(q-1)}^{ge}
\nonumber \\
    & \ \ \ \ 
        - \frac{\dot{\beta}_{ge}^* \beta_{ge} + \dot{\beta}_{ge} \beta_{ge}^*}{2} \lambda_{nmpq}^{ge},
\end{align}
\begin{align}
    \dot{\lambda}_{nmpq}^{gf}
    &= \dot{\rho}_{nmgf}^{\mathsf{P}}
            d_{p,q} 
            e^{-\ci \Im(\alpha_f \alpha_g^*)}
        - \ci \frac{\mathrm{d} \Im(\alpha_f \alpha_g^*)}{\mathrm{d} t} 
            \lambda_{nmpq}^{gf}
\nonumber \\
    & \ \ \ \ 
        + \sqrt{p} \dot{\beta}_{gf} \lambda_{nm(p-1)q}^{gf}
        - \sqrt{q} \dot{\beta}_{gf}^* \lambda_{nmp(q-1)}^{gf}
\nonumber \\
    & \ \ \ \ 
        - \frac{\dot{\beta}_{gf}^* \beta_{gf} + \dot{\beta}_{gf} \beta_{gf}^*}{2} \lambda_{nmpq}^{gf},
\end{align}
\begin{align}
    \dot{\lambda}_{nmpq}^{ef}
    &= \dot{\rho}_{nmef}^{\mathsf{P}}
            d_{p,q} 
            e^{-\ci \Im(\alpha_f \alpha_e^*)}
        - \ci \frac{\mathrm{d} \Im(\alpha_f \alpha_e^*)}{\mathrm{d} t} 
            \lambda_{nmpq}^{ef}
\nonumber \\
    & \ \ \ \ 
        + \sqrt{p} \dot{\beta}_{ef} \lambda_{nm(p-1)q}^{ef}
        - \sqrt{q} \dot{\beta}_{ef}^* \lambda_{nmp(q-1)}^{ef}
\nonumber \\
    & \ \ \ \ 
        - \frac{\dot{\beta}_{ef}^* \beta_{ef} + \dot{\beta}_{ef} \beta_{ef}^*}{2} \lambda_{nmpq}^{ef}.
\end{align}
The three equations take the same forms; we show the simplification of Eq.(\ref{eq:lambda_nmpq_ge}) as an example. Since we have a differential equation for $\rho_{nmge}^{\mathsf{P}}$, the first term on the RHS of Eq.(\ref{eq:lambda_nmpq_ge}) reduces to
\begin{align}
    &\dot{\rho}_{nmge}^{\mathsf{P}}
            d_{p,q} 
            e^{-\ci \Im(\alpha_e \alpha_g^*)}
\nonumber\\[1.5mm]
    &= \Big[
            \ci \Tilde{\omega}_{eg}'
            -\ci \Delta_{\text{rd}} (n-m) 
            + \ci \chi_{\text{qr}} m
\nonumber \\
    & \ \ \ \   \ \ \ \ 
            - \gamma_{2,ge} 
            - \kappa (n + m) / 2
            - \Gamma_{\text{m},ge} / 2
        \Big]
        \lambda_{nmpq}^{ge}
\nonumber \\[1.5mm]
    & \ \ \ \ 
        + \kappa \sqrt{(n+1)(m+1)}
            \lambda_{(n+1)(m+1)pq}^{ge}
\nonumber \\
    & \ \ \ \ 
        - \frac{2\kappa \beta_{ge}^*}{3} 
            \sqrt{n+1}
            \lambda_{(n+1)mpq}^{ge}
\nonumber \\
    & \ \ \ \ 
        + \frac{2\kappa \beta_{ge}}{3} 
            \sqrt{m+1}
            \lambda_{n(m+1)pq}^{ge}.
\end{align}
To simplify the second term on the RHS of Eq.(\ref{eq:lambda_nmpq_ge}), we first compute
% \begin{align}
%     \dot{\alpha}_e \alpha_g^*
%     &= \Big[
%             - \ci (\Delta_{\text{rd}} + \chi_{\text{qr}} - \ci \kappa/2) \alpha_e
%             + \ci \epsilon
%         \Big] \alpha_g^*
% \nonumber \\
%     &= - \ci (\Delta_{\text{rd}} + \chi_{\text{qr}}) 
%                 \alpha_e \alpha_g^*
%             - \kappa/2 \alpha_e \alpha_g^*
%             + \ci \epsilon \alpha_g^*,
% \end{align}
% \begin{align}
%     \alpha_e \dot{\alpha_g}^*
%     &= \alpha_e
%         \Big[
%             - \ci (\Delta_{\text{rd}} - \ci \kappa/2) \alpha_g
%             + \ci \epsilon
%         \Big]^*
% \nonumber \\
%     &= \ci \Delta_{\text{rd}} \alpha_e \alpha_g^* 
%             - \kappa/2 \alpha_e \alpha_g^*
%             - \ci \epsilon^* \alpha_e,
% \end{align}
% and
\begin{align}
    &\dot{\alpha}_e \alpha_g^* + \alpha_e \dot{\alpha_g}^*
\nonumber \\
    &= \Big[
            - \ci (\Delta_{\text{rd}} + \chi_{\text{qr}} - \ci \kappa/2) \alpha_e
            + \ci \epsilon
        \Big] \alpha_g^*
\nonumber \\
    & \ \ \ \ 
        + \alpha_e
        \Big[
            - \ci (\Delta_{\text{rd}} - \ci \kappa/2) \alpha_g
            + \ci \epsilon
        \Big]^*
\nonumber \\
    &= - \ci \chi_{\text{qr}} \alpha_e\alpha_g^*
        - \kappa \alpha_e \alpha_g^*
        + \ci ( \epsilon \alpha_g^* - \epsilon^* \alpha_e).
\end{align}
Then,
\begin{align}
    &- \ci \frac{\mathrm{d} \Im(\alpha_e \alpha_g^*)}{\mathrm{d} t} 
        \lambda_{nmpq}^{ge}
\nonumber \\
    &= - \ci \Im \!
        \Big[
            - \ci \chi_{\text{qr}} \alpha_e\alpha_g^*
            - \kappa \alpha_e \alpha_g^*
            + \ci ( \epsilon \alpha_g^* - \epsilon^* \alpha_e)
        \Big]
        \lambda_{nmpq}^{ge}
\nonumber \\
    &= \ci 
        \Big[
            \chi_{\text{qr}} \Re(\alpha_e \alpha_g^*)
            + \kappa \Im(\alpha_e \alpha_g^*)
            - \frac{\epsilon^* \beta_{ge} + \epsilon \beta_{ge}^*}{2}
        \Big]
        \lambda_{nmpq}^{ge}.
\end{align}
We keep the terms involving $\lambda_{nm(p-1)q}^{ge}$ and $\lambda_{nmp(q-1)}^{ge}$ untouched and directly go to the last term on the RHS of Eq.(\ref{eq:lambda_nmpq_ge}). Using
\begin{align}
    \dot{\beta}_{ge} \beta_{ge}^*
    &= (\dot{\alpha}_g - \dot{\alpha}_e)\beta_{ge}^*
\nonumber \\
    &= - \ci \Delta_{\text{rd}} |\beta_{ge}|^2 - \frac{\kappa |\beta_{ge}|^2}{2} + \ci \chi_{\text{qr}} \alpha_{e}\beta_{ge}^*,
\end{align}
we obtain
\begin{align}
    &- \frac{\dot{\beta}_{ge}^* \beta_{ge} + \dot{\beta}_{ge} \beta_{ge}^*}{2} \lambda_{nmpq}^{ge}
\nonumber \\
    &= \left[ 
            \frac{\kappa |\beta_{ge}|^2}{2}
            + \chi_{\text{qr}} \Im(\alpha_e \alpha_g^*)
        \right] 
        \lambda_{nmpq}^{ge}.
\end{align}
Finally, combining all the pieces yields
\begin{align}
    \dot{\lambda}_{nmpq}^{ge}
    &= \Big[
            \ci \Bar{\omega}_{eg}
            - \gamma_{2,ge} 
            + \chi_{\text{qr}} \Im(\alpha_e \alpha_g^*)
\nonumber \\
    & \ \ \ \ \ \ \ \ 
            -\ci \Delta_{\text{rd}} (n-m) 
\nonumber \\
    & \ \ \ \ \ \ \ \ 
            + \ci \chi_{\text{qr}} m
            - \kappa (n + m) / 2
        \Big]
        \lambda_{nmpq}^{ge}
\nonumber \\
    & \ \ \ \ 
        + \kappa \sqrt{(n+1)(m+1)}
            \lambda_{(n+1)(m+1)pq}^{ge}
\nonumber \\
    & \ \ \ \ 
        - \frac{2\kappa \beta_{ge}^*}{3} 
            \sqrt{n+1}
            \lambda_{(n+1)mpq}^{ge}
\nonumber \\
    & \ \ \ \ 
        + \frac{2\kappa \beta_{ge}}{3} 
            \sqrt{m+1}
            \lambda_{n(m+1)pq}^{ge},
\end{align}
where the net frequency difference between $\ket{e}$ and $\ket{g}$ is found to be
\begin{align}
    \bar{\omega}_{eg}
    &= (\Tilde{\omega}_{e}' - \Tilde{\omega}_{g}')
        + \chi_{\text{qr}} \Re(\alpha_e \alpha_g^*)
\nonumber \\
    & \ \ \ \ 
        + \kappa \Im(\alpha_e \alpha_g^*)
        - \Big(
                \epsilon^* \beta_{ge} 
                + \epsilon \beta_{ge}^* 
            \Big) / 2
\nonumber \\
    &= \Big[ \omega_{\text{q}} 
            + \Delta_{e,1}
            - \kappa \Im(\alpha_e \alpha_g^*)
            - \Delta_{g,1} 
        \Big]
\nonumber \\
    &\ \ \ \ 
        + \chi_{\text{qr}} \Re(\alpha_e \alpha_g^*)
        + \kappa \Im(\alpha_e \alpha_g^*)
\nonumber \\
    & \ \ \ \ 
        - \Big(
                \epsilon^* \beta_{ge} 
                + \epsilon \beta_{ge}^* 
            \Big) / 2
\nonumber \\
    &= \tilde{\omega}_{\text{q}}
        + \chi_{\text{qr}} \Re(\alpha_e \alpha_g^*).
\end{align}
Applying the same procedure to the other two equations, we find
\begin{align}
    \dot{\lambda}_{nmpq}^{gf}
    &= \Big[
            \ci \Bar{\omega}_{fg}
            - \gamma_{2,gf} 
            + 2 \chi_{\text{qr}} \Im(\alpha_f \alpha_g^*)
\nonumber \\
    & \ \ \ \ \ \ \ \ 
            -\ci \Delta_{\text{rd}} (n-m) 
\nonumber \\
    & \ \ \ \ \ \ \ \ 
            + \ci 2 \chi_{\text{qr}} m
            - \kappa (n + m) / 2
        \Big]
        \lambda_{nmpq}^{gf}
\nonumber \\
    & \ \ \ \ 
        + \kappa \sqrt{(n+1)(m+1)}
            \lambda_{(n+1)(m+1)pq}^{gf}
\nonumber \\
    & \ \ \ \ 
        - \frac{2\kappa \beta_{gf}^*}{3} 
            \sqrt{n+1}
            \lambda_{(n+1)mpq}^{gf}
\nonumber \\
    & \ \ \ \ 
        + \frac{2\kappa \beta_{gf}}{3} 
            \sqrt{m+1}
            \lambda_{n(m+1)pq}^{gf}
\end{align}
and
\begin{align}
    \dot{\lambda}_{nmpq}^{ef}
    &= \Big[
            \ci \Bar{\omega}_{fe}
            - \gamma_{2,ef} 
            + 2 \chi_{\text{qr}} \Im(\alpha_f \alpha_e^*)
\nonumber \\
    & \ \ \ \ \ \ \ \ 
            - \ci \Delta_{\text{rd}} (n - m) 
\nonumber \\
    & \ \ \ \ \ \ \ \ 
            + \ci \chi_{\text{qr}} (2m - n)
            - \kappa (n + m) / 2
        \Big]
        \lambda_{nmpq}^{ef}
\nonumber \\[1.5mm]
    & \ \ \ \ 
        + \kappa \sqrt{(n+1)(m+1)}
            \lambda_{(n+1)(m+1)pq}^{ef}
\nonumber \\
    & \ \ \ \ 
        - \frac{2\kappa \beta_{ef}^*}{3} 
            \sqrt{n+1}
            \lambda_{(n+1)mpq}^{ef}
\nonumber \\
    & \ \ \ \ 
        + \frac{2\kappa \beta_{ef}}{3} 
            \sqrt{m+1}
            \lambda_{n(m+1)pq}^{ef}
\end{align}
with the net frequency differences
\begin{align}
    \Bar{\omega}_{fg}
    &= 2 \tilde{\omega}_{\text{q}} + \alpha_{\text{q}}
        + 2\chi_{\text{qr}} \Re(\alpha_f \alpha_g^*),
\\
    \Bar{\omega}_{fe}
    &= \tilde{\omega}_{\text{q}} + \alpha_{\text{q}}
        + \chi_{\text{qr}} \Re(\alpha_f \alpha_e^*).
\end{align}

Since we are in the transformed frame, the photon population is initially displaced to the vacuum state, i.e., ${\lambda}_{nmpq}^{ab} \propto \rho^{\mathsf{P}}_{nm ab}= 0$. In addition, there is no mechanism to excite $\lambda_{nmpq}^{ab}$ with $n,m,p,q >0$ because the three displacement operators are designed to keep the photon number zero in the displaced frame \cite{PhysRevLett.70.2273}; hence, $\rho_{\mathcal{S}, ab} = \lambda_{0000}^{ab}$ (see Eq.(\ref{eq:rho_s_ge_and_lambda_nmmn_ge})) and 
\begin{align} \label{eq:lambda_0000_ge}
    \dot{\rho}_{\mathcal{S}, ge}
    &= \dot{\lambda}_{0000}^{ge}
\nonumber \\
    &= \Big[
            \ci \Bar{\omega}_{eg}
            - \gamma_{2,ge} 
            + \chi_{\text{qr}} \Im(\alpha_e \alpha_g^*)
        \Big]
        \lambda_{0000}^{ge},
\\ \label{eq:lambda_0000_gf}
    \dot{\rho}_{\mathcal{S}, gf}
    &= \dot{\lambda}_{0000}^{gf}
\nonumber \\
    &= \Big[
            \ci \Bar{\omega}_{gf}
            - \gamma_{2,gf} 
            + 2 \chi_{\text{qr}} \Im(\alpha_f \alpha_g^*)
        \Big]
        \lambda_{0000}^{gf},
\\ \label{eq:lambda_0000_ef}
    \dot{\rho}_{\mathcal{S}, ef}
    &= \dot{\lambda}_{0000}^{ef}
\nonumber \\
    &= \Big[
            \ci \Bar{\omega}_{ef}
            - \gamma_{2,ef}
            + \chi_{\text{qr}} \Im(\alpha_f \alpha_e^*)
        \Big]
        \lambda_{0000}^{ef}.
\end{align}
As explained in the main text, by assuming the frequency shifts are much smaller than the bare frequencies, we can approximate the qutrit as a Markovian system, thus writing down Eq.(\ref{eq:effective_qutrit_me_main}). Although a qutrit is used in the derivation, the result can be easily generalized to a general qudit. The reader who goes through the steps for the qutrit case should have no problem making the generalization.

\section{\label{Appendix:D}Derivation of the Qutrit Effective Stochastic Master Equation in the Diffusive Limit}
\subsection{Heuristic Derivation of the Qudit SME}
Given the quantum channel defined in Eq.(\ref{eq:qutrit_state_update_rule_main}) of the main text, we look for a stochastic differential equation by sending $\Delta t$ to 0. We start with $\eta = 1$ so that we do not need to worry about averaging over the unobserved information; of course, such an assumption is unphysical so we will need to relax it later. To introduce random processes that capture the measurement noise, we first examine various moments of the measurement outcomes $I_k = I(t_{k+1})$ and $Q_k = Q(t_{k+1})$ within $[t_k, t_k + \Delta t)$ conditioned on the qutrit state at $t_k$. To begin with, the conditional expectation of $I_k$ and $Q_k$ are
\begin{align}
    \mathbb{E}
        \Bigsl[
            I_k \, | \, 
            \hat{\rho}_{\mathcal{S}}(t_{k})
        \Bigsr] 
    &= \iint 
            \mathrm{d} I' 
            \mathrm{d} Q' \, 
            I' 
            f(I',Q'|\hat{\rho}_{\mathcal{S}}(t_k))
\nonumber \\
    &= \sqrt{\eta \kappa \Delta t} 
        \Tr[\hat{\rho}_{\mathcal{S}}(t_k) \hat{L}_{I}(t_k)]
\nonumber \\
    &= \mathcal{O}(\sqrt{\Delta t}),
\end{align}
\begin{align}
    \mathbb{E}
        \Bigsl[
            Q_k \, | \, 
            \hat{\rho}_{\mathcal{S}}(t_{k})
        \Bigsr] 
    &= \iint 
            \mathrm{d} I' 
            \mathrm{d} Q' \, 
            Q' 
            f(I',Q'|\hat{\rho}_{\mathcal{S}}(t_k))
\nonumber \\
    &= \sqrt{\eta \kappa \Delta t} 
        \Tr[\hat{\rho}_{\mathcal{S}}(t_k) \hat{L}_{Q}(t_k)]
\nonumber \\
    &= \mathcal{O}(\sqrt{\Delta t}),
\end{align}
where we have defined
\begin{align}
    \hat{L}_I (t)
    &= \Bar{I}_g(t) \hat{\Pi}_g
        + \Bar{I}_e(t) \hat{\Pi}_e
        + \Bar{I}_f(t) \hat{\Pi}_f,
\\
    \hat{L}_Q(t) 
    &= \Bar{Q}_g(t) \hat{\Pi}_g
        + \Bar{Q}_e(t) \hat{\Pi}_e
        + \Bar{Q}_f(t) \hat{\Pi}_f.
\end{align}
More importantly, we also have
\begin{align}
    \mathbb{E}
        \Bigsl[
            I_k^2 \, | \, 
            \hat{\rho}_{\mathcal{S}}(t_{k})
        \Bigsr]
    &= \sum_{a} \rho_{\mathcal{S},aa}
            \left(
                \eta \kappa \Delta t \Bar{I}_a^2 
                + \frac{1}{4}
            \right)
\nonumber \\
    &= \eta \kappa \Delta t \Tr(\hat{\rho}_{\mathcal{S}} \hat{L}_{I}^{\dagger} \hat{L}_{I})
        + \frac{1}{4}
\nonumber \\
    &= \mathcal{O}(1),
\end{align}
\begin{align}
    \mathbb{E}
        \Bigsl[
            Q_k^2 \, | \, 
            \hat{\rho}_{\mathcal{S}}(t_{k})
        \Bigsr]
    &= \sum_{a} \rho_{\mathcal{S},aa}
            \left(
                \eta \kappa \Delta t \Bar{Q}_a^2 
                + \frac{1}{4}
            \right)
\nonumber \\
    &= \eta \kappa \Delta t \Tr(\hat{\rho}_{\mathcal{S}} \hat{L}_{Q}^{\dagger} \hat{L}_{Q})
        + \frac{1}{4}
\nonumber \\
    &= \mathcal{O}(1),
\end{align}
\begin{align}
    \mathbb{E}\Bigsl[
            I_k Q_k \, | \, 
            \hat{\rho}_{\mathcal{S}}(t_{k})
        \Bigsr]
    &= \eta \kappa \Delta t
            \Tr(\hat{\rho}_{\mathcal{S}} \hat{L}_{I} \hat{L}_{Q})
\nonumber \\
    &= \mathcal{O}(\Delta t).
\end{align}
We observe that the second moments of $I_k$ and $Q_k$ are nonvanishing as $\Delta t \rightarrow 0$ but the correlation between $I_k$ and $Q_k$ would vanish for small $\Delta t$, which implies that measurements of $I_k$ and $Q_k$ are related to two independent random processes. We can keep computing higher moments, such as
\begin{align}
    &\mathbb{E}\Bigsl[
            I_k^2 Q_k \, | \, 
            \hat{\rho}_{\mathcal{S}}(t_{k})
        \Bigsr]
\nonumber \\
    &= \frac{1}{4} \sqrt{\eta \kappa \Delta t}
            \Tr(\hat{\rho}_{\mathcal{S}} \hat{L}_{Q})
        + \eta \kappa \Delta t
            \Tr(\hat{\rho}_{\mathcal{S}} \hat{L}_{I}^2 \hat{L}_{Q})
\nonumber \\
    &= \mathcal{O}(\sqrt{\Delta t}),
\end{align}
\begin{align}
    &\mathbb{E}\Bigsl[
            I_k Q_k^2 \, | \,
            \hat{\rho}_{\mathcal{S}}(t_{k})
        \Bigsr]
\nonumber \\
    &= \frac{1}{4} \sqrt{\eta \kappa \Delta t}
            \Tr(\hat{\rho}_{\mathcal{S}} \hat{L}_{I})
        + \eta \kappa \Delta t
            \Tr(\hat{\rho}_{\mathcal{S}} \hat{L}_{I} \hat{L}_{Q}^2)
\nonumber \\
    &= \mathcal{O}(\sqrt{\Delta t}),
\end{align}
but it's clear that terms containing $I_k^2$ and $Q_k^2$ are not negligible and should be examined carefully as $\Delta t \rightarrow 0$. 

In the diffusive limit, we introduce two random processes, $W_I(t)$ and $W_Q(t)$, related to $I(t)$ and $Q(t)$ by
\begin{gather} \label{eq:Wiener_I_def}
    I_k 
    = \sqrt{\eta \kappa \Delta t} 
            \Tr[ 
                \hat{\rho}_{\mathcal{S}}(t_k)
                \hat{L}_{I} (t_k)
            ]
        + \frac{\Delta W_{I}(t_k)}{2\sqrt{\Delta t}},
\\ \label{eq:Wiener_Q_def}
    Q_k 
    = \sqrt{\eta \kappa \Delta t} 
            \Tr[ 
                \hat{\rho}_{\mathcal{S}}(t_k)
                \hat{L}_{Q} (t_k)
            ]
        + \frac{\Delta W_{Q}(t_k)}{2\sqrt{\Delta t}},
\end{gather} 
where $\Delta W_{I}(t_k) =  W_{I}(t_{k+1}) -  W_{I}(t_k)$ and $\Delta W_{Q}(t_k) =  W_{Q}(t_{k+1}) -  W_{Q}(t_k)$. From the moments of $I_k$ and $Q_k$, one can easily verify that, up to the first order in $\Delta t$, 
\begin{gather}
    \mathbb{E}\Bigsl[
            \Delta W_{I}(t_k) \, | \, 
            \hat{\rho}_{\mathcal{S}}(t_{k})
        \Bigsr]
    = \mathbb{E}\Bigsl[
            \Delta W_{Q}(t_k) \, | \, 
            \hat{\rho}_{\mathcal{S}}(t_{k})
        \Bigsr]
    = 0,
\\
    \mathbb{E}\Bigsl[
            \Delta W_{I}(t_k) \Delta W_{Q}(t_k) \, | \, 
            \hat{\rho}_{\mathcal{S}}(t_{k})
        \Bigsr]
    = 0.
\end{gather}
\begin{equation}
    \mathbb{E}\Bigsl[
            (\Delta W_{I}(t_k))^2 \, | \, 
            \hat{\rho}_{\mathcal{S}}(t_{k})
        \Bigsr]
    = \mathbb{E}\Bigsl[
            (\Delta W_{Q}(t_k))^2 \, | \, 
            \hat{\rho}_{\mathcal{S}}(t_{k})
        \Bigsr]
    = \Delta t,
\end{equation}
Hence, $W_{I}$ and $W_{Q}$ can be treated as two independent Wiener processes. Moreover, since the moments of the Wiener increments are independent of the qutrit state, we can drop the conditioning above. However, note that the qutrit state depends on the past trajectory of $W_{I}$ and $W_{Q}$, which is why $\hat{\rho}_{\mathcal{S}}$ should always be interpreted as the conditional states. Putting it differently, according to Eq.(\ref{eq:Wiener_I_def}) and (\ref{eq:Wiener_Q_def}), we notice that the history of $I$ and $Q$ are determined once we have specified a realization of $W_I$ and $W_Q$; therefore, we can generate the quantum trajectories of the qutrit during the dispersive measurement by simulating all possible realizations of $W_I$ and $W_Q$.

With the preparation above, we are ready to derive the stochastic master equation in the diffusive limit from Eq.(\ref{eq:qutrit_state_update_rule_main}). In the limit as $\Delta t \rightarrow 0$, we can expand Eq.(\ref{eq:qutrit_state_update_rule_main}) to first order in $\Delta t$ with the caution that the Wiener processes follow the It\^{o}'s rule, i.e., $(\Delta W_{I})^2 = (\Delta W_{Q})^2 = \Delta t$ and $\Delta W_{I} \Delta W_{Q} = 0$ as $\Delta t \rightarrow 0$. In addition, the calculation is considerably simplified with the following observation: The Kraus operator
\begin{align}
    &\hat{K}_{IQ}(t_k)
\nonumber \\
    &= \mathscr{N}_{k} 
        \sum_{a \in \{g,e,f\}} 
            \exp
            \bigg \{- \Big[ I - \sqrt{\eta \kappa \Delta t} \, \Bar{I}_a(t_k) \Big]^2 
\nonumber \\
    & \ \ \ \  \ \ \ \  \ \ \ \  \ \ \ \  \ \ \ \  \ \ \ \  \ \ \ \ 
            - \Big[ Q - \sqrt{\eta \kappa \Delta t} \, \Bar{Q}_a(t_k) \Big]^2
            \Bigg\}
            \hat{\Pi}_a
\nonumber \\
    &\approx 
        \Tilde{\mathscr{N}}_{k} 
        \exp \Bigg\{
            - \left[
                    I - \sqrt{\eta \kappa \Delta t} 
                    \sum_{a}      
                        \Bar{I}_a(t_k) 
                        \hat{\Pi}_a
                \right]^2 
\nonumber \\
    & \ \ \ \  \ \ \ \  \ \ \ \  \ \ \ \ 
            - \left[ 
                    Q - \sqrt{\eta \kappa \Delta t} 
                    \sum_{a} 
                        \Bar{Q}_a(t_k)
                        \hat{\Pi}_a
                \right]^2
        \Bigg\}
\nonumber \\
    &=
        \Tilde{\mathscr{N}}_{k} 
        \exp \Bigg\{
            - \Big[
                    I - \sqrt{\eta \kappa \Delta t} 
                    \hat{L}_I(t_k)
                \Big]^2 
\nonumber \\
    & \ \ \ \  \ \ \ \  \ \ \ \  \ \ \ \ 
            - \Big[ 
                    Q - \sqrt{\eta \kappa \Delta t} 
                    \hat{L}_Q(t_k)
                \Big]^2
        \Bigg\}
\end{align}
to the first order in $\Delta t$. Since $\Bigsl[\hat{L}_I, \hat{L}_Q\Bigsr] = 0$, we immediately have
\begin{align}
    \hat{K}_{IQ}(t_k)
    &\approx 
    \Tilde{\mathscr{N}}_{k} 
        \exp{
            - \Big[
                    I - \sqrt{\eta \kappa \Delta t} 
                    \hat{L}_I(t_k)
                \Big]^2 
        }
\nonumber \\
    & \ \ \ \  \ \
        \times \exp{
            - \Big[
                    Q - \sqrt{\eta \kappa \Delta t} 
                    \hat{L}_Q(t_k)
                \Big]^2
        },
\end{align}
which is a great simplification and a verification that the values of $I$ and $Q$ are uncorrelated in the diffusive limit. Next, we replace $I$ and $Q$ using Eq.(\ref{eq:Wiener_I_def}) and (\ref{eq:Wiener_Q_def}) so that the Kraus operator at $t_k$ is implicitly fixed by a realization of $W_I$ and $W_Q$:
\begin{align}
    &\hat{K}_{I_k Q_k}(t_k)
\nonumber \\
    &= \Tilde{\mathscr{N}}_{k} 
        \exp \bigg[
                \sqrt{\eta \kappa} \hat{L}_I \Delta W_{I}(t_k)
                + 2 \eta \kappa \Delta t 
                    \Bigsl \langle
                        \hat{L}_{I}(t_k)
                    \Bigsr \rangle
                    \hat{L}_{I}(t_k)
\nonumber \\
    & \ \ \ \ \ \ \ \ \ \ \ \ \ \ 
                - \eta \kappa \Delta t \hat{L}_{I}^2(t_k)
                - \eta \kappa \Delta t \Bigsl \langle
                        \hat{L}_{I}(t_k)
                    \Bigsr \rangle^2
                - \sqrt{\Delta t} /2
            \bigg]
\nonumber \\
    & \ \ \ \ \ \
        \times \exp \bigg[
                \sqrt{\eta \kappa} \hat{L}_Q \Delta W_{Q}(t_k)
                + 2 \eta \kappa \Delta t 
                    \Bigsl \langle
                        \hat{L}_{Q}(t_k)
                    \Bigsr \rangle
                    \hat{L}_{Q}(t_k)
\nonumber \\
    & \ \ \ \ \ \ \ \ \ \ \ \ \ \ 
                - \eta \kappa \Delta t \hat{L}_{Q}^2(t_k)
                - \eta \kappa \Delta t \Bigsl \langle
                        \hat{L}_{Q}(t_k)
                    \Bigsr \rangle^2
                - \sqrt{\Delta t} /2
            \bigg]
\nonumber \\
    &= \Tilde{N}_{k}
        \exp \bigg[
                \sqrt{\eta \kappa} \hat{L}_I \Delta W_{I}(t_k)
\nonumber \\[-1mm]
    & \ \ \ \ \ \ \ \ \ \ \ \ \ \ 
                + 2 \eta \kappa \Delta t 
                    \Bigsl \langle
                        \hat{L}_{I}(t_k)
                    \Bigsr \rangle
                    \hat{L}_{I}(t_k)
                - \eta \kappa \Delta t \hat{L}_{I}^2(t_k)
        \bigg]
\nonumber \\
    & \ \ \ \ \ \
        \times \exp \bigg[
                \sqrt{\eta \kappa} \hat{L}_Q \Delta W_{Q}(t_k)
\nonumber \\[-1mm]
    & \ \ \ \ \ \ \ \ \ \ \ \ \ \ 
                + 2 \eta \kappa \Delta t 
                    \Bigsl \langle
                        \hat{L}_{Q}(t_k)
                    \Bigsr \rangle
                    \hat{L}_{Q}(t_k)
                - \eta \kappa \Delta t \hat{L}_{Q}^2(t_k)
        \bigg],
\end{align}
where we have used It\^{o}'s rule and lumped all scalar terms into $\Tilde{N}_k$. We have also adopted the notation $\Bigsl \langle \hat{A} \Bigsr \rangle = \Tr\Bigsl[\hat{\rho}_{\mathcal{S}}(t) \hat{A}\Bigsr]$. In addition, note that we do not need to know the value of $\Tilde{N}_k$ because it appears in both the numerator and denominator of Eq.(\ref{eq:qutrit_state_update_rule_main}) and will thus be canceled.

At this point, there is not much simplification possible, but the math is also quite straightforward. After some algebra and several applications of It\^{o}'s rule, we arrive at 
\begin{align}
    &\hat{\rho}_{\mathcal{S}}(t_{k+1})
\nonumber \\
    &= \frac{\hat{K}_{I_k Q_k}(t_k) \hat{\rho}_{\mathcal{S}}(t_k) \hat{K}_{I_k Q_k}^{\dagger}(t_k)}{\Tr[\hat{K}_{I_k Q_k} (t_k)\hat{\rho}_{\mathcal{S}}(t_k) \hat{K}_{I_k Q_k}^{\dagger}(t_k)] }
\nonumber \\
    &= \hat{\rho}_{\mathcal{S}}(t_{k})
        + \eta \kappa
            \bigg[ 
                \hat{L}_{I}(t_k) 
                    \hat{\rho}_{\mathcal{S}}(t_k)  
                    \hat{L}_{I}(t_k)
\nonumber \\
    & \ \ \ \ \ \ \ \ \ \ \ \ \ \ \ \ 
                - \frac{1}{2} 
                    \hat{L}_{I}^2(t_k) 
                    \hat{\rho}_{\mathcal{S}}(t_k)
                - \frac{1}{2} 
                    \hat{\rho}_{\mathcal{S}}(t_k)
                    \hat{L}_{I}^2(t_k) 
            \bigg]
            \Delta t
\nonumber \\
    & \ \ \ \
        + \sqrt{\eta \kappa} 
            \bigg[ 
                \hat{L}_{I}(t_k) 
                    \hat{\rho}_{\mathcal{S}}(t_k) 
                + \hat{\rho}_{\mathcal{S}}(t_k)
                    \hat{L}_{I}(t_k) 
\nonumber \\
    & \ \ \ \ \ \ \ \  \ \ \ \ \ \ \ \ 
                - 2 \Bigsl \langle
                        \hat{L}_{I}(t_k)
                    \Bigsr \rangle 
                    \hat{\rho}_{\mathcal{S}}(t_k)
            \bigg]
            \Delta W_{I}(t_k)
\nonumber \\
    & \ \ \ \ 
        + \eta \kappa
            \bigg[ 
                \hat{L}_{Q}(t_k) \hat{\rho}_{\mathcal{S}}(t_k)
                    \hat{L}_{Q}(t_k)
\nonumber \\
    & \ \ \ \ \ \ \ \  \ \ \ \ 
                - \frac{1}{2} 
                    \hat{L}_{Q}^2(t_k) 
                    \hat{\rho}_{\mathcal{S}}(t_k)
                - \frac{1}{2} 
                    \hat{\rho}_{\mathcal{S}}(t_k)
                    \hat{L}_{Q}^2(t_k) 
            \bigg]
            \Delta t
\nonumber \\
    & \ \ \ \ 
        + \sqrt{\eta \kappa} 
            \bigg[ 
                \hat{L}_{Q}(t_k) 
                    \hat{\rho}_{\mathcal{S}}(t_k) 
                + \hat{\rho}_{\mathcal{S}}(t_k)
                    \hat{L}_{Q}(t_k) 
\nonumber \\
    & \ \ \ \ \ \ \ \  \ \ \ \ \ \ \ \ 
                - 2 \Bigsl \langle
                        \hat{L}_{Q}(t_k)
                    \Bigsr \rangle 
                    \hat{\rho}_{\mathcal{S}}(t_k)
            \bigg]
            \Delta W_{Q}(t_k)
\end{align}
to the first order in $\Delta t$. By replacing $\Delta t$ and $\Delta W(t)$ by differential $\mathrm{d} t$ and $\mathrm{d} W(t)$, respectively, and let $\mathrm{d} \hat{\rho}_{\mathcal{S}}(t) = \hat{\rho}_{\mathcal{S}}(t+\Delta t) - \hat{\rho}_{\mathcal{S}}(t)$ as $\Delta t \rightarrow 0$, we finally obtain the stochastic master equation
\begin{align}
    &\mathrm{d} \hat{\rho}_{\mathcal{S}}(t)
\nonumber \\
    &= \eta \kappa 
        \mathcalboondox{D}
                \Bigsl[
                    \hat{L}_{I}(t)
                \Bigsr]
            \hat{\rho}_{\mathcal{S}} (t) \mathrm{d} t
        + \eta \kappa 
        \mathcalboondox{D}
                \Bigsl[
                    \hat{L}_{Q}(t)
                \Bigsr]
            \hat{\rho}_{\mathcal{S}} (t) \mathrm{d} t
\nonumber \\
    &\ \ \ \ 
        + \sqrt{\eta \kappa} 
            \Big[
                    \hat{L}_{I}(t)
                        \hat{\rho}_{\mathcal{S}}(t)
                    + \hat{\rho}_{\mathcal{S}}(t)
                        \hat{L}_{I}(t)
\nonumber \\
    &\ \ \ \ \ \ \ \ \ \ \ \ \ \ \ \ \ \ \ \ \ \ \ \ \ \ \ \ 
                    - 2 \Bigsl \langle
                            \hat{L}_{I} (t)
                        \Bigsr \rangle
                        \hat{\rho}_{\mathcal{S}}(t)
                \Big]
                \mathrm{d} W_{I}(t)
\nonumber \\
    &\ \ \ \ 
        + \sqrt{\eta \kappa} 
            \Big[
                    \hat{L}_{Q}(t)
                        \hat{\rho}_{\mathcal{S}}(t)
                    + \hat{\rho}_{\mathcal{S}}(t)
                        \hat{L}_{Q}(t)
\nonumber \\
    &\ \ \ \ \ \ \ \ \ \ \ \ \ \ \ \ \ \ \ \ \ \ \ \ \ \ \ \ 
                    - 2 \Bigsl \langle
                            \hat{L}_{Q} (t)
                        \Bigsr \rangle
                        \hat{\rho}_{\mathcal{S}}(t)
                \Big]
                \mathrm{d} W_{Q}(t).
\end{align}
In an experiment, we do not have control over $W_I$ and $W_Q$; instead, we observe current/voltage-like quantities of the form
\begin{align}
    V_{I,k}
    \doteq \frac{2 I_k}{\sqrt{\Delta t}}
    &= \sqrt{\eta \kappa} 
            \Bigsl \langle
                2 \hat{L}_{I} (t_k)
            \Bigsr \rangle
        + \frac{\Delta W_{I}(t_k)}{\Delta t},
\\
    V_{Q,k}
    \doteq \frac{2 Q_k}{\sqrt{\Delta t}}
    &= \sqrt{\eta \kappa} 
            \Bigsl \langle
                2 \hat{L}_{Q} (t_k)
            \Bigsr \rangle
        + \frac{\Delta W_{Q}(t_k)}{\Delta t},
\end{align} 
which, in the continuous limit, become
\begin{align}\label{eq:voltage_I_def}
    V_{I}(t)
    &= \sqrt{\eta \kappa} 
            \Bigsl \langle
                2 \hat{L}_{I} (t)
            \Bigsr \rangle
        + \xi_{I}(t),
\\ \label{eq:voltage_Q_def}
    V_{Q}(t)
    &= \sqrt{\eta \kappa} 
            \Bigsl \langle
                2 \hat{L}_{Q} (t)
            \Bigsr \rangle
        + \xi_{Q}(t),
\end{align} 
where $\xi_{I}(t) = \dot{W}_I(t)$ and $\xi_{Q}(t) = \dot{W}_Q(t)$ are classical white-noise signals defined by its expectation and autocorrelation
\begin{align}
    &\mathbb{E}[\xi_{I}(t)] 
    = \mathbb{E}[\xi_{Q}(t)] 
    = \mathbb{E}[\xi_{I}(t)\xi_{Q}(t')] 
    = 0,
\\
    &\mathbb{E}[\xi_{I}(t)\xi_{I}(t')]
    = \mathbb{E}[\xi_{Q}(t)\xi_{Q}(t')]
    = \delta(t-t').
\end{align}
Since the ensemble average of the white noise is zero, Eq.(\ref{eq:voltage_I_def}) and (\ref{eq:voltage_Q_def}) formally justify the reason why we can determine the qutrit state, which is encoded in $\Bigsl \langle \hat{L}_{I} (t) \Bigsr \rangle$ and $\Bigsl \langle \hat{L}_{Q} (t) \Bigsr \rangle$, by making measurement of $V_I$ and $V_Q$.

There is still one detail to be corrected, which is the fact that we have set $\eta = 1$ in the above derivation. Following the standard treatment of detection inefficiency (i.e., $0 \leq \eta < 1$), we introduce two new Wiener processes $W_{I}'(t)$ and $W_{Q}'(t)$ such that the efficiencies associated with them are both set to be $(1-\eta)$. In addition, the four Wiener processes should be independent (think of $(W_I, W_I')$ and $(W_Q, W_Q')$ as two Poisson branching processes but in the diffusive limit). Consequently, the stochastic master equation has four stochastic terms
\begin{align}
    &\mathrm{d} \hat{\rho}_{\mathcal{S}}
\nonumber \\
    &= \eta \kappa 
        \mathcalboondox{D}
                \Bigsl[
                    \hat{L}_{I}
                \Bigsr]
            \hat{\rho}_{\mathcal{S}} \mathrm{d} t
        + (1-\eta) \kappa 
        \mathcalboondox{D}
                \Bigsl[
                    \hat{L}_{I}
                \Bigsr]
            \hat{\rho}_{\mathcal{S}} \mathrm{d} t
\nonumber \\
    & \ \ \ \ 
        + \eta \kappa 
        \mathcalboondox{D}
                \Bigsl[
                    \hat{L}_{Q}
                \Bigsr]
            \hat{\rho}_{\mathcal{S}} \mathrm{d} t
        + (1-\eta) \kappa 
        \mathcalboondox{D}
                \Bigsl[
                    \hat{L}_{Q}
                \Bigsr]
            \hat{\rho}_{\mathcal{S}} \mathrm{d} t
\nonumber \\
    &\ \ \ \ 
        + \sqrt{\eta \kappa} 
            \Big[
                    \hat{L}_{I}
                        \hat{\rho}_{\mathcal{S}}
                    + \hat{\rho}_{\mathcal{S}}
                        \hat{L}_{I}
                    - 2 \Bigsl \langle
                            \hat{L}_{I}
                        \Bigsr \rangle
                        \hat{\rho}_{\mathcal{S}}
                \Big]
                \mathrm{d} W_{I}
\nonumber \\
    &\ \ \ \ 
        + \sqrt{\eta \kappa} 
            \Big[
                    \hat{L}_{Q}
                        \hat{\rho}_{\mathcal{S}}
                    + \hat{\rho}_{\mathcal{S}}
                        \hat{L}_{Q}
                    - 2 \Bigsl \langle
                            \hat{L}_{Q}
                        \Bigsr \rangle
                        \hat{\rho}_{\mathcal{S}}
                \Big]
                \mathrm{d} W_{Q}
\nonumber \\
    &\ \ \ \ 
        + \sqrt{(1-\eta) \kappa} 
            \Big[
                    \hat{L}_{I}
                        \hat{\rho}_{\mathcal{S}}
                    + \hat{\rho}_{\mathcal{S}}
                        \hat{L}_{I}
                    - 2 \Bigsl \langle
                            \hat{L}_{I}
                        \Bigsr \rangle
                        \hat{\rho}_{\mathcal{S}}
                \Big]
                \mathrm{d} W_{I}
\nonumber \\
    &\ \ \ \ 
        + \sqrt{(1-\eta) \kappa} 
            \Big[
                    \hat{L}_{Q}
                        \hat{\rho}_{\mathcal{S}}
                    + \hat{\rho}_{\mathcal{S}}
                        \hat{L}_{Q}
                    - 2 \Bigsl \langle
                            \hat{L}_{Q}
                        \Bigsr \rangle
                        \hat{\rho}_{\mathcal{S}}
                \Big]
                \mathrm{d} W_{Q}'.
\end{align}
where the information encoded in the last two terms is assumed to be lost in the measurement; thus, marginalizing $W_{I}'(t)$ and $W_{Q}'(t)$ yields
\begin{align}
    \!\! \mathrm{d} \hat{\rho}_{\mathcal{S}}
    &= \kappa 
        \mathcalboondox{D}
                \Bigsl[
                    \hat{L}_{I}
                \Bigsr]
            \hat{\rho}_{\mathcal{S}} \mathrm{d} t
        + \kappa 
        \mathcalboondox{D}
                \Bigsl[
                    \hat{L}_{Q}
                \Bigsr]
            \hat{\rho}_{\mathcal{S}} \mathrm{d} t
\nonumber \\
        & \ \ \ \ 
        + \sqrt{\eta \kappa} 
            \Big[
                    \hat{L}_{I}
                        \hat{\rho}_{\mathcal{S}}
                    + \hat{\rho}_{\mathcal{S}}
                        \hat{L}_{I}
                    - 2 \Bigsl \langle
                            \hat{L}_{I}
                        \Bigsr \rangle
                        \hat{\rho}_{\mathcal{S}}
                \Big]
                \mathrm{d} W_{I}
\nonumber \\
    & \ \ \ \ 
        + \sqrt{\eta \kappa} 
            \Big[
                    \hat{L}_{Q}
                        \hat{\rho}_{\mathcal{S}}
                    + \hat{\rho}_{\mathcal{S}}
                        \hat{L}_{Q}
                    - 2 \Bigsl \langle
                            \hat{L}_{Q}
                        \Bigsr \rangle
                        \hat{\rho}_{\mathcal{S}}
                \Big]
                \mathrm{d} W_{Q}.
\end{align}
Note that despite reducing the amount of information gained in an imperfect measurement, the qubit still dephases as if the efficiency is 1.

So far, we have been ignoring all the other decoherence channels of the qutrit. We can generalize the stochastic master equation by simply adding the qutrit decay and pure dephasing terms. In addition, by realizing that
\begin{align}
    & \kappa 
        \mathcalboondox{D}
                \Bigsl[
                    \hat{L}_{I}
                \Bigsr]
            \hat{\rho}_{\mathcal{S}} 
        + \kappa 
        \mathcalboondox{D}
                \Bigsl[
                    \hat{L}_{Q}
                \Bigsr]
            \hat{\rho}_{\mathcal{S}}
\nonumber\\
    &= \frac{\kappa |\beta_{ge}|^2}{4} 
            \mathcalboondox{D}
                \Bigsl[
                    \hat{\sigma}_{z,ge}
                \Bigsr]
            \hat{\rho}_{\mathcal{S}} 
        + \frac{\kappa |\beta_{gf}|^2}{4} 
            \mathcalboondox{D}
                \Bigsl[
                    \hat{\sigma}_{z,gf}
                \Bigsr]
            \hat{\rho}_{\mathcal{S}} 
\nonumber \\
    & \ \ \ \ 
        + \frac{\kappa |\beta_{ef}|^2}{4} 
            \mathcalboondox{D}
                \Bigsl[
                    \hat{\sigma}_{z,ef}
                \Bigsr]
            \hat{\rho}_{\mathcal{S}} 
% \nonumber\\
%     &= \frac{\Gamma_{\text{m}, ge}}{4} 
%                 \mathcalboondox{D}
%                     \Bigsl[
%                         \hat{\sigma}_{z,ge}
%                     \Bigsr]
%                 \hat{\rho}_{\mathcal{S}} 
%             + \frac{\Gamma_{\text{m}, gf}}{4} 
%                 \mathcalboondox{D}
%                     \Bigsl[
%                         \hat{\sigma}_{z,gf}
%                     \Bigsr]
%                 \hat{\rho}_{\mathcal{S}} 
\nonumber \\
    & \ \ \ \ 
            + \frac{\Gamma_{\text{m}, ef}}{4} 
                \mathcalboondox{D}
                    \Bigsl[
                        \hat{\sigma}_{z,ef}
                    \Bigsr]
                \hat{\rho}_{\mathcal{S}},
\end{align}
we obtain
\begin{align}
    \!\!\!\! \mathrm{d} \hat{\rho}_{\mathcal{S}}
    &= \bigg(
        - \frac{\ci}{\hbar}
            \Big[ 
                \hat{H}_{\text{q,eff}}, 
                \hat{\rho}_{\mathcal{S}}
            \Big] 
        + \gamma_{1,ge} 
            \mathcalboondox{D}
                \Bigsl[
                    \hat{\sigma}_{ge}
                \Bigsr] 
            \hat{\rho}_{\mathcal{S}} 
\nonumber \\
    &  \ \ \ \ \ \ \ \ 
        + \gamma_{1,gf} 
            \mathcalboondox{D}
                \Bigsl[
                    \hat{\sigma}_{gf}
                \Bigsr] 
            \hat{\rho}_{\mathcal{S}} 
        + \gamma_{1,ef} 
            \mathcalboondox{D}
                \Bigsl[
                    \hat{\sigma}_{ef}
                \Bigsr] 
            \hat{\rho}_{\mathcal{S}} 
\nonumber \\
    & \ \ \ \ \ \ \ \ 
        + \frac{\gamma_{\phi,ge}}{2} 
            \mathcalboondox{D}
                \Bigsl[
                    \hat{\sigma}_{z,ge}
                \Bigsr]
            \hat{\rho}_{\mathcal{S}}
        + \frac{\gamma_{\phi,gf}}{2} 
            \mathcalboondox{D}
                \Bigsl[
                    \hat{\sigma}_{z,gf}
                \Bigsr]
            \hat{\rho}_{\mathcal{S}}
\nonumber \\
    &  \ \ \ \ \ \ \ \ 
        + \frac{\gamma_{\phi,ef}}{2} 
            \mathcalboondox{D}
                \Bigsl[
                    \hat{\sigma}_{z,ef}
                \Bigsr]
            \hat{\rho}_{\mathcal{S}}
        \bigg)
        \mathrm{d} t
\nonumber \\
    &  \ \ \ \ 
        + \bigg(
            \frac{\Gamma_{\text{m}, ge}}{4} 
                \mathcalboondox{D}
                    \Bigsl[
                        \hat{\sigma}_{z,ge}
                    \Bigsr]
                \hat{\rho}_{\mathcal{S}} 
            + \frac{\Gamma_{\text{m}, gf}}{4} 
                \mathcalboondox{D}
                    \Bigsl[
                        \hat{\sigma}_{z,gf}
                    \Bigsr]
                \hat{\rho}_{\mathcal{S}} 
\nonumber \\
    &  \ \ \ \ \ \ \ \ 
            + \frac{\Gamma_{\text{m}, ef}}{4} 
                \mathcalboondox{D}
                    \Bigsl[
                        \hat{\sigma}_{z,ef}
                    \Bigsr]
                \hat{\rho}_{\mathcal{S}} 
        \bigg)
        \mathrm{d} t
\nonumber\\
    &  \ \ \ \
            + \sqrt{\eta \kappa} 
            \mathcalboondox{M}
                    \Bigsl[
                        \hat{L}_{I}
                    \Bigsr]
                \hat{\rho}_{\mathcal{S}} 
                \mathrm{d} W_{I}
        + \sqrt{\eta \kappa} 
            \mathcalboondox{M}
                    \Bigsl[
                        \hat{L}_{Q}
                    \Bigsr]
                \hat{\rho}_{\mathcal{S}} 
                \mathrm{d} W_{Q}.
\end{align}

Furthermore, if we use the expression for $\alpha_g$, $\alpha_e$, and $\alpha_f$ at steady state, i.e., Eq.(\ref{eq:steady_state_coherent_amp_g})-(\ref{eq:steady_state_coherent_amp_f}),
we can show that
\begin{align}
    &\Gamma_{\text{m}, ge} (+\infty)
    = 2 \Gamma_{\text{d}, ge}(+\infty)
\nonumber \\
    &= \frac{\kappa |\epsilon|^2 \chi_{\text{qr}}}{[\Delta_{\text{rd}}^2 + (\kappa/2)^2][(\Delta_{\text{rd}} + \chi_{\text{qr}})^2 + (\kappa/2)^2]},
\\[2mm]
    &\Gamma_{\text{m}, gf} (+\infty)
    = 2 \Gamma_{\text{d}, gf}(+\infty)
\nonumber \\
    &= \frac{\kappa |\epsilon|^2 \chi_{\text{qr}}}{[\Delta_{\text{rd}}^2 + (\kappa/2)^2][(\Delta_{\text{rd}} + 2\chi_{\text{qr}})^2 + (\kappa/2)^2]},
\\[2mm]
    &\Gamma_{\text{m}, ef} (+\infty)
    = 2 \Gamma_{\text{d}, ef}(+\infty)
\nonumber \\
    &= \frac{\kappa |\epsilon|^2 \chi_{\text{qr}}}{[(\Delta_{\text{rd}} + \chi_{\text{qr}})^2 + (\kappa/2)^2][(\Delta_{\text{rd}} + 2\chi_{\text{qr}})^2 + (\kappa/2)^2]},
\end{align}
and, thus, at steady state, the heterodyne measurement indeed induces a dephasing at $\Gamma_{\text{d}, ab}$ between any two energy levels of the qutrit. In summary, by replacing $\Gamma_{\text{m}, ab}(t)$ by $2\Gamma_{\text{d}, ab}(t)$ in the decoherence terms, we arrive at the effective qutrit stochastic master equation
\begin{align}
    \mathrm{d} \hat{\rho}_{\mathcal{S}}
    &= \bigg(
        - \frac{\ci}{\hbar}
            \Big[ 
                \hat{H}_{\text{q,eff}}, 
                \hat{\rho}_{\mathcal{S}}
            \Big] 
        + \gamma_{1,ge} 
            \mathcalboondox{D}
                \Bigsl[
                    \hat{\sigma}_{ge}
                \Bigsr] 
            \hat{\rho}_{\mathcal{S}} 
\nonumber \\
    &  \ \ \ \ \ \ \ \ 
        + \gamma_{1,gf} 
            \mathcalboondox{D}
                \Bigsl[
                    \hat{\sigma}_{gf}
                \Bigsr] 
            \hat{\rho}_{\mathcal{S}} 
        + \gamma_{1,ef} 
            \mathcalboondox{D}
                \Bigsl[
                    \hat{\sigma}_{ef}
                \Bigsr] 
            \hat{\rho}_{\mathcal{S}} 
\nonumber \\
    &  \ \ \ \ \ \ \ \ 
        + \frac{\gamma_{\phi,ge} + \Gamma_{\text{d},ge}}{2} 
            \mathcalboondox{D}
                \Bigsl[
                    \hat{\sigma}_{z,ge}
                \Bigsr]
            \hat{\rho}_{\mathcal{S}}
\nonumber \\
    &  \ \ \ \ \ \ \ \ 
        + \frac{\gamma_{\phi,gf}+ \Gamma_{\text{d},gf}}{2} 
            \mathcalboondox{D}
                \Bigsl[
                    \hat{\sigma}_{z,gf}
                \Bigsr]
            \hat{\rho}_{\mathcal{S}}
\nonumber \\
    &  \ \ \ \ \ \ \ \ 
        + \frac{\gamma_{\phi,ef}+ \Gamma_{\text{d},ef}}{2} 
            \mathcalboondox{D}
                \Bigsl[
                    \hat{\sigma}_{z,ef}
                \Bigsr]
            \hat{\rho}_{\mathcal{S}}
        \bigg)
        \mathrm{d} t
\nonumber\\ \label{eq:SME_final_appendix}
    &  \ \ \ \ 
        + \sqrt{\eta \kappa} 
            \mathcalboondox{M}
                    \Bigsl[
                        \hat{L}_{I}
                    \Bigsr]
                \hat{\rho}_{\mathcal{S}} 
                \mathrm{d} W_{I}
        + \sqrt{\eta \kappa} 
            \mathcalboondox{M}
                    \Bigsl[
                        \hat{L}_{Q}
                    \Bigsr]
                \hat{\rho}_{\mathcal{S}} 
                \mathrm{d} W_{Q}.
\end{align}

\subsection{Derivation of the Qudit SME using the Displaced Frame}\label{subsection:full_SME_derivation}

In general, a resonator subject to the heterodyne detection (in the diffusive limit) is described by the SME \cite{wiseman2010quantum}
\begin{align}
    \mathrm{d} \hat{\rho}_{\mathcal{R}} 
    &= - \ci \Big[ 
                \hat{H}_{\mathcal{R}},
                \hat{\rho}_{\mathcal{R}} 
            \Big]
            \mathrm{d} t 
        + \mathcalboondox{D}
                    \Bigsl[
                        \hat{a}
                    \Bigsr]
            \hat{\rho}_{\mathcal{R}} 
            \mathrm{d} t 
\nonumber\\ \label{eq:SME_res_heterodyne}
    & \ \ \ \
        + \sqrt{\eta \kappa} 
            \mathcalboondox{M}
                    \Bigsl[
                        \hat{a}
                    \Bigsr]
            \hat{\rho}_{\mathcal{R}} 
            \mathrm{d} W_I
        + \sqrt{\eta \kappa} 
            \mathcalboondox{M}
                    \Bigsl[
                        - \ci \hat{a}
                    \Bigsr]
            \hat{\rho}_{\mathcal{R}} 
            \mathrm{d} W_Q,
\end{align}
where $\hat{H}_{\mathcal{R}}$ is the resonator Hamiltonian and $\eta$ includes the halved efficiency due to the IQ demodulation. As before, $W_I$ and $W_Q$ are two independent (real) Wiener processes. In addition, the associated heterodyne records are governed by
\begin{align}
    V_{I} (t) 
    &= \sqrt{\eta \kappa} 
        \Bigsl\langle 
            \hat{a} + \hat{a}^{\dagger}
        \Bigsr\rangle (t) 
        + \xi_{I}(t) ,
\\
    V_{Q} (t) 
    &= \sqrt{\eta \kappa} 
        \Bigsl\langle 
            - \ci (\hat{a} - \hat{a}^{\dagger})
        \Bigsr\rangle (t) 
        + \xi_{Q}(t) ,
\end{align}
where $\xi_{I} = \dot{W}_I$ and $\xi_{Q} = \dot{W}_Q$. 

Eq.(\ref{eq:SME_res_heterodyne}) is only the resonator part; the decoherence of the qutrit and the qutrit-resonator coupling can be added directly, resulting in
\begin{align}
    \mathrm{d}\hat{\rho}
    &= \bigg(
        - \frac{\ci}{\hbar}
            \Big[ 
                \hat{H}_{\text{eff}}, 
                \hat{\rho}
            \Big]
        + \kappa 
            \mathcalboondox{D}\big[\hat{a}\big] 
            \hat{\rho}
        + \gamma_{1,ge} 
            \mathcalboondox{D}
                \Bigsl[
                    \hat{\sigma}_{ge}
                \Bigsr] 
            \hat{\rho}
\nonumber \\
    & \ \ \ \  \ \ \ \
        + \gamma_{1,gf} 
            \mathcalboondox{D}
                \Bigsl[
                    \hat{\sigma}_{gf}
                \Bigsr] 
            \hat{\rho}
        + \gamma_{1,ef} 
            \mathcalboondox{D}
                \Bigsl[
                    \hat{\sigma}_{ef}
                \Bigsr] 
            \hat{\rho}
\nonumber \\
    & \ \ \ \  \ \ \ \
        + \frac{\gamma_{\phi,ge}}{2} 
            \mathcalboondox{D}
                \Bigsl[
                    \hat{\sigma}_{z,ge}
                \Bigsr]
            \hat{\rho}
        + \frac{\gamma_{\phi,gf}}{2} 
            \mathcalboondox{D}
                \Bigsl[
                    \hat{\sigma}_{z,gf}
                \Bigsr]
            \hat{\rho}
\nonumber \\ 
    & \ \ \ \  \ \ \ \
        + \frac{\gamma_{\phi,ef}}{2} 
            \mathcalboondox{D}
                \Bigsl[
                    \hat{\sigma}_{z,ef}
                \Bigsr]
            \hat{\rho}
    \bigg) \mathrm{d}t
\nonumber \\ \label{eq:qutrit_res_full_sme}
    & \ \ \ \
        + \sqrt{\eta \kappa} 
            \mathcalboondox{M}
                    \Bigsl[
                        \hat{a}
                    \Bigsr]
            \hat{\rho} \, 
            \mathrm{d} W_I
        + \sqrt{\eta \kappa} 
            \mathcalboondox{M}
                    \Bigsl[
                        - \ci \hat{a}
                    \Bigsr]
            \hat{\rho} \,
            \mathrm{d} W_Q.
\end{align}
For simplicity, we have denoted $\hat{\rho} = \hat{\rho}_{\mathcal{SR}}$ as the state of the combined system conditioned on the measurement record. To find an effective qutrit SME, we use the displacement operator introduced in Eq.(\ref{eq:dispaced_frame_transformation}). Since we have already solved the time evolution of the density operator in the displaced frame without the heterodyne measurement, we only need to deal with terms that come from the two measurement superoperators:
\begin{align}
    \mathrm{d}\hat{\rho}^{\mathsf{P}}
    &= \Big[ 
            \text{RHS of Eq.(\ref{eq:displaced_frame_me_full})}
        \Big] \mathrm{d} t
\nonumber \\
    &\ \ \ \ 
        + \sqrt{\eta \kappa}
            \Big[
            \Bigsl(
                \hat{a} \hat{\rho}^{\mathsf{P}}
                + \hat{\rho}^{\mathsf{P}} \hat{a}
            \Bigsr)
            + \Bigsl(
                \hat{\Pi}_{\alpha}  
                    \hat{\rho}^{\mathsf{P}}
                + \hat{\rho}^{\mathsf{P}} 
                    \hat{\Pi}_{\alpha}^{\dagger}
            \Bigsr)
\nonumber \\
    &\ \ \ \  \ \ \ \  \ \ \ \  \ \ \ \ 
            - \Bigsl \langle 
                    \hat{a}
                    + \hat{a}^{\dagger} 
                \Bigsr \rangle^{\mathsf{P}}
                \hat{\rho}^{\mathsf{P}}
            - \Bigsl \langle 
                    \hat{\Pi}_{\alpha} 
                    + \hat{\Pi}_{\alpha}^{\dagger} 
                \Bigsr \rangle^{\mathsf{P}}
                \hat{\rho}^{\mathsf{P}}
            \Big] 
            \mathrm{d} W_I
\nonumber \\
    &\ \ \ \ 
        + \sqrt{\eta \kappa}
            \Big[
            -\ci \Bigsl(
                \hat{a} \hat{\rho}^{\mathsf{P}}
                - \hat{\rho}^{\mathsf{P}} \hat{a}
            \Bigsr)
            - \ci \Bigsl(
                \hat{\Pi}_{\alpha}  
                    \hat{\rho}^{\mathsf{P}}
                - \hat{\rho}^{\mathsf{P}} 
                    \hat{\Pi}_{\alpha}^{\dagger}
            \Bigsr)
\nonumber \\ \label{eq:qutrit_res_full_sme_displaced}
    &\ \ \ \  \ \ \ \  \ \ \ \  \ \ \ \ 
            + \ci \Bigsl \langle 
                    \hat{a}
                    - \hat{a}^{\dagger} 
                \Bigsr \rangle^{\mathsf{P}}
                \hat{\rho}^{\mathsf{P}}
            + \ci \Bigsl \langle 
                    \hat{\Pi}_{\alpha} 
                    - \hat{\Pi}_{\alpha}^{\dagger} 
                \Bigsr \rangle^{\mathsf{P}}
                \hat{\rho}^{\mathsf{P}}
            \Big] 
            \mathrm{d} W_Q
\end{align}
Note that $\langle \hat{c}\rangle^{\mathsf{P}} = \Tr\big(\hat{c}\hat{\rho}^{\mathsf{P}}\big)$ is the expectation value of $\hat{c}$ in the displaced frame.

Following the same procedure of tracing out the resonator subspace shown in Appendix \ref{Appendix:C}, we obtain, for example, 
\begin{align}
    &\mathrm{d} \rho_{\mathcal{S},gg}
\nonumber \\
    &= \bigg(
            \sum_{n}
                \dot{\rho}^{\mathsf{P}}_{nngg}
        \bigg) 
        \mathrm{d} t
\nonumber \\
    &= \gamma_{1, ge} \rho_{\mathcal{S},ee} 
            \mathrm{d} t
        + \gamma_{1, gf} \rho_{\mathcal{S},ff} 
            \mathrm{d} t
\nonumber \\
    & \ \ \ \ 
        + 2 \sqrt{\eta \kappa} 
            \Bigg[
                \Re(\alpha_g) 
                - \!\! 
                    \sum_{a\in\{g,e,f\}}
                    \!\! 
                    \Re(\alpha_g) 
                    \rho_{\mathcal{S},aa}
            \Bigg] \rho_{\mathcal{S},gg} \mathrm{d} W_I
\nonumber \\ \label{eq:qutrit_population_gg_with_measurement}
    & \ \ \ \ 
        + 2 \sqrt{\eta \kappa} 
            \Bigg[
                \Im(\alpha_g) 
                - \!\! 
                    \sum_{a\in\{g,e,f\}}
                    \!\! 
                    \Im(\alpha_g) 
                    \rho_{\mathcal{S},aa}
            \Bigg] \rho_{\mathcal{S},gg} \mathrm{d} W_Q,
\end{align}
where we have used the fact that $\rho^{\mathsf{P}}_{nm ab} = 0$ with $n,m>0$ cannot be excited in the displaced frame to remove terms of the form
\begin{equation}
    \sum_{n} \sqrt{n+1} \rho_{(n+1)naa}^{\mathsf{P}}
    \ \ \text{ and } \ \ 
    \sum_{n} \sqrt{n+1} \rho_{n(n+1)aa}^{\mathsf{P}}.
\end{equation}
The decay terms in Eq.(\ref{eq:qutrit_population_gg_with_measurement}) are exactly the same as the ones in Eq.(\ref{eq:qutrit_population_gg}) and the newly appeared stochastic terms match the matrix element
$$\bra{g} \Big( \sqrt{\eta \kappa} \mathcalboondox{M}\Bigsl[\hat{L}_{I} \Bigsr] \hat{\rho}_{\mathcal{S}} \mathrm{d} W_{I} + \sqrt{\eta \kappa} \mathcalboondox{M} \Bigsl[ \hat{L}_{Q} \Bigsr] \hat{\rho}_{\mathcal{S}} \mathrm{d} W_{Q} \Big) \ket{g}.$$ It is not hard to see that the derivatives of $\rho_{\mathcal{S},ee}$ and $\rho_{\mathcal{S},ff}$ take a similar form.

Next, the stochastic differential equations of the off-diagonal terms of $\hat{\rho}_{\mathcal{S}}$ can be computed similarly. To avoid complicated algebra, we need to apply, again, the fact that ${\lambda}_{nmpq}^{ab} \propto \rho^{\mathsf{P}}_{nm ab}= 0$ with $n,m,p,q >0$, which yields
\begin{align}
    &\mathrm{d} \rho_{\mathcal{S},ge} 
\nonumber \\
    &= \Big[ \text{RHS of Eq.(\ref{eq:lambda_0000_ge})}\Big] \mathrm{d} t
\nonumber \\
    & \ \ 
        + \!\sqrt{\eta \kappa} 
            \Bigg[ 
                (\alpha_g + \alpha_e^*)
                - \!\!\!
                    \sum_{a\in\{g,e,f\}}
                    \!\!\!
                    2 \Re(\alpha_g) 
                    \rho_{\mathcal{S},aa}
            \Bigg] \rho_{\mathcal{S},ge} \mathrm{d} W_I
\nonumber \\ \label{eq:rho_S_ge_sme}
    & \ \  
        + \!\sqrt{\eta \kappa} 
            \Bigg[ 
                \ci (- \alpha_g + \alpha_e^*)
                - \!\!\! 
                    \sum_{a\in\{g,e,f\}}
                    \!\!\! 
                    2 \Im(\alpha_g) 
                    \rho_{\mathcal{S},aa}
            \Bigg] \rho_{\mathcal{S},ge} \mathrm{d} W_Q.
\end{align}
Hence, by starting from the SME of the combined system, we have shown that the qutrit acquires the measurement-induced dephasing rate $\Gamma_{\text{d}, ge}$ after taking the ensemble average (i.e., ignoring the stochastic terms in Eq.(\ref{eq:rho_S_ge_sme})). Note, however, that the stochastic terms in Eq.(\ref{eq:rho_S_ge_sme}) do not equal $$\bra{g} \Big( \sqrt{\eta \kappa} \mathcalboondox{M}\Bigsl[\hat{L}_{I} \Bigsr] \hat{\rho}_{\mathcal{S}} \mathrm{d} W_{I} + \sqrt{\eta \kappa} \mathcalboondox{M} \Bigsl[ \hat{L}_{Q} \Bigsr] \hat{\rho}_{\mathcal{S}} \mathrm{d} W_{Q} \Big) \ket{e}.$$ Instead, we observe that $$\bra{g} \Big( \sqrt{\eta \kappa} \mathcalboondox{M}\Bigsl[\hat{L}_{I}' \Bigsr] \hat{\rho}_{\mathcal{S}} \mathrm{d} W_{I} + \sqrt{\eta \kappa} \mathcalboondox{M} \Bigsl[ \hat{L}_{Q}' \Bigsr] \hat{\rho}_{\mathcal{S}} \mathrm{d} W_{Q} \Big) \ket{e}$$ will produce the desired terms if we let
\begin{align}
    \hat{L}_I' (t)
    &= \alpha_g(t) \hat{\Pi}_g
        + \alpha_e(t) \hat{\Pi}_e
        + \alpha_f(t) \hat{\Pi}_f ,
\\
    \hat{L}_Q' (t) 
    &= - \ci \alpha_g(t) \hat{\Pi}_g
        -\ci \alpha_e(t) \hat{\Pi}_e
        - \ci \alpha_f(t) \hat{\Pi}_f.
\end{align}
Moreover, these new measurement operators are naturally compatible with Eq.(\ref{eq:qutrit_population_gg_with_measurement}). One can simulate the effective qutrit SME using $\hat{L}_I'$ and $\hat{L}_Q'$, but the quantum trajectories will look qualitatively the same as the ones simulated with $\hat{L}_I$ and $\hat{L}_Q$.

The difference between $\Bigsl(\hat{L}_I, \hat{L}_Q\Bigsr)$ and $\Bigsl( \hat{L}_I', \hat{L}_Q'\Bigsr)$ lies in the definition of the two quadratures in the full SME. If we use
\begin{align}
    &\mathrm{d} \hat{\rho}_{\mathcal{R}} 
    = - \ci \Big[ 
                \hat{H}_{\mathcal{R}},
                \hat{\rho}_{\mathcal{R}} 
            \Big]
            \mathrm{d} t 
        + \mathcalboondox{D}
                    \Bigsl[
                        \hat{a}
                    \Bigsr]
            \hat{\rho}_{\mathcal{R}} 
            \mathrm{d} t 
\nonumber\\  \label{eq:SME_res_heterodyne_2}
    &
        + \sqrt{\eta \kappa} 
            \mathcalboondox{M}
                    \Bigg[
                        \frac{\hat{a} + \hat{a}^{\dagger}}{2}
                    \Bigg]
            \hat{\rho}_{\mathcal{R}} 
            \mathrm{d} W_I
        + \sqrt{\eta \kappa} 
            \mathcalboondox{M}
                    \Bigg[
                        \frac{\hat{a} - \hat{a}^{\dagger}}{2\ci}
                    \Bigg]
            \hat{\rho}_{\mathcal{R}} 
            \mathrm{d} W_Q
\end{align}
instead of Eq.(\ref{eq:SME_res_heterodyne}), Eq.(\ref{eq:rho_S_ge_sme}) changes to
\begin{align}
    &\mathrm{d} \rho_{\mathcal{S},ge} 
\nonumber \\
    &= \Big[ \text{RHS of Eq.(\ref{eq:lambda_0000_ge})}\Big] \mathrm{d} t
\nonumber \\
    & 
        + \!\sqrt{\eta \kappa} 
            \Bigg[ 
                \Re(\alpha_g + \alpha_e)
                - \!\! 
                    \sum_{a\in\{g,e,f\}}
                    \!\! 
                    2\Re(\alpha_g) 
                    \rho_{\mathcal{S},aa}
            \Bigg] \rho_{\mathcal{S},ge} \mathrm{d} W_I
\nonumber \\ 
    &  
        + \!\sqrt{\eta \kappa} 
            \Bigg[ 
                \Im(\alpha_g + \alpha_e)
                - \!\! 
                    \sum_{a\in\{g,e,f\}}
                    \!\! 
                    2\Im(\alpha_g) 
                    \rho_{\mathcal{S},aa}
            \Bigg] \rho_{\mathcal{S},ge} \mathrm{d} W_Q
\nonumber \\
    &= \Big[ \text{RHS of Eq.(\ref{eq:lambda_0000_ge})}\Big] \mathrm{d} t
\nonumber \\ \label{eq:rho_S_ge_sme_2}
    & \ \ 
        + \bra{g} \Big( \sqrt{\eta \kappa} \mathcalboondox{M}\Bigsl[\hat{L}_{I} \Bigsr] \hat{\rho}_{\mathcal{S}} \mathrm{d} W_{I} + \sqrt{\eta \kappa} \mathcalboondox{M} \Bigsl[ \hat{L}_{Q} \Bigsr] \hat{\rho}_{\mathcal{S}} \mathrm{d} W_{Q} \Big) \ket{e}.
\end{align}
Combining the stochastic differential equations of diagonal and off-diagonal terms gives the effective qutrit SME in Eq.(\ref{eq:sme_final_form}) of the main text, provided that we ignore the measurement-induced frequency shifts. The effective qudit SME with $D>3$ can also be derived following the same logic.

Furthermore, the heterodyne records used in the SME of the combined system should be modified to depend only on the qutrit operators. To wit, we use the displaced frame and the fact that $\rho^{\mathsf{P}}_{nm ab} = 0$ with $n,m>0$ again:
\begin{align}
    V_{I} 
    &= \sqrt{\eta \kappa} 
        \Bigsl \langle 
            \hat{a} + \hat{a}^{\dagger}
        \Bigsr \rangle^{\mathsf{P}} 
        + \sqrt{\eta \kappa} 
        \Bigsl \langle 
            \hat{\Pi}_{\alpha} + \hat{\Pi}_{\alpha}^{\dagger}
        \Bigsr \rangle^{\mathsf{P}} 
        + \xi_{I}
\nonumber \\
    &= \sqrt{\eta \kappa} 
        \Bigsl \langle 
            \hat{\Pi}_{\alpha} + \hat{\Pi}_{\alpha}^{\dagger}
        \Bigsr \rangle^{\mathsf{P}} 
        + \xi_{I}
\nonumber \\
    &= \sqrt{\eta \kappa} 
        \sum_{n} 
        \sum_{a \in \{g,e,f\}}
            (\alpha_a + \alpha_a^*) 
            \rho_{nnaa}^{\mathsf{P}}
        + \xi_{I}
\nonumber \\ \label{eq:voltage_I_def_corrected}
    &= \sqrt{\eta \kappa} 
        \Bigsl \langle 
             2 \hat{L}_{I} 
        \Bigsr \rangle
        + \xi_{I}.
\end{align}
Note that the expectation value in the last line of Eq.(\ref{eq:voltage_I_def_corrected}) is computed with respect to the qutrit state only, i.e., $\Bigsl \langle 2 \hat{L}_{I} \Bigsr \rangle = \Tr(2 \hat{L}_{I} \hat{\rho}_{\mathcal{S}})$. The other quadrature follows a similar stochastic differential equation and, in summary, we have 
\begin{equation}
    V_{I} (t) 
    = \sqrt{\eta \kappa} 
        \Bigsl \langle 
             2 \hat{L}_{I}(t)  
        \Bigsr \rangle
        + \xi_{I}(t) ,
\end{equation}
\begin{equation}
    V_{Q} (t) 
    = \sqrt{\eta \kappa} 
        \Bigsl \langle 
             2 \hat{L}_{Q}(t)  
        \Bigsr \rangle
        + \xi_{Q}(t) ,
\end{equation}
which is exactly Eq.(\ref{eq:voltage_I_def}) and (\ref{eq:voltage_Q_def}).

For simplicity, we have assumed that $\phi = 0$, i.e., the cable delay is ignored. One can replace $\hat{a}$ and $\alpha_a$ with $\hat{a}e^{-\ci \phi}$ and $\alpha_a e^{-\ci \phi}$, respectively, in all the equations above to account for a constant phase shift, which is included in the main text for generality.

% For heterodyne detection, it's often custom to use a complex Wiener process
% \begin{equation}
%     \mathrm{d}Z = \frac{dW_{I} + \ci dW_{Q}}{\sqrt{2}},
% \end{equation}
% which satisfies the properties $\mathrm{d}Z^* \mathrm{d}Z = \mathrm{d}t$ and $\mathrm{d}Z^2 = 0$. One can verify that the last two terms in Eq.(\ref{eq:SME_final_appendix}) is equivalent to
% \begin{equation}
    
% \end{equation}

% \nocite{*}

\bibliography{apssamp}% Produces the bibliography via BibTeX.

\end{document}